# A New Method for Measuring the Half-life of $^{10}$C

**Kent Kwan Ho Leung**

A thesis submitted in fulfilment of the requirements
for the degree of Masters of Science in Physics,
The University of Auckland, 2007.

# Abstract

A new measurement of the half-life of the positron emitter $^{10}$C has been made. The $^{10}$C decay belongs to the set of superallowed $0^+ \rightarrow 0^+$ $\beta$-decays whose $\mathcal{F}t$ values offer stringent tests of the Conserved Vector Current (CVC) hypothesis and which can be used in calculating $|V_{ud}|$, the largest element of the Cabibbo-Kobayashi-Maskawa (CKM) quark mixing matrix of the Standard Model. The experimental method used four plastic scintillator detectors in a 3-fold coincidence mode in order to detect the unique $2 \times 511\text{keV} + 718\text{keV}$ $\gamma$-ray decay signature of $^{10}$C. The half-life values obtained from the data were strongly dependent on the detector count rates, but a final value of $19.303 \pm 0.013$ s was determined by extrapolating the dependence to zero count rate. This study highlights the significance of systematic errors due to high count rates in precision half-life experiments and raises doubts about both the two previously accepted values of the half-life of $^{10}$C.

# Acknowledgments

First and foremost I would like to thank my supervisor, Paul Barker, without whom this thesis would not have been possible. He has been incredibly patient and understanding throughout the year. I have learned so much through his guidance and for that I am truly indebted.

I would like to thank Aidan Byrne who was most kind and helpful during my stay at the Australian National University. The support from my co-supervisor, Irene Barnett, was most welcomed. In addition, I am grateful for the assistance provided by the staff of the physics workshops, both here in Auckland and in Canberra.

I also wish to express my gratitude to all my friends and family. In particular I would like to thank Helen Nathan who has been wonderful and supportive throughout, my parents for always being there for me and backing me every step of the way, and finally, Fabienne Haupert and Michael Hopper for being great friends.

# Contents

# List of Figures

# List of Tables

# Abbreviations

| | |
|---|---|
| ADC | Analog to Digital Converter |
| ANU | Australian National University |
| BNC | Bayonet Neill-Concelman or Bayonet Nut Connector |
| CAMAC | Computer Automated Measurement And Control |
| CRD | Count Rate Dependent |
| CVC | Conserved Vector Current |
| DAQ | Data Acquisition |
| NMR | Nuclear Magnetic Resonance |
| NIM | Nuclear Instrumentation Module |
| PDF | Probability Density Function |
| PDG | Particle Data Group |
| TAC | Time to Amplitude Converter |

# Chapter 1

# Introduction

## 1.1  Motivation

This thesis describes a new method of measuring the half-life of the $^{10}$C positron ($\beta^+$) emitter which decays via:

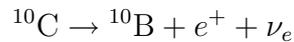

This decay contains a branch between a pair of nuclear analog states of spin parity, $J^\pi = 0^+$, and isospin triplet T=1 (see figure 1.1). These decays have been under continuous study for the past five decades as they provide access to clean tests of some of the fundamental assumptions in the weak interaction section of the Standard Model. This has led to ongoing efforts to improve the precision of both the experiment and theory required in performing these tests.

## 1.2  Application of Nuclear Beta Decays

### 1.2.1  The Conserved Vector Current (CVC) Hypothesis

The Conserved Vector Current (CVC) hypothesis asserts that the vector coupling constant for semileptonic weak interactions, $G_V$, is a true constant. That is, it is not renormalised by strong interactions that may take place in a nuclear medium. The weak interaction's conserved vector current is analogous to the electromagnetic interaction's conserved vector current which implies that the proton's electric charge is not changed when it is in a nuclear medium (the term "vector" refers to the Lorentz transformation properties of the operator corresponding to the interaction). Nuclear $\beta$ decays therefore provide a wide range of varying nuclear conditions with which to test the CVC hypothesis.

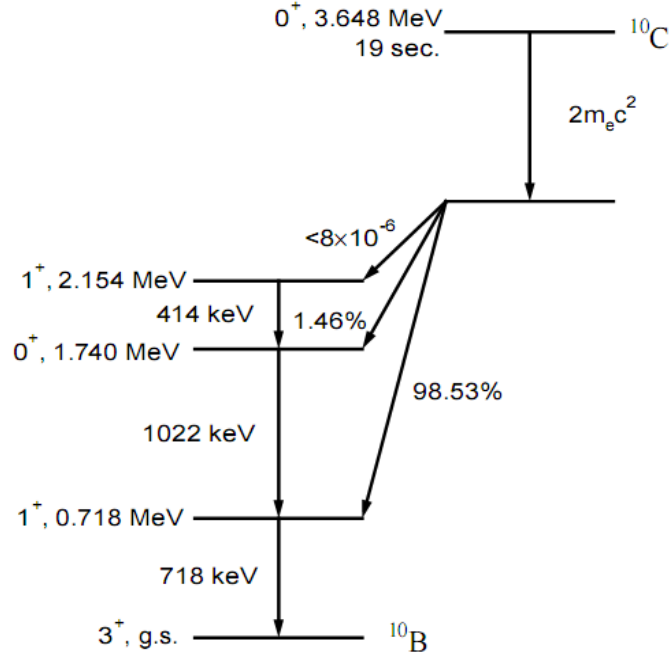

Figure 1.1: The Decay Scheme of $^{10}$C

### 1.2.2 The $ft$-value and Its Relation to $G_V$

The strength of a $\beta$-decay is measured in terms of its "$ft$-value" (or less commonly referred to as the comparative half-life). As the name suggests, the $ft$-value is obtained from the product of the statistical rate function $f$ and the partial half-life $t$.

The statistical rate function $f$ is an integral over phase space defined by:

$$f = \int_1^{W_0} pW \left(W_0 - W\right)^2 F\left(Z, W\right) S\left(Z, W\right) dW \qquad (1.1)$$

where $W$ is the $\beta$-particle energy (in electron rest-mass units), $p = (W^2 - 1)^{1/2}$ is the $\beta$-particle momentum, $Z$ is the charge of the daughter nucleus, $F(Z, W)$ is the Fermi function (which incorporates the gross effect of the Coulomb field of the daughter nucleus on the emitted $\beta$-particle), and $S(Z, W)$ is the shape-correction function. $W_0$ is the maximum value of $W$ and is related to the experimentally determined transition energy $Q_{EC}$ by:

$$W_0 = \frac{Q_{EC}}{m_e c^2} - 2 \tag{1.2}$$

For superallowed transitions, $S(Z, W)$ takes into account the effects due to the nuclear charge-density distribution, existence of second-forbidden transitions and screening effects from the atomic electrons. This is required to reduce the uncertainty from theoretical contributions in $f$ to a level much lower than that from experimental contributions[1] (which come solely from $Q_{EC}$).

The partial half-life $t$ is defined by:

$$t = \frac{t_{1/2}}{R} \left(1 + P_{EC}\right) \tag{1.3}$$

where $t_{1/2}$ and $R$ are, respectively, the experimentally determined half-life and branching ratio of the transition, and $P_{EC}$ is the calculated electron-capture fraction.

The $ft$-value, for a general $\beta$-decay, is related to a mixture of both the vector coupling constant $G_V$ and the axial-vector coupling constant $G_A$ by:

$$ft = \frac{K}{G_V^2 \langle M_V \rangle^2 + G_A^2 \langle M_A \rangle^2} \tag{1.4}$$

where $\langle M_V \rangle^2$ and $\langle M_A \rangle^2$ are the transition matrix elements of the vector and axial-vector $\beta$ decay operators and $K$ is a constant given by:

$$\frac{K}{(\hbar c)^6} = \frac{2\pi^3 \hbar \ln 2}{(m_e c^2)^5} = (8120.271 \pm 12) \times 10^{-10} \text{GeV}^{-4}\text{s} \tag{1.5}$$

For $\beta^+$ decays, the vector (or Fermi) transition refers to the case when the emitted leptons, the positron $e^+$ and the neutrino $\nu_e$ (both spin 1/2 particles), combine to form the anti-parallel spin singlet (S=0) state. The axial-vector (or Gamow-Teller) transition refers to the emitted leptons combining to form the parallel spin triplet (S=1) state. In the case of a superallowed $J^\pi = 0^+ \rightarrow 0^+$ transition, where $\Delta J = 0$, the conservation of angular momentum requires that the possibility of axial-vector transitions vanish (i.e. $\langle M_A \rangle^2 = 0$). The absence of axial-vector coupling makes studying these nuclear decays a very clean way of determining $G_V$.

The transition operator for vector beta decay is the isospin ladder operator, that is, it acts to increase the z-component of the isospin by one unit (e.g. $^{10}C(T_Z = -1) \rightarrow {}^{10}B(T_Z = 0)$), so that the transition matrix elements are simply the Clebsch-Gordan coefficient of angular momentum algebra. Therefore, in the case of $T = 0$ analog states, we get $\langle M_V \rangle^2 = 2$.

For superallowed $0^+ \rightarrow 0^+$ transitions, equation 1.4 becomes:

$$ft = \frac{K}{2G_V^2} \tag{1.6}$$

Thus, there exists a direct relationship between the $ft$-value and the vector coupling constant and only the half-life, branching ratio and transition energy are required to be measured experimentally to calculate $G_V$.

### 1.2.3  Corrections Required for Obtaining the $\mathcal{F}t$-value

Before the $ft$-values from the superallowed $0^+ \rightarrow 0^+$ beta decays can be compared, minor corrections are required to be taken into account. Firstly, there is the Coulomb or isospin-symmetry-breaking correction denoted by $\delta_C$. This is required because the charge difference between the neutron and proton means that the nuclear wave functions of the initial and final states are not identical (that is to say, the isospin symmetry between the initial and final states is broken). Physically, this comes from two effects: a) the degree of configuration mixing in the nuclear shell-model wave function varies among members of the isospin $T = 1$ triplet (i.e. transitions with $T_Z = -1 \rightarrow 0$ are different to $T_Z = 0 \rightarrow +1$), and b) a proton is typically less bound in a nucleus than a neutron; thus its radial wave function extends further from the nucleus. This reduces the radial overlap of the initial and final nuclei from the normally assumed value of one[2]. These effects (by far the larger contribution is that from (a)) combine to reduce the isospin transition matrix element:

$$\langle M_V \rangle^2 \to 2(1 - \delta_C) \tag{1.7}$$

Secondly, there are also radiative corrections which result from electromagnetic interactions with the spectator nucleons and, for example, the emission of bremsstrahlung photons (which go undetected in the experiment) by the positrons. These can be split into a transition-dependent part, denoted by $\delta_R$, and a transition-independent part, denoted by $\Delta_R^V$. These act to modify $ft$ and $G_V$ (in equation 1.6) by:

$$\begin{aligned} ft &\to ft(1 + \delta_R) \\ G_V &\to G_V(1 + \Delta_R^V) \end{aligned} \tag{1.8}$$

The transition-independent part of the radiative corrections $\Delta_R^V$ ($\sim2\%$) is the larger of the two. This correction comes primarily from short-distance loop effects (such as the exchange of a $W$ boson or a photon between the proton and positron) and is thus nucleus-independent. Note that $\Delta_R^V$ does not effect the test of the validity of the CVC hypothesis since the same correction is applied to all nuclear beta decays.

The transition-dependent part of the radiative corrections $\delta_R$ can be further splitted into two terms:

$$\delta_R = \delta_R' + \delta_{NS} \tag{1.9}$$

of which the first, $\delta_R'$, is a function of the $\beta$-particle energy only and the charge of the final nucleus Z; it therefore depends on the particular nuclear transition but is independent on the nuclear structure. The second term, $\delta_{NS}$, is dependent on the details of nuclear structure.

The $\mathcal{F}t$-value is defined to incorporate all the corrections discussed previously. Recalling that $G_V$ is the vector coupling constant and $K$ is a constant given by equation 1.5, $\mathcal{F}t$ is related to $G_V$ by:

$$\mathcal{F}t \equiv ft(1 + \delta_R')(1 + \delta_{NS} - \delta_C) = \frac{K}{2G_V^2(1 + \Delta_R^V)} \tag{1.10}$$

Note that the correction terms are separated in a different way than previously described. The equation is written in this form to emphasise the different dependence of the corrections. The first term in brackets on the left-hand side, $1 + \delta'_R$, is independent of nuclear structure, whereas the second term, $1 + \delta_{NS} - \delta_C$, is dependent of nuclear structure and requires calculations based on the details of the shell-model. The correction term, $1 + \Delta^V_R$, the same value for all $\beta^+$-decays, is kept on the right-hand side of the equation as it depends only on the details of the weak interaction.

## 1.3    Current Status of the $\mathcal{F}t$-values

A complete survey from 2005 of all half-life, decay-energy and branching-ratio measurements as well as theoretical corrections related to the set of superallowed $0^+ \rightarrow 0^+$ transitions is given in Ref.[1]. In this survey, some values had to be updated with improved calibration procedures while others were simply rejected (with justified causes). The reasons for rejecting values of half-life measurements are discussed in section 1.4.

The currently accepted $\mathcal{F}t$ values of the 12 best-studied superallowed $0^+ \rightarrow 0^+$ transitions are shown in figure 1.2. The uncertainties from experimental and theoretical input factors that contribute to each of the nine best determined $\mathcal{F}t$ values, otherwise known as the "error budget", are shown in figure 1.3.

The main contribution to the error of the $^{10}$C $\mathcal{F}t$ value is from measurements of the branching ratio. Although probably the most measured branching ratio of the superallowed $0^+ \rightarrow 0^+$ set, there exists a large experimental uncertainty in determining the small fraction of the superallowed branch (only $\sim$1.5%) to the total decay (see section 3.3 for more details on the $^{10}$C decay). The next largest contribution is from the half-life, with only two previous measurements deemed to be acceptable, followed by that from the decay-energy.

### 1.3.1    Confirmation of the CVC Hypothesis

Due to the minor corrections required (see equation 1.10), the $\mathcal{F}t$ values from superallowed $0^+ \rightarrow 0^+$ transitions should be constant if the CVC hypothesis holds. It is visually evident that this is true (figure 1.2). The weighted average and $1\sigma$ error of the 12 most precisely determined $\mathcal{F}t$ values in 2005 from Ref.[1] are:

$$\overline{\mathcal{F}t} = 3072.7 \pm 0.8\text{s}$$

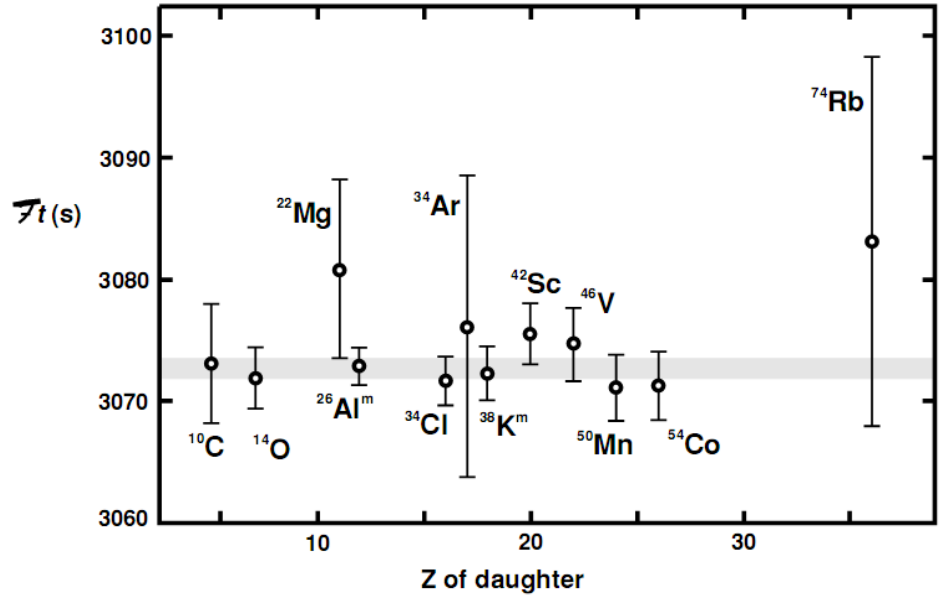

Figure 1.2: The status, in 2005 from Ref.[1], of the $\mathcal{F}t$ values plotted as a function of the charge of the daughter nucleus for the best studied superallowed $0^+ \rightarrow 0^+$ $\beta^+$ decays. The shaded horizontal band represents the weighted average $\overline{\mathcal{F}t}$ and $1\sigma$ uncertainty.

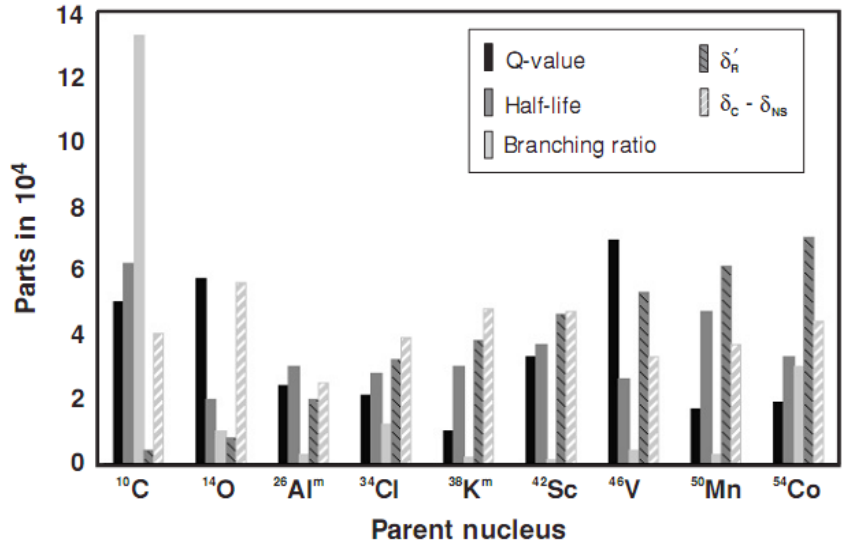

Figure 1.3: Histogram of the uncertainties contributing to the $\mathcal{F}t$ values of the nine precise studied superallowed $0^+ \rightarrow 0^+$ $\beta^+$ decays (from Ref.[1]).

The normalised chi-squared $\chi^2_\nu = 0.42$. This confirms the CVC hypothesis to a high precision of 0.03%.

## 1.3.2   CKM Matrix Unitarity

The quark eigenstates of the weak interaction contain a mixture of the quark eigenstates of the strong interaction (or mass eigenstates). In the Standard Model, the explanation of this comes from spontaneous symmetry breaking by the Higgs mechanism. The relationship between the two sets of eigenstates is given by the Cabibbo-Kobayashi-Maskawa (CKM) quark mixing matrix V, where, by convention, mixing involves the lightest members of the three quark families:

$$\begin{pmatrix} d' \\ s' \\ b' \end{pmatrix} = \begin{pmatrix} V_{ud} & V_{us} & V_{ub} \\ V_{cd} & V_{cs} & V_{cb} \\ V_{td} & V_{ts} & V_{tb} \end{pmatrix} \begin{pmatrix} d \\ s \\ b \end{pmatrix} \tag{1.11}$$

where $(d', s', b')$ are the weak eigenstates and $(d, s, b)$ the mass eigenstates. Note that none of the values in the matrix is given by the theory. They all must be determined experimentally.

The CKM matrix is expected to be unitary in the Standard Model. The most demanding test of this comes from the top row, i.e. $|V_{ud}|^2 + |V_{us}|^2 + |V_{ub}|^2 = 1$. If the matrix turned out to be apparently non-unitary, this would signify the existence of new physics beyond the Standard Model such as the existence of a fourth generation of quarks.

The first element of the CKM matrix, $V_{ud}$, gives the amount of mixing between the first generation (up and down) quarks. An illustrative way of viewing the effects of this mixing is to compare the strength of neutron decay to that of muon decay[3]:

$$n(udd) \rightarrow p(uuu) + e^- + \bar{\nu}_e$$
$$\mu^- \rightarrow e^- + \bar{\nu}_e + \bar{\nu}_\mu$$

The strength of muon decay, a purely leptonic process, is given by the intrinsic strength of the weak interaction $G_F$ (the Fermi coupling constant). The neutron decay processes, or in general, nuclear $\beta$-decays (from the CVC hypothesis), are semi-leptonic processes as they involve both quarks and leptons. The strength of a semi-leptonic process deviates from $G_F$ due to the mixing between the quarks. In the case of the superallowed $0^+ \rightarrow 0^+$ $\beta$-decay strength, where only the first generation quarks are involved, $G_V$ is modified by:

$$G_V = G_F V_{ud} \quad \Rightarrow \quad |V_{ud}|^2 = \left(\frac{G_V}{G_F}\right)^2 \tag{1.12}$$

where $G_V$ is related to $\mathcal{F}t$ from equation 1.10 by:

$$G_V^2 = \frac{K}{2\mathcal{F}t(1 + \Delta_R^V)} \tag{1.13}$$

**Current Status**

Using the Particle Data Group (PDG) value for the pure-leptonic weak interaction coupling constant from muon decay of[4]:

$$\frac{G_F}{(hc)^3} = (1.16639 \pm 0.00001) \times 10^{-5} \text{GeV}^{-2}$$

together with $\overline{\mathcal{F}t}$ from superallowed $0^+ \rightarrow 0^+$ $\beta$-decay in 2005 from Ref.[1], the value of the first element of the CKM matrix is:

$$|V_{ud}| = 0.9738 \pm 0.004 \qquad (2005) \tag{1.14}$$

When the previous value was combined with the PDG's recommended values for that time[4] of $|V_{us}| = 0.2200 \pm 0.0026$ and $|V_{ub}| = 0.00367 \pm 0.00047$,

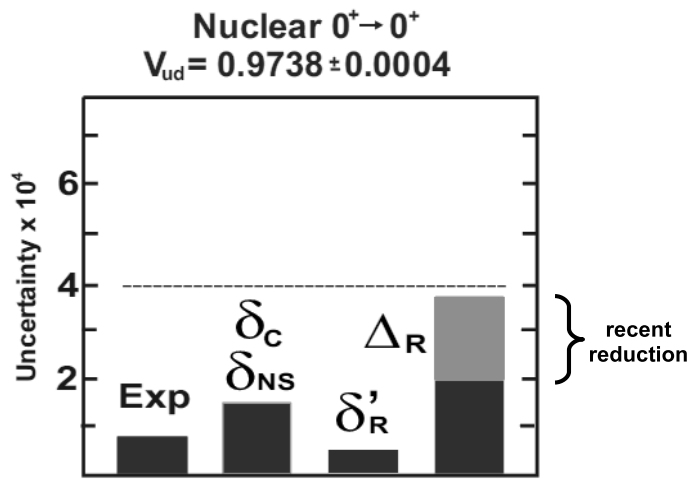

Figure 1.4: Histogram of the errors contributing to the first element of the CKM matrix $V_{ud}$ from superallowed $0^+ \rightarrow 0^+$ $\beta$-decays (from Ref.[5]).

the unitarity sum was found to be:

$$|V_{ud}|^2 + |V_{us}|^2 + |V_{ub}|^2 = 0.9966 \pm 0.0014 \qquad (2005) \qquad (1.15)$$

This failed the unitary test by 2.4 standard deviations. A nagging problem that has persisted for over a decade. The uncertainty contribution from the 2005 result was mainly from $|V_{us}|^2$ ($\sigma_{|V_{us}|^2} = 0.0011$) followed by that from $|V_{ud}|^2$ ($\sigma_{|V_{ud}|^2} = 0.0008$).

The error budget contributing to the value of $|V_{ud}|^2$ in 2005 is shown in figure 1.4. The largest contribution to the uncertainty comes from the nucleus-independent radiative correction $\Delta_R^V$. This has prompted a recent recalculation of this term (Ref.[6]) which has reduced the uncertainty to half of its previous value. This result, with the addition of new experimental measurements since 2005, gives the currently best value for $|V_{ud}|$ as[7]:

$$|V_{ud}| = 0.97370 \pm 0.00027 \qquad (2007) \qquad (1.16)$$

The $|V_{us}|$ value has also been recently improved. In 2005, two new measurements of this quantity gave a result that was found to be higher than the previous accepted result by $2.5\sigma$ [3]. This has prompted two more mea-surements of $|V_{us}|$, with them both agreeing with the higher value. This has increased the accepted value of $|V_{us}|$ by $\sim 2\%$ to the new value[7] in 2007 of $|V_{us}| = 0.2257 \pm 0.0021$.

The status of the unitarity of the top row of the CKM matrix in 2007 thus becomes:

$$|V_{ud}|^2 + |V_{us}|^2 + |V_{ub}|^2 = 0.9991 \pm 0.0010 \qquad (2007) \qquad (1.17)$$

where the largest contribution of uncertainty is still from $|V_{us}|^2$ ($\sigma_{|V_{us}|^2} = 0.0009$) followed by that from $|V_{ud}|^2$ ($\sigma_{|V_{ud}|^2} = 0.0005$). Note that in both cases, the uncertainties are dominated by contribution from theory rather than from experiment. It is clear that the unitarity condition is well satisfied by this value.

Unfortunately, there still remains a debate over the form factor $f_+(0)$ that modifies the value of $|V_{us}| \rightarrow f_+(0)|V_{us}|$. In reaching the value in equation 1.17, an older $f_+(0)$ value (preferred by the PDG) was used despite recent calculations. If one of the new calculations had been used instead[7], the unitarity sum would again have fallen short by two standard deviations. The nagging problem of the unitary sum falling below unity experienced over the last decade has possibly therefore not yet been solved.

### 1.3.3  Association with Neutron and Pion Beta Decay

Other possibilities for determining $G_V$ and hence $|V_{ud}|$ are from neutron decay or pion beta decay. Both processes are free from the nuclear structure dependent corrections ($\delta_{NS}$ and $\delta_C$) that make a large contribution to the uncertainty of the $\mathcal{F}t$ values and $|V_{ud}|$ from nuclear beta decay. Unfortunately there are many experimental difficulties involved in measuring both decays.

Neutron decays ($n \rightarrow p + e^- + \bar{\nu}_e$), unlike nuclear $0^+ \rightarrow 0^+$ $\beta$-decays, contain a mixture of both vector (Fermi) and axial-vector (Gamow-Teller) transitions. Measurements of the free neutron lifetime as well as of the parameter $\lambda$ (where $\lambda = G_A/G_V$), from what are called "beta asymmetry" experiments, are thus required to determine the $ft$ value. The experimental difficulties are exemplified by the most recent measurement of the neutron half-life which disagrees from the average of all previous measurements by 6.5 standard deviations[5]. If the new value is to be discarded (as recommended by the PGD[7]) then one obtains the $|V_{ud}|$ value from neutron decay as:

$$|V_{ud}| = 0.9746 \pm 0.0019 \qquad \text{(neutron decay)} \qquad (1.18)$$

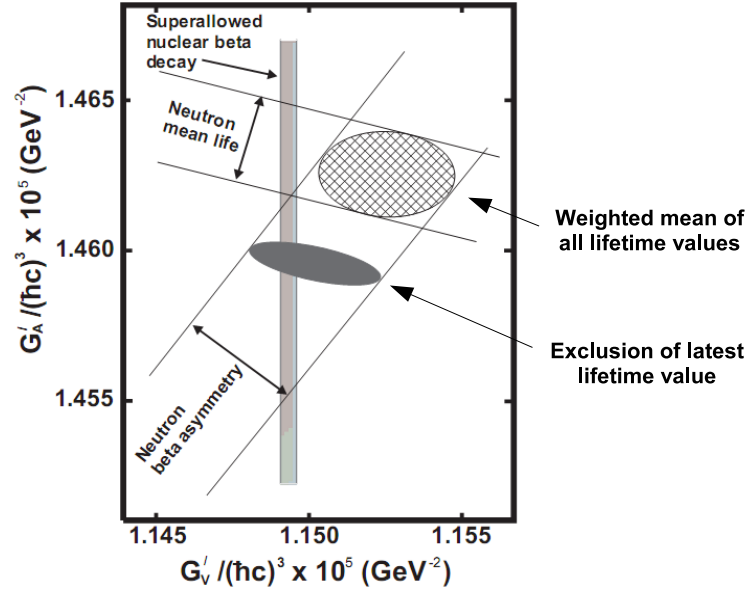

Figure 1.5: Comparison of the possible values of $G_A$ and $G_V$ obtained from neutron decay and nuclear beta decay experiments. The limits of the solid bar and shaded ovals are $1\sigma$ errors.

If the new value is included in the overall average of the neutron lifetime, then a value of $|V_{ud}| = 0.9765 \pm 0.0020$ is obtained[5]. Comparisons of the two results with that from nuclear beta decay are illustrated in figure 1.5.

Pion beta decay ($\pi^+ \rightarrow \pi^0 + e^+ + \nu_e$) is a $0^- \rightarrow 0^-$ pure vector transition are therefore no separate measurement of the axial-vector component is required. Unfortunately the pion beta decay branch is only a very small fraction ($\sim 10^{-8}$) of the total pion decay. This leads to these experiments contributing the largest uncertainty in determining $|V_{ud}|$. A recent experiment which succeeded in reducing the uncertainty in the branching ratio by a factor of 6 managed to reach a precision of only 0.5%. Using this new value, the PDG obtains[7]:

$$|V_{ud}| = 0.9749 \pm 0.0019 \qquad \text{(pion beta decay)} \qquad (1.19)$$

Although both these techniques contain "cleaner" theoretical corrections in calculating $|V_{ud}|$, the experimental difficulties have not yet been sufficiently overcome to give the results adequate precision.

### 1.3.4 Future Directions

The unitarity problem of the top row of the CKM matrix is currently focused on the value of $|V_{us}|$. This has prompted many new experimental and theoretical efforts to improve the accuracy of this value in recent times. Soon the focus will be shifted back to improving the accuracy of the largest element of the CKM matrix, $|V_{ud}|$. It is important to note that even if a value was found to satisfy the unitary sum within uncertainty, improving the precision of our knowledge of the CKM matrix elements would still be important. This is because the Standard Model is currently believed to be not the complete theory primarily due to its failure to incorporate the gravitation interaction, the existence of 29 free parameters required to be determined by experiment and its lack of explanation for the existence of cosmological dark matter. Therefore, it is not a question of whether the CKM matrix is unitary, but rather, to what degree it is not unitary. Improvement in the precision of the unitary sum thus allows an upper limit to be set on the size of effects from beyond Standard Model physics.

The CVC hypothesis has been confirmed for nuclear beta decays through the constancy of the $\mathcal{F}t$ values being determined to a precision of 0.03%. Nonetheless, it is not unreasonable to entertain the possibility of a systematic effect that would shift the overall value of $\overline{\mathcal{F}t}$ and thus $|V_{ud}|$. An excellent test for the existence of such an effect would come from comparisons of the $|V_{ud}|$ values from neutron decay or pion beta decay. Unfortunately, the current precisions from these two methods are an order of magnitude poorer than that from nuclear beta decay. Any direct comparisons with reasonable precision are not likely to be made in the near future. Therefore it may be necessary to search for systematic errors within the results of the different nuclear beta decays.

Systematic effects from theoretical contributions are difficult to overcome. As an example, the theoretical uncertainty of the nuclear structure dependent correction terms ($\delta_{NS}$ and $\delta_C$) is dominated by the systematic difference between calculations using a Woods-Saxon model or a Hartree-Fock model for the nuclear wave-functions. It might therefore be necessary to instead reduce all uncertainties in the experimental parameters and then infer the existence of systematic effects in the theoretical corrections by looking at unaccounted deviation from constancy of the individual $\mathcal{F}t$ values. This would act to improve the current understanding of the processes occurring in nuclei.

The future direction of high precision nuclear beta decay $\mathcal{F}t$-value measurements may therefore be to open up one's imagination and search for the effects of as many kinds of systematic errors as possible. One way of doing so could be to use a completely new method of measuring a parameter, for example, the recent use of on-line Penning traps for measurements of $Q_{EC}$ values (Ref.[8]). Alternatively, one could critically analyse all aspects of the methodology used for previous measurements (Ref.[9]).

$^{10}$C makes for a particularly interesting case study for such a purpose. It has the lowest atomic number (Z) in the set of superallowed beta decays such that the Z-dependent effect from the Coulomb correction ($\delta_C$) has the least uncertainty. It is also one of the two decays (along with $^{14}$O) from the set of the nine best determined superallowed beta decays with an isospin z-component transition of $T_Z = -1 \rightarrow 0$. The corrections dependent on this are known to be systematically different to those for the other group of $T_Z = 0 \rightarrow +1$ transitions. It is important to reduce the experimental uncertainty in the measurements of the $^{10}$C decay.

## 1.4   Previous $^{10}$C Half-life Measurements

Prior to 1974, there existed four measurements of the $^{10}$C half-life but they formed a mutually inconsistent set[10]. Since then, not including the current study, only one other measurement has been made. In the latest 2005 survey[1], only two of the five values were accepted (namely those from Refs. [11] and [12]). This has appeared to rectify the inconsistency problem. Of the values rejected (from Refs. [13], [14] and [15]), the first two were discarded because of a more recently discovered statistical bias in the fitting of half-lives using linear regression techniques. This problem can be easily overcome by using a "maximum-likelihood" (or equivalent) analysis procedure (see section 4.3). The last value was discarded because the coauthor later considered that pile-up had been inadequately accounted for. This problem will be much more difficult to avoid.

Pile-up occurs when two (or more) uncorrelated pulses coming from a detector overlap in time. This results in the signal being summed in the electronics, causing a distortion in the shapes and amplitudes of all the subsequent pulses involved. Since the amount of pile-up increases with count rate, a systematic error will be introduced in the half-life measurement. Although the problem of pile-up has been recognised in the past (for example in Ref.[15] and more so in Ref.[16]), and in fact observed (for example in Ref.[12]), the full extent of the problem did not become apparent until a simple test was performed in Ref.[9].

The pile-up test involved a germanium detector. The change in shape of a 1064keV $\gamma$ peak, from a $^{207}$Bi source fixed in position, was observed in the spectra for increasing total count rates. This was achieved by varying the distance of a $^{137}$Cs source, which produces 662keV $\gamma$s, relative to the detector. The results of the test led the authors to abandon the use of the germanium detector and opt instead for the use of plastic scintillation detectors which offer pulses around 300 times shorter in time.

Of the $^{14}$O half-life measurements accepted in the 2005 survey (Ref.[1]) and one other made since then (Ref.[17]), 3 have been with plastic scintillation detectors and 5 with germanium detectors. If the probability densities of each of the half-life values (presented by a normalised Gaussian distribution with the half-life as the mean and $1\sigma$ error as the standard deviation) are plotted along with the their sum, then an ideograph is formed. This allows a visual representation of the mutual consistency between the values. Figure 1.6 shows a comparison between the "ideographs" calculated separately for both the plastic scintillation and germanium detector measurements. The difference between the two summed probability density peaks is ∼0.05s. The currently accepted value (i.e. the weighted mean) is 70.620±0.013 s. Therefore the shift is ∼4 standard deviations larger than the uncertainty of the combined result; hinting at a systematic difference between measurements made with the two types of detectors. This provides motivation for the current experiment to measure the $^{10}$C half-life using plastic scintillators.

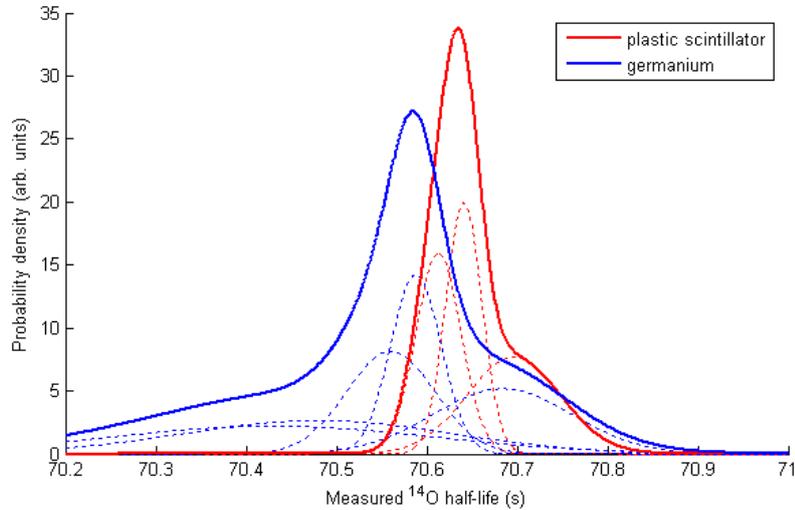

Figure 1.6: Ideographs of the $^{14}$O half-life measurements using plastic scintillators and germanium detectors.

## 1.5    Thesis Guide

The order of the chapters in this thesis is approximately chronological with respect to the work done. One interpretation of how the different parts of the thesis link together logically is given below:

The $^{10}$C used in this experiment was produced through the $^{10}$B(p, n)$^{10}$C reaction at the 14UD Pelletron accelerator facility at the Australian National University (ANU), Canberra. Details of the accelerator, targets and beam energies used can be found in section 4.1 of Chapter 4.

The trade-off in using plastic scintillators is a lack of energy resolution making it almost impossible to separate activity from $^{10}$C and other contaminants from a single detector energy spectrum. The contaminant causing the biggest problem is the long-lived positron emitter $^{11}$C which is produced from $^{11}$B impurities in our targets. The method for distinguishing between activity from $^{10}$C and from unwanted contaminants is to use a 3-fold (or "3 out of 4") time coincidence system with four equal-sized plastic scintillation detectors. Details of the construction process of the four detectors, the testing of the "3 out of 4" coincidence system and details of the associated (or "beam room") electronics used for determining time coincidence are given in Chapter 3.

The events from the detectors, after being processed by the beam room electronics, are collected by the data acquisition system. This system, controlled by the user through a computer terminal, serves two primary functions. Firstly, it automates the timing of the beam bombardment and of the data collection cycle. Secondly, it collects data in an event mode format, that is, the pulse-height and the time from the start of the data collection period are recorded for each event. The development of this system and the thorough tests performed to justify its use in a high precision nuclear experiment is given in Chapter 2.

Description of the decay curves obtained and the methodology used in the extraction of the $^{10}$C half-life are described in sections 4.2 and 4.3 of Chapter 4. Details of the data analysis required in obtaining a final value of the half-life are found in Chapter 5. Finally, a discussion of the significance of the value obtained and possible future directions for this experiment are given in Chapter 6.

# Chapter 2

# The Data Acquisition System

## 2.1 Introduction

The development of the event mode data acquisition system used in the experiment is described in this chapter. It consists of a computer controller, running software called "Kmax", communicating with a CAMAC (Computer Automated Measurement And Control) crate which facilitates the execution of automated routines using electronic hardware such as ADCs, scalers and so forth. The goal of the system is to allow the user to input desired parameters, initiate the start of a new experimental *run* (see section 2.2 for terminology), and have the data collected, displayed and recorded automatically.

The collection of event mode data, that is time and energy information collected for each event, has only been made possible in recent years due to improvements in the electronic devices used in the data acquisition system. The use of event mode data has been shown to be highly valuable in the critical analysis process required in high precision half-life measurements. It allows time and energy cuts to be made on the data at any time, removing the need for *a priori* hardware cut-off settings which may result in a loss of useful data. By making appropriate time and energy cuts, tasks such as removing contaminant activity from a decay curve or searching for count rate dependent effects that may not have been considered ahead of the experiment are possible (see Chapters 4 and 5).

The system developed is an extensive upgrade of a previous system established by Barnett[2] in 1995. The catalyst for the change is the dependence of the old system on the NuBus interface which has become obsolete. As older machines died out, it became much too hard to find compatible replacements.

The almost inevitable upgrade of the CAMAC Crate Controller (to one using the USB interface) also meant that much desired increases in the speed and of the buffer size of the system were gained. The final outcome is not only a more flexible system, but also one that is capable of collecting event mode data at much higher rates than was previously possible.

## 2.2   Basic Experimental Process

It is necessary at this point to describe the bare bones details of our experimental procedure and to define the terminology used in this thesis.

$^{10}$C decays with a 20s half-life hence it is not a naturally existing isotope of carbon. In this experiment, it is produced through the $^{10}B(p, n)^{10}C$ reaction. The sample of $^{10}$B sits on a thin gold foil and is called *the target*.

A single cycle of measurement is called a *pass*. A collection of successive passes is called a *run*. A pass can be broken up in the following sequence of events:

- The target is moved to the *beam on position*. It is now ready for the beam to strike. There exists a thick layer of gold, called the *beam stopper*, directly behind the target to absorb the remaining beam of high energy protons.

- The beam of high energy protons bombards the target for around 3 half-lives. This is called the *beam on time*. The beam is switched off at the end of this period.

- The target is moved to the *beam off position* where it is now placed in front of the detectors. The delay before the start of data collection is called the *delay time*.

- Recording of event mode data lasts for approximately 15 half-lives. This is called the *beam off time*. The target remains in the beam off position unless another pass is initiated, in which case, the cycle is repeated.

The target and detectors are located in the Beam Room, which is over 50m away from the Control Room where the data acquisition system (and also the user) sits. Appropriately, this spatial separation also serves to separate the two aspects of the detection system, namely, the detection system and the data acquisition system. It is the latter with which this chapter is concerned.

## 2.3 System Description

### 2.3.1 Computer Controller and CAMAC

The core of the system is the CMC100 Crate Controller manufactured by Cheesecote Mountain CAMAC. This device (occupying the right-hand most slot of a CAMAC crate) is connected to the computer, running the Kmax software package from Sparrow Corporation, by means of a USB 2.0 link. This combination allows communication of the computer (or host) with the various hardware devices on the CAMAC crate such as ADCs, scalers and so forth.

The CMC100 and the Kmax v8 software were purchased as a package from Sparrow Corporation. Although Kmax v8 is multi-platform based (being a Java program), only Macintosh OSX drivers have been developed; for this reason a Powerbook G4 laptop was also purchased.

Before the purchases we were told that combining the three would allow a straight-forward reproduction of the primary features of the old system. This did not turn out to be the case due to incomplete driver and firmware development in key areas of the CMC100 with Kmax operation. It was only after two months of constant communication with Sparrow that solutions and work-arounds were completed. The final working software, called "*E-Lifetime*", requires Kmax v8.2.2 and CMC100 with USB version D, FPGA version 31, DLL version 8.

E-lifetime requires other devices on the CAMAC crate. The exact device models stated below are not explicitly required. Replacement with other devices which have the same functions is certainly feasible and possibly advantageous. However, as we shall see in the rest of this chapter, many tests must be undertaken before one can be confident of the inerrant use of a given device.

The devices used on the CAMAC crate are:

**LeCroy 3514 ADC:** An Analog-to-Digital Converter with a short and constant dead time operated in strobe mode. When a strobe pulse triggers the digitisation of an analog pulse, a system request in the form of a LAM (described in section 2.4.1), is produced requesting attention from the crate controller.

**KineticSystems 3615 Hex Scaler:** A module containing six independent 24-bit counters capable of count rates up to 50MHz. The crate controller can perform separate read and clear commands on each of the counters.

**Homemade Sequencer:** When used in combination with the (also) homemade "BNC Output Box", it allows the crate controller to produce DC levels, lasting down to milliseconds in duration, along six independent outputs.

## 2.3.2   Control Room Electronics

Several additional Nuclear Instrumentation Module (NIM) devices, collectively called the Control Room Electronics, are used to process three inputs and produce three outputs. The inputs come from the Beam Room and provide the data to be collected. The outputs allow communication with any external systems controlling the experiment (see figure 2.3.1). The processes performed by the Control Room Electronics are described below.

- Movement of the target to the beam on and beam off position is controlled by two positive gate pulses of around one millisecond in duration. They are produced using the Homemade Sequencer in conjunction with the BNC Output Box. Before the pulses are sent down the long cables to the Control Room, they are buffered using Single Channel Analysers in case the Homemade Sequencer is not capable of driving 50m of cable.

Figure 2.1: Circuit Diagram of the Control Room Electronics

- The period for which the beam is activated is controlled by a negative DC level produced by a DC level timer. The initiation of the negative DC level is triggered by the target beam on position movement pulse, delayed by half a second. The timer is set to stop the negative DC level half a second before the target beam off position movement. This is to ensure that the target is not moving while the beam is on.

- The current time value is stored on the KSC 3615 Hex Scaler to allow it to be read by the crate controller. The time is incremented by a homemade High Precision Oscillator where the output passes through a single channel analyser to convert the positive TTL to negative pulses recognisable by the scaler.

- Information about the amount of beam current striking the gold beam stopper, fixed behind the target in the beam on position, is recorded by a channel on the KSC 3615 Hex scaler.

- An analog pulse of height proportional to energy to be recorded provides the input to the LeCroy 3514 ADC.

- The positive sloping edge of a TTL pulse triggers the acquisition of the voltage across the ADC's input when in strobe mode. The strobe pulses are passed through a Gate & Delay Generator so that the timing of the strobe edge can be adjusted to match the peak of the ADC analog input pulse.

## 2.4   CAMAC Operating with The CMC100

### 2.4.1   CAMAC Description

In the CAMAC[18] standard, communication between modules and the crate controller is established using The Dataway, a series of inter-connected and individual lines connected to the back of every module. Execution of commands, conveyance of status information and transfer of data are achieved by setting up appropriate signals down the lines of the Dataway. Only the key elements of the Dataway required in understanding the elementary principles of CAMAC operation will be described here.

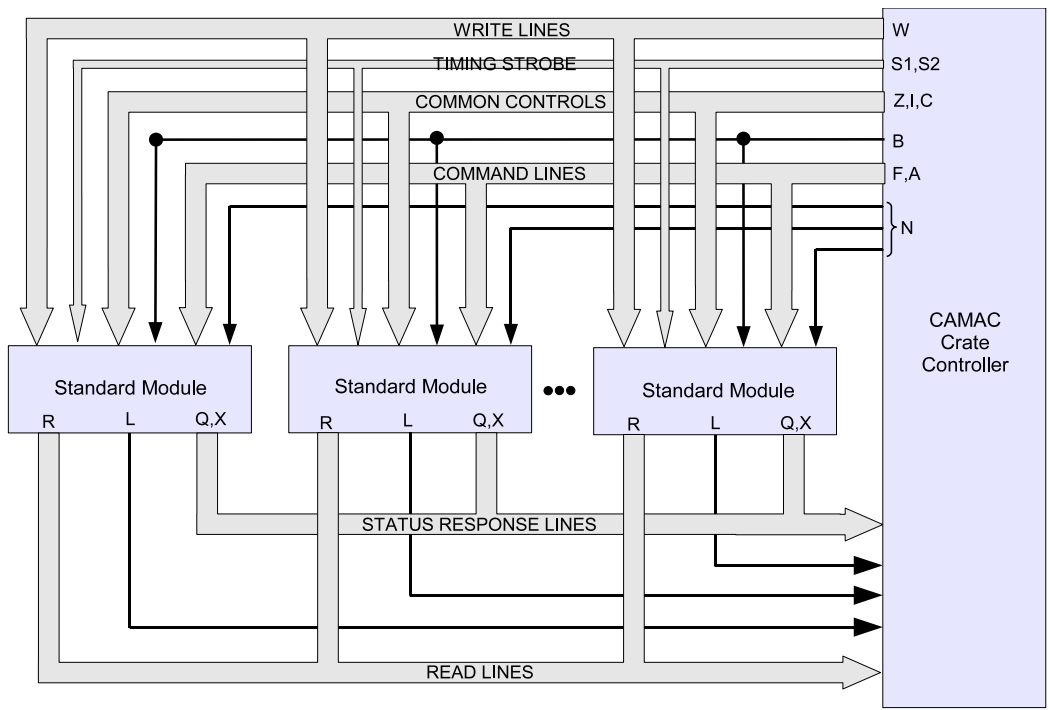

Figure 2.2: Block Diagram of the CAMAC Dataway

## Description of a Command

A CAMAC command is initiated by the crate controller by addressing a module's individual station number (N) line. The type of task to be performed is specified using the sub-address (A) and function (F) lines common to all modules. A command of this type is thus commonly called a "NAF command".

Upon acceptance of a command, a module will produce a signal down the response (X) line and begin to act on the command. Also, an accompanying signal down the busy (B) line is generated during a command to indicate to all units that an operation is currently in progress.

A command may or may not involve the transfer of data between the crate controller and the modules. Data is transferred to and from a module using the common write (W) and read (R) buses respectively. There are 24 lines on each of the buses and hence up to 24 bits may be transferred during a single operation.

### The Status of a Device

In CAMAC standard, a device may request attention from the crate controller through the production of the Look-At-Me (LAM) signal. Any unit may generate a LAM down its individual (L) line if there is no busy (B) signal present. The use of LAM signals (or simply called "LAMs") is required to implement automated data acquisition.

For example, in our system, LAMs are generated by the LeCroy 3514 ADC upon receiving a strobe pulse followed by the digitisation of the voltage. This LAM then causes the unit to halt, then wait for attention from the controller. The response comes in the form of the reading of the digitised value and clearing of the LAM, readying the device for the next pulse.

Another line that can be used to transmit status information of a device is the response (Q) line. This allows the transfer of binary information regarding a selected feature of the module. The signal on the Q line remains even after the command requesting the information has finished.

### Common Controls

There are three common control signals that can be used to control all units on the CAMAC crate without the need for individual addresses. They are the initialise (Z), the inhibit (I) and the clear (C) commands. Initialisation causes all units to return to their basic states. This resets all registers and disables all LAMs where possible. Inhibit stops all activity on the Dataway thus halts all processes that might already be occurring such as data collection. Clear causes only the registers on all the units to be cleared.

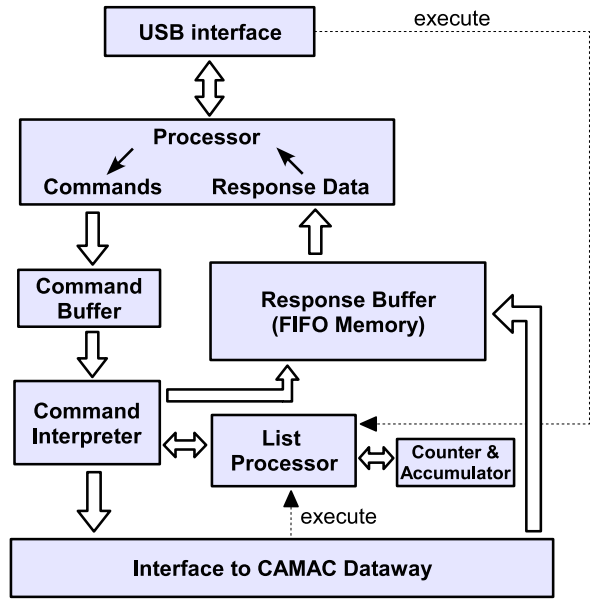

Figure 2.3: A Block Diagram of the CMC100

## 2.4.2 The CMC100 Crate Controller

The CMC100[19] accepts commands and data as 32 bit words from the USB v2.0 (v1.1 is supported but is restricted by lower speeds) interface. Both are temporarily queued in the Command Buffer while they are sequentially sent to the Command Interpreter for immediate execution. Some commands are internal to the CMC100 with the responses produced internally as well, hence they do not require the use of the CAMAC Dataway. Regardless of where they are produced, all response words are sent to a one mega-word sized FIFO (First-In-First-Out) Response Buffer where they are stored until read. Note: there has been a slight change in the response processing due to bugs encountered, see section 2.6 for more details.

A different and much faster mode of the execution of commands requires the use of the List Processor. Multiple short programs, up to 512 words in size, can be sent to the list processor to be stored; ready for future use. The programs can include useful commands such as conditionals, loops, and delays. Furthermore, a 20-bit counter and a 32-bit accumulator register are accessible to the programs.

Execution of the program in the List Sequencer can be initiated by either a command from the host or as a response to a LAM or an external NIM pulse. The mode of thisbehavior is determined by the bit pattern on the Control Register. Also available is a LAM Mask Register used to store a bit pattern that determines whether LAMs from a particular module are to be acknowledged by the CMC100.

The execution of the commands and the triggering of the list by a LAM follow a set priority. List triggering occurs only after the current and any queued commands in the Command Buffer have finished execution. While the list is running, any commands from the host are queued until after the completion of the program.

## 2.5   Kmax

Kmax is an environment used for instrument control, data acquisition and data analysis. It contains high level support for event-by-event data acquisition, data sorting and histogramming. Programs within Kmax, called *toolsheets*, allow the user to create an application specific to the task required.

Kmax v8 and its programs are based on the Java Platform which functions as follows: the source code contained in the toolsheet is compiled into byte code (a binary `.class` file) by a platform independent Java compiler. This is interpreted at runtime by the Java Virtual Machine (JVM) which is platform specific software. Therefore, this allows a Java program written on any machine to be executed on any other, provided an appropriate JVM is installed (hence the "*write once, run anywhere*" slogan from the creators of the Java Language, Sun Microsystems).

### 2.5.1   The Toolsheet Environment

The Toolsheet Environment consists of the Device Panel, the Editor Panel and the Control Panel:

**The Device Panel** is used to specify the slot locations of the devices used and which Module Description Resource (MDR) files are loaded. It is only necessary to specify the devices that Kmax has direct interactions with, for instance just the CMC100.

**The Editor Panel** is where the source code for the program is written. The code uses the standard Java Runtime Environment (JRE), allowing the use of standard Java libraries as well as the special Kmax extension package. This package allows: the implementation of the Kmax Runtime interface, the use of device drivers (see below), the access of the Toolsheet environment, and the interaction with the widgets in the Control Panel.

**The Control Panel** is where the toolsheet's user interface is constructed and displayed (see figure 2.4). Kmax provides a set of predefined objects (such as buttons, check boxes, text fields, histograms, *et cetera*) called widgets, to be created in this panel. The behaviour of these widgets is then defined in the Editor Panel. Additional panels, similar in function to this one, can be created by the user.

### 2.5.2 The CMC100 Drivers

The interaction of a Java program with hardware of a computer has to be done using native languages (such as C or C++). Most common devices are supported by the JVM, but for more unusual devices such as the CMC100, a specially created Java Native Interface (JNI) is required.

A JNI is a framework that allows Java Programs to access code written in a native language. To do this, the CMC100 driver package is split into two files; one written in Java and the other, a native library file, written in C. The Java component (also known as the Java Wrapper) provides an interface to the native code from Kmax. Thus, successful implementations of a given function in both the Java Wrapper and the native library are required.

## 2.6 The Basic MCA Program

A Multi-Channel Analyser (MCA) program is a good example of an application of the many fundamental aspects of a DAQ system. Consequently, the devel-opment of the Basic MCA toolsheet serves as a good tool for understanding and testing the elementary operations of our DAQ system.

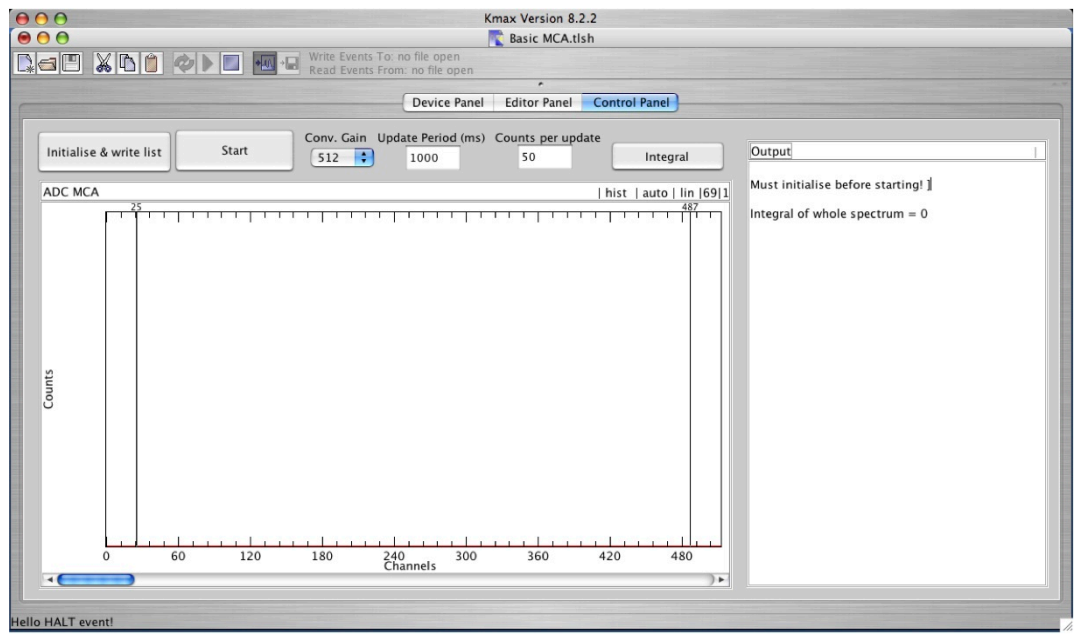

Figure 2.4: The Control Panel of the Basic MCA toolsheet

A description of the development of the Basic MCA toolsheet, following the design shown in figure 2.5, is given here. This design was chosen because it can be easily altered to conform to the needs of E-Lifetime. Possible minor alterations that may offer improvements to this design are discussed briefly in section 2.6.4.

## 2.6.1   Implementation

The Basic MCA toolsheet runs as follows:

- When any toolsheet is first loaded in Kmax, it needs to be compiled and executed by the user. Only then is the toolsheet active. (Note: there is a setting to automatically complete these steps when Kmax is started but this makes the loading of a toolsheet sluggish.)

- The desired ADC conversion gain and the time interval at which data is to be updated are entered by the user.

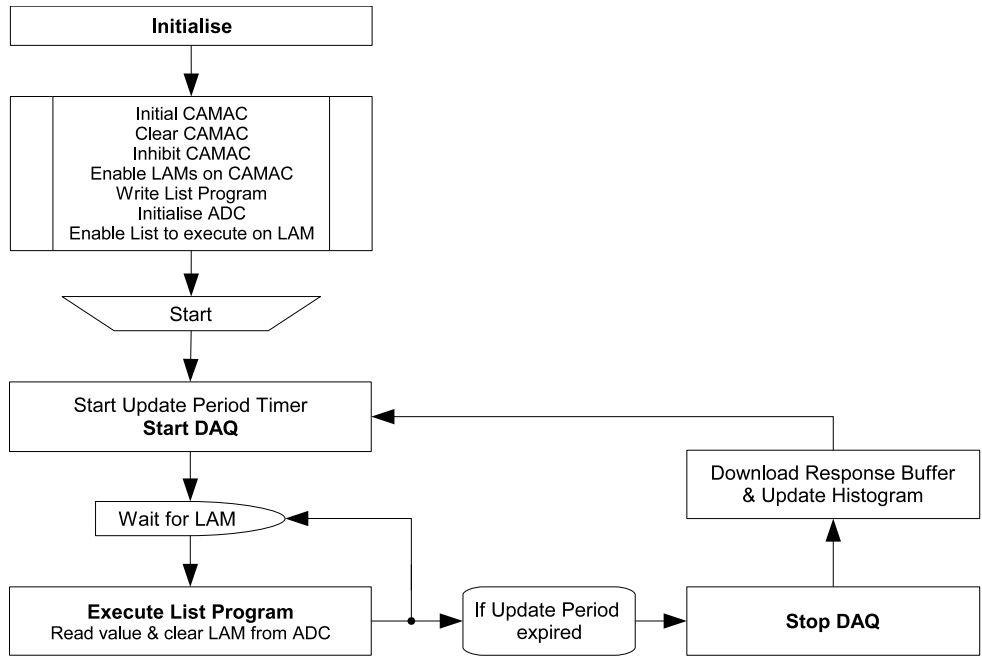

Figure 2.5: Flowchart of the Basic MCA toolsheet

- The "Initialise & write list" button should then be pressed to set up all the devices (note the setting of the inhibit) as well as write the program to the List Processor. The system is then ready for data acquisition to begin.

- When "Start" is pressed, a timer lasting for the Update Interval begins to tick and DAQ is enabled by the removal of the inhibit.

- The ADC will generate a LAM whenever a pulse is digitised. This then triggers the execution of the list program which contains a "Read value & Clear LAM from ADC" command. Once the LAM is cleared at the end of this command, the ADC becomes ready to acquire the next pulse.

- The above DAQ process will continue until the Update Interval timer expires, in which case DAQ is stopped by the setting of the inhibit. The Response Buffer contents are then downloaded and the data is updated to a histogram. The Update Period timer and the DAQ are then restarted.

- The data collection and update loop continues until "Stop" is clicked.

This causes the DAQ to stop and the buffer downloaded and the histogram updated one last time.

Note that the two key defining features of this implementation are the use of the CMC100's built in "execute list on LAM" feature and the use of the Inhibit to toggle data collection.

After implementation and testing of the above program, it was discovered that problems downloading the Response Buffer content existed. But more importantly, it was also found that starting of the DAQ caused the program to crash. After persistent contact with Sparrow Corporation, updates to the firmware and drivers were eventually made to eliminate the problems plaguing these elementary operations (see below).

## 2.6.2   Starting & Stopping List Sequencer problem

After pin-pointing (by producing output after every operation) the cause of the crashes down to the "enable DAQ" procedure. It was observed that crashes were more frequent for higher data rates. When a crash occurred, Kmax froze and the Host LED on the CMC100 stayed lit while the N, LAM & Trigger LEDs flashed continuously.

The Host LED indicates communication is currently in progress with the host; its normal behavior is to flash whenever a host command is executed. The N, LAM & Trigger LEDs indicate that a NAF command has been activated, that there is a presence of a LAM and that the list has been triggered respectively. The simultaneous flashing of all three signified that a value was being read by the list sequencer which was triggered by an ADC LAM.

To discover the meaning of the continuously lit host light, a simple test was performed. A long delay was set to execute in the list sequencer program. While this was running, a command was manually triggered to be sent to the device from the host. It was observed that the Host LED became lit (it normally just flashes) at the instant that the host command was sent. It continued to stay lit until the list program had finished execution. This signified that a continuously lit Host LED indicated that there existed a host command queued for execution after the current list program. This is the expected priority of the CMC100 events described in section 2.4.2.

It was further observed that the Response Buffer was returning empty even if it was certain that values had been collected by it (indicated by the N, LAM & Trigger LEDs). It was discovered that any command executed by the CMC100, including trivial commands such as setting an inhibit, before reading the Response Buffer cause it to be cleared.

**The Cause of the Crashes**

Piecing together the information, it was concluded[20] that when a host command was performed in Kmax, a "Clear Response Buffer" command always followed. This explained the unwanted clearing of the Response Buffer. It was then deduced that the crashing was caused by a LAM, which triggering list execution, sliding in between the two commands. The second of the commands, expected by the program to always immediately follow the first, becomes queued which causes Kmax to timeout and crash. This would also explain the Host LED coming on as well as the increased rate of crashing for higher data rates.

It is worth noting that Sparrow had known about the problems associated with starting and stopping list execution and had been trying to solve them for several months. No mention of the specific cause was given to me in our conversations until I had mentioned our analysis. This has led me to believe that it was our work that identified the specific cause of the problem, contributing to the eventual solution.

**Atomic Commands**

After further discussion with Sparrow[21], it was discovered that the CMC100 commands were meant to be *atomic*. An atomic command, in general, is a set of commands combined so that they appear as a single command to the rest of the system. The set of commands either succeeds or fails all at once hence no in-between state is accessible. This protects the system from entering an invalid state which may cause problems. Atomicity of the CMC100 commands is achieved by satisfying two conditions:

(a) Until all immediate commands in a CMC100 atomic command are completed, no other commands should take place. This allows the group to be viewed as a single command.

(b) The Response Buffer, which may get used during immediate commands, is to be cleared at the end of an atomic command. This returns the system to the state that it was in before any commands had taken place.

All NAF commands produce a response to the CMC100 Response Buffer, which may simply be a non-useful zero for the cases when the read lines are not used. This explains the "Clear Response Buffer" command at the end of every command in order for (b) to be achieved.

Unfortunately (a) is not always satisfied, as demonstrated by the Basic MCA Program test. From certain sections of the code from the native library (only obtainable from Sparrow by special request) it was clear to see that the "Clear Response Buffer" command and the NFA commands were being executed separately by the host. This allowed the possibility of another command sliding in between.

**The Solution**

The previous implementation does not take of the advantage of the Command Buffer in the priority of execution described in section 2.4.2. The solution and current implementation by Sparrow is therefore to send both the NFA and the "Clear Response Buffer" command as a block to the Command Buffer. The commands queued in the Command Buffer executes uninterruptedly, preventing the triggering of the list program, until it is empty. This new implementation satisfies the conditions of atomicity.

The problem of the Response Buffer always clearing, which may lead to loss of useful data, is solved by altering the response data path by the inclusion of the Alternate Response Buffer (see figure 2.6). In this new design, response words are tagged according to whether they were produced from the Command Buffer or the List Sequencer which determines whether they are sent to Alternate Response Buffer or the standard Response Buffer respectively. This allows the performance of necessary host commands, such as starting and stopping data acquisition, without disrupting the actual data collected.

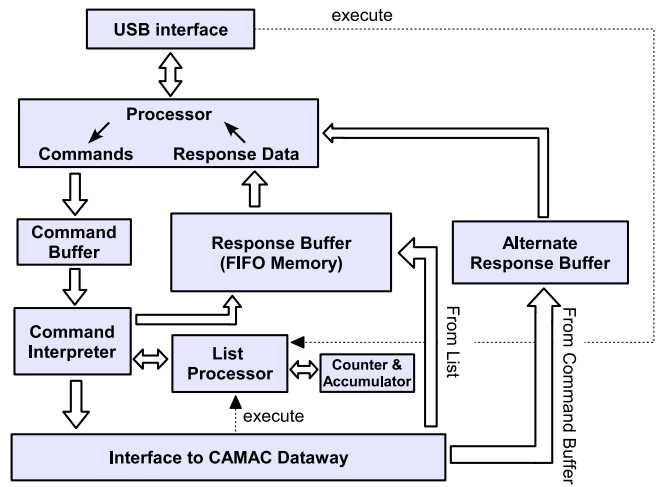

Figure 2.6: The new CMC100 Block Diagram with the inclusion of the Alternate Response Buffer

### 2.6.3   Buffer Read Problems

It is important to know the Response Buffer count before reading the buffer. Without this information, one would need to specify an over-estimate of the amount of data to be read (this would have to be the size of the whole buffer which is one mega-word or four mega-bytes to avoid problems), and then analyse the returned over-sized array to decide which sections of it contain useful information. This is cumbersome.

In the original system, the ability to read the Response Buffer count had not been implemented and, as it turned out, was not even possible. The modification described previously provided a path for status information to be returned without having to read the Response Buffer itself. With the help of this feature, the ability to read the Response Buffer count was included by Sparrow in the same update.

What is returned from the "Read Response Buffer count" command is not always usable though. The buffer count returned is only correct if it is greater than 639 words. The reason for this is not well understood. In the situation where the counts are below 639 counts, the way to work around the problem is to read the entirety of the maximum possible count of 639 words and then decide which of these are useful. Although not perfect, the worst case scenario of having to transfer 639 more words than required is significantly more tolerable than the previous scenario of having to transfer the whole 1 mega-word buffer every time.

### 2.6.4   Possible Alternatives

As mentioned at the start of the section, there may be minor changes that might improve the performance of the Basic MCA Program. I use the term "minor" because the key feature of this program, namely the use of the List Processor executing upon the detection of a LAM, is fundamental to the success of any fast DAQ system using the CMC100.

The main determining factor of a good MCA is the amount of dead time inherent to the system. A main contributor to the dead time in this system is the stopping and starting of the DAQ required for each data update, estimated to take on the order of tens of milli-seconds (compared with the 6 micro-second dead time per event determined by the hardware specifications, see section 2.8) done typically every second.

An alternative method of starting or stopping DAQ is to enable or disable list execution by writing a new bit pattern to the Control Register. This command is internal to the CMC100 and does not require the use of the CAMAC Dataway as opposed to the setting of an inhibit. This may reduce the dead time of starting and stopping DAQ by a few micro-seconds. An actual test still needs to be performed.

A further reduction of the dead time could be achieved if the Response Buffer content was to be read "on-the-fly". But due to the priority and atomicity of the CMC100 commands, the collection of the next value can not take place while the host is reading the Response Buffer count and downloading the buffer contents. So no true "on-the-fly" operation is possible. The closest thing is the removal of the starting and stopping DAQ commands, which take only a few micro-seconds. To gain this dead time reduction however, the buffer count read and content download commands will both need to be sent to the Command Buffer together while still allowing the output of the first to be input to the second command. This is needed to prevent problems during high data rates caused by the triggering of the list processor sliding in between the commands.

This significance of the last paragraph is to show that the current design of the DAQ procedures in the Basic MCA Program is very sound. To improve the performance by even a small amount, a considerable amount of effort would be required. This DAQ procedure thus provides a good framework for the final E-lifetime program to be built upon.

## 2.7 E-lifetime

### 2.7.1 The User Interface

The E-lifetime Control Panel is shown in figure 2.7. Apart from the toolbar at the top which provides access to the Kmax core operations such as compiling and running the toolsheet, all the other widgets are created additionally and therefore can be altered by the user.

The Parameters sub-pane allows the user to adjust the Beam On, Delay and Beam Off times and the number of passes (see section 2.2 for terminology) to be used in the run. The ADC conversion gain and the Hexcounter read type (see section 2.8.2) are also selected in this sub-pane. The "Save Data" checkbox needs to be ticked if Event Mode Data is to be recorded. The "Min Beam On Proton" and "Max Beam on Proton" textfields sets the condition on the proton beam status before a warning is displayed. These values are defined as the integrated beam current striking the thick gold backing (which sits behind the target in the path of the beam) during the Beam On and Beam Off times. The "Time Interval size" needs to match the setting on the High Precision Oscillator for correct horizontal scaling of the Time Projection histogram.

The Status sub-pane allows the user to see the current stage of the pass cycle. The current pass number, the Beam On and Beam Off Proton inte-grated current and number of events in the last pass are also shown here. A useful countdown of the time remaining in the long Beam Off period was later included.

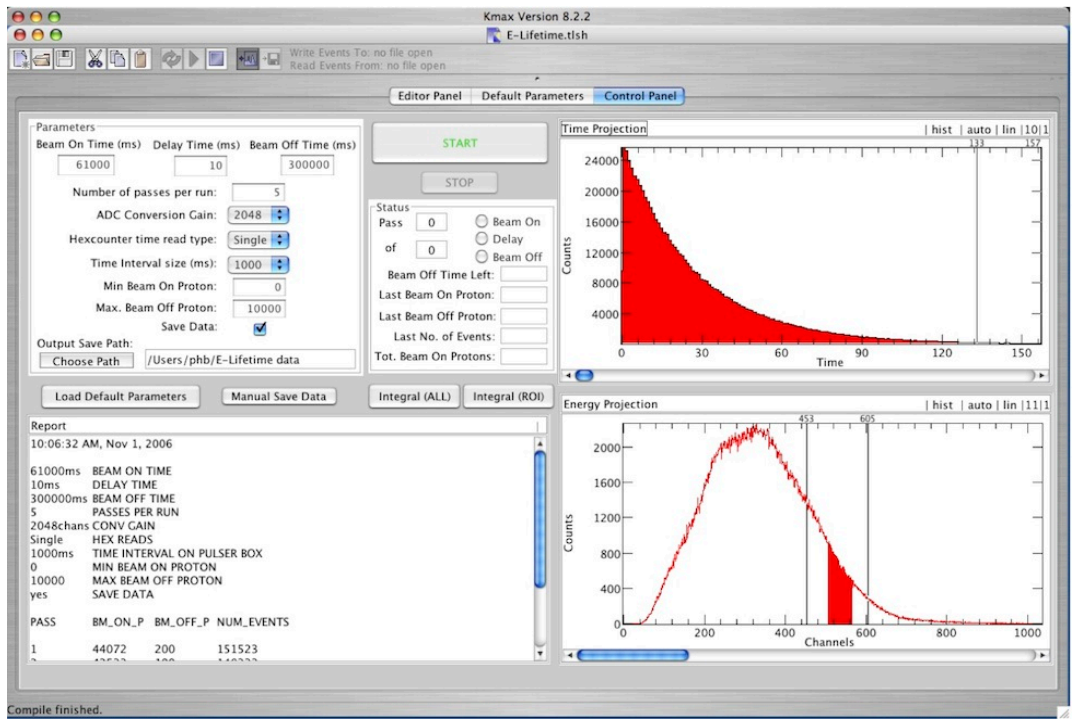

Figure 2.7: The Control Panel of the E-Lifetime toolsheet

The Report window, as the name suggests, provides an updated report of the run details. The starting time and parameters used for a run are inserted at the start of the report; followed by tabulated Beam On Proton, Beam Off Proton and Number of Events information from previous passes. If the user requests additional passes, information of this is also displayed in the report.

The Time Projection and Energy Projection windows display the cumulated time and energy histograms of the Event Mode Data. Zooming, expanding and log scale are supported by these Kmax Histogram widgets. The integral of the whole energy spectrum, or just the selected Region-of-Interest (ROI) can also be outputted to the Report window.

If the Save Data check box is ticked, at the completion of a run, the user will be prompted for a run name for the current Event Mode Data file and cor-responding Report, Time and Energy histogram files. If a run was terminated before completion, or if the Save Data check-box was not ticked, the user can still save available data by using the "Manual Save Data" button. The file path for the saved files is displayed at the bottom of the Parameters sub-pane.

The Default Parameters are loaded when the Toolsheet is first started, they can also be loaded at any time by clicking the "Load Default Parameters" button. The default parameters, stored in the "`E-Lifetime parameters file.txt`" found in the Kmax root directory, can be altered within Kmax by using the Default Parameters panel.

### 2.7.2 The Program Details

The goal of E-lifetime is to provide a user interface and automated program for controlling experimental runs. This requires the performance of data acquisition, target movement and processing of status information at appropriate times during the Beam On, Delay and Beam Off periods. This is achieved by combining Kmax and Java software features with the CMC100 and CAMAC crate hardware. The logic of the E-lifetime program is most clearly displayed in the flow diagram of figure 2.8 (see *Appendix A* for full source code).

The DAQ procedure of the Basic MCA toolsheet is modified to collect Event Mode Data. This is done by adding a command to read the time value on the HexScaler in the List Processor program. But this also requires additional modifications to the data processing steps since the buffer contents now contains both a time and an energy value for every event.

#### Event Mode Data Writing

Event Mode Data is extracted from the Response Buffer contents by alternately assigning values as time and energy data. A check of whether the Response Buffer count is a multiple of two is always done. If this is found not to be the case the program is stopped and an appropriate error message is displayed.

Event Mode data is written to a file on disk using the `BufferedWriter` class which is part of the standard `java.io` package. The characters are written in the platform's default encoding (UTF-16 for MAC OS X) using a default 512

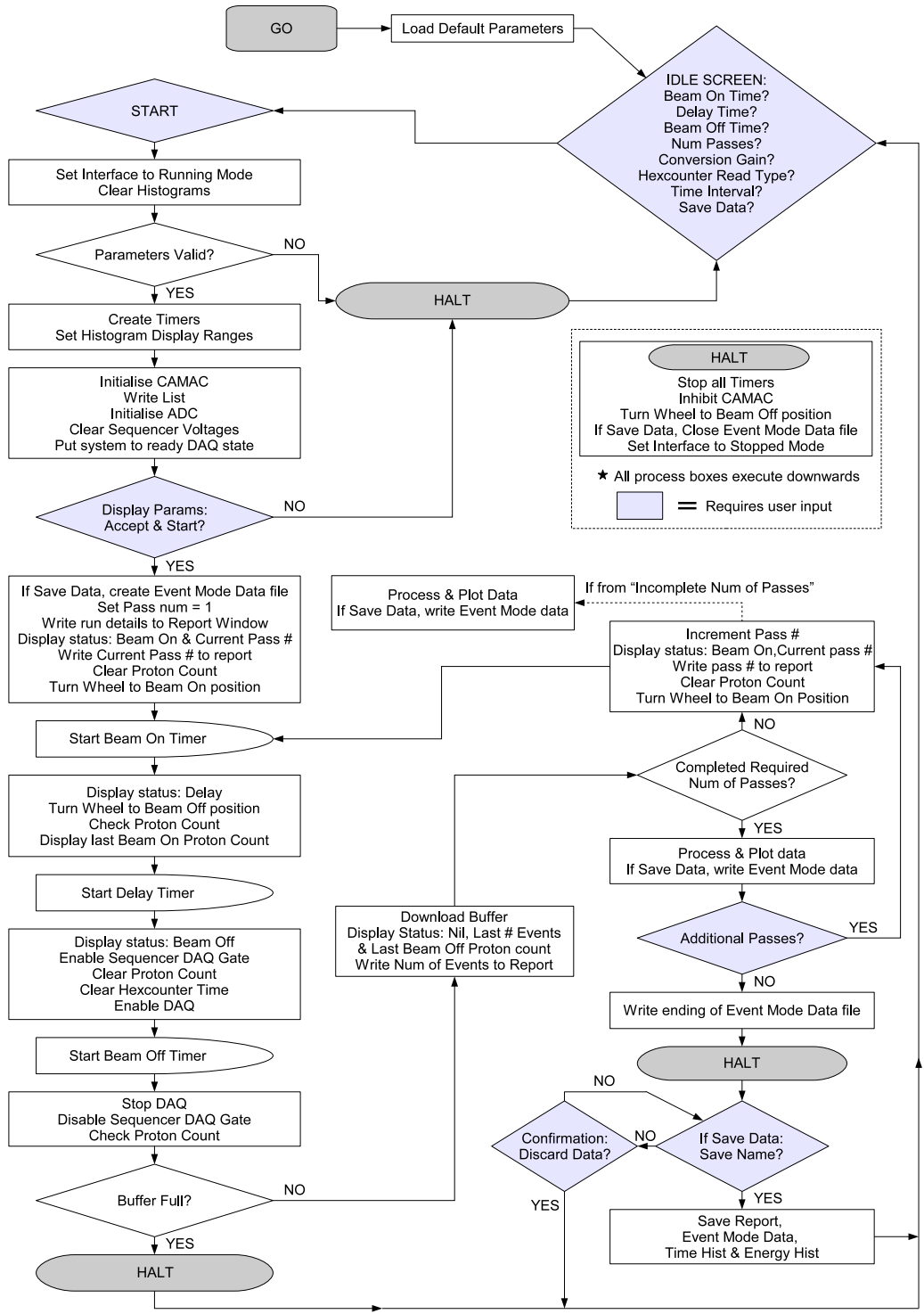

Figure 2.8: The Flow Diagram of the E-Lifetime toolsheet

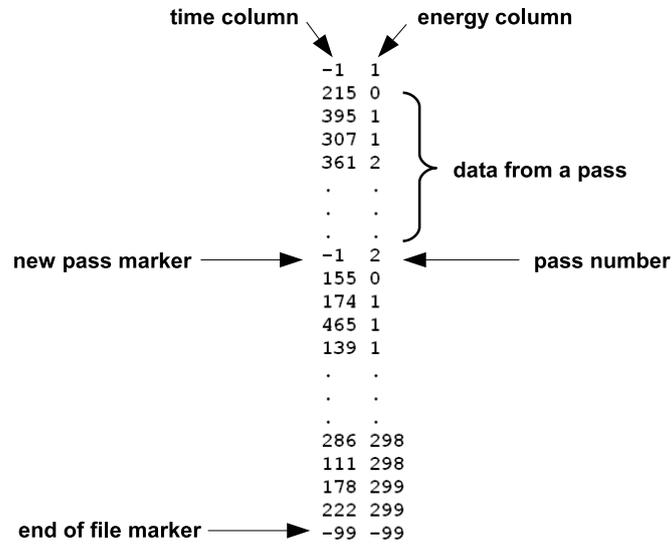

Figure 2.9: An example of the format of an Event Mode Data file

byte (128 character) sized buffer.

An example of the format of an Event Mode Data file is shown in figure 2.9. Each entry in the Event Mode data file exists on its own line.. The start of a new pass is indicated by ("`-1 <pass no.>`"), while the end of the file is indicated by ("`-99 -99`"). This convention mimics that of the old system and was chosen so that files can be easily imported into data analysis software.

The Response Buffer filled with Event Mode data takes several seconds to write to disk. This usually takes place during the Beam On time therefore it is imperative for the Beam On Time to be much longer than this. Although not a problem for the $^{10}$C experiment, it might become a problem for experiments requiring a short Beam On time. To increase the speed of the Event Mode data writing, the size of Java's write buffer should be increased from its default value of 512 bytes.

**The Histograms & Report**

The provision of predefined widgets in Kmax makes the creation of a user interface simple. The Report window uses the Textpane widget, whereas the Time and Energy projection histograms use the Histogram widget.

The Textpane widget is quite simple. Tt provides a window for displaying large portions of text which is useful for error messages or status information to be conveyed from the program. The Histogram widget is much more complex. It provides built in functions for plotting of histogram data in a variety of styles and colours, as well as Region-Of-Interest features such as integration and centroid search.

Data is displayed on the histograms by first adding it to the Kmax Event Buffer tagged with an ID to specify which histogram it should be displayed in. The Event Buffer then needs to be sorted and the histogram updated. The Event Buffer needs to be cleared at the end of every update, otherwise repeats of the same data will be added.

It is worth mentioning at this point that the default Event Buffer size of 512 words was found to be much too small. In order to avoid repeatedly adding and sorting small blocks of data, the size allocated for this buffer was increased to one million words. This is done by changing the settings of the Toolsheet properties.

The saving of the Report and histogram data uses the built-in in export functions implemented in the widgets. These were chosen so that the saved files could be easily imported back into the respective widgets. The Report file is simply a text file (in ASCII) containing exactly what was in the report window. The saved histograms, however, are in a special Kmax histogram data format (details of which can be found in Ref.[22]).

**Error checking features**

The E-lifetime toolsheet was designed to include lots of error checking features to protect the system from fatigued users after a long day of experiments. For instance, when the user clicks "Start", the experiment does not begin right away, instead a check of the parameters inputted is made. Things that are checked include the possibility of negative values or letters in the places where positive integers are expected or the Beam On and Beam Off times are too short due to the user accidentally using units of seconds as opposed to milli-seconds. After all of this is has been completed, a final confirmation prompt is displayed before the run is started.

Once a run is underway, many of the interface features that may cause problems when accidentally pressed, such as the Start button and the controls on the Parameters sub-pane, are disabled. The only option left available to the user is to stop the current run.

If "Stop" is clicked during a run, data collection is halted and in addition, loose ends are tidied up. One of these is to make sure that the target is back in the Beam Off position which is very important if a run is stopped due to the failure of the beam shutting off.

When a run has completed, the option to not save the data can not be achieved simply by pressing the mouse. Instead, the user is forced to clear the Save Name textfield in the prompt with the keyboard and, furthermore, required to respond to a confirmation prompt. This prevents the loss of valuable data by mistake.

## 2.8 Detailed test of all aspects of the DAQ

### 2.8.1 The LeCroy ADCs

The LeCroy 3514 ADC used is a newly purchased second-hand unit from O. E. Technologies. Before the purchase, we had the chance to test two of these units, as well as two other LeCroy 3511 ADCs. In the end, it was decided that both the 3514 models were to be purchased in order to have a back-up device.

Detailed testing of all the functions of the ADC was performed. This included the Lower-Level, Upper-Level Discriminator and Zero Level settings, the conversion gain, operation in strobe and peak detect modes, strobe triggering on positive and negative sloping edges, input pulse polarity, Busy out delay, Control Register writing, response to CAMAC common controls and clearing and testing of LAMs.

**ADC Dead Time**

The most crucial of these tests is of the ADC dead time which is the time spent digitising a pulse and during which all other pulses are ignored. This test was done using a HP8010A Pulse Generator to produce double pulses with variable separation times. At a fixed rate from the pulser unit, the separation time of the pulses was reduced and the count rate collected by the ADC observed using the Basic MCA program.

A step function is expected to be seen in a plot of count rate versus pulse separation, but as shown in the plots of figure 2.10, a strange dip was observed. This was soon discovered to be caused by an incorrect use of the ADC in the List Processor.

It is imperative to have the LAM that caused the current list execution to be cleared before the end of the list program, otherwise the list will execute repetitively. At first, two commands were placed in the list processor, the "Read ADC value" followed by the "Clear ADC LAM" command. It was later found that the "Read ADC value" command already automatically performed the LAM clearing operation. Hence, the unnecessary "Clear ADC LAM" command only served to interrupt the acquisition process of the second pulse. This caused either the partial acquisition or, for the short duration that the "Clear ADC LAM" command were in progress, no acquisition at all of the second pulse. Since this testing was done early on in the system development, the strange behavior was rectified for E-Lifetime by simply removing the "Clear ADC LAM" command from the list.

The strange dip in the plots of figure 2.10 should then be ignored and the dead time should be interpreted to extend to the first rise in counts only. Incidentally, the time between that and the first fall in counts gave a measurement of the cycle time between commands (see section 2.8.4 for more information).

The differences between the dead time of the two ADC models and different conversion gains agrees well with the stated values in the ADC manuals[23]. Note that the absolute values measured consist of not only the ADC dead times, but also the time taken for the CMC100 to respond on the LAM and for the List Processor to read the value from the ADC. Again, see section 2.8.4 for more information.

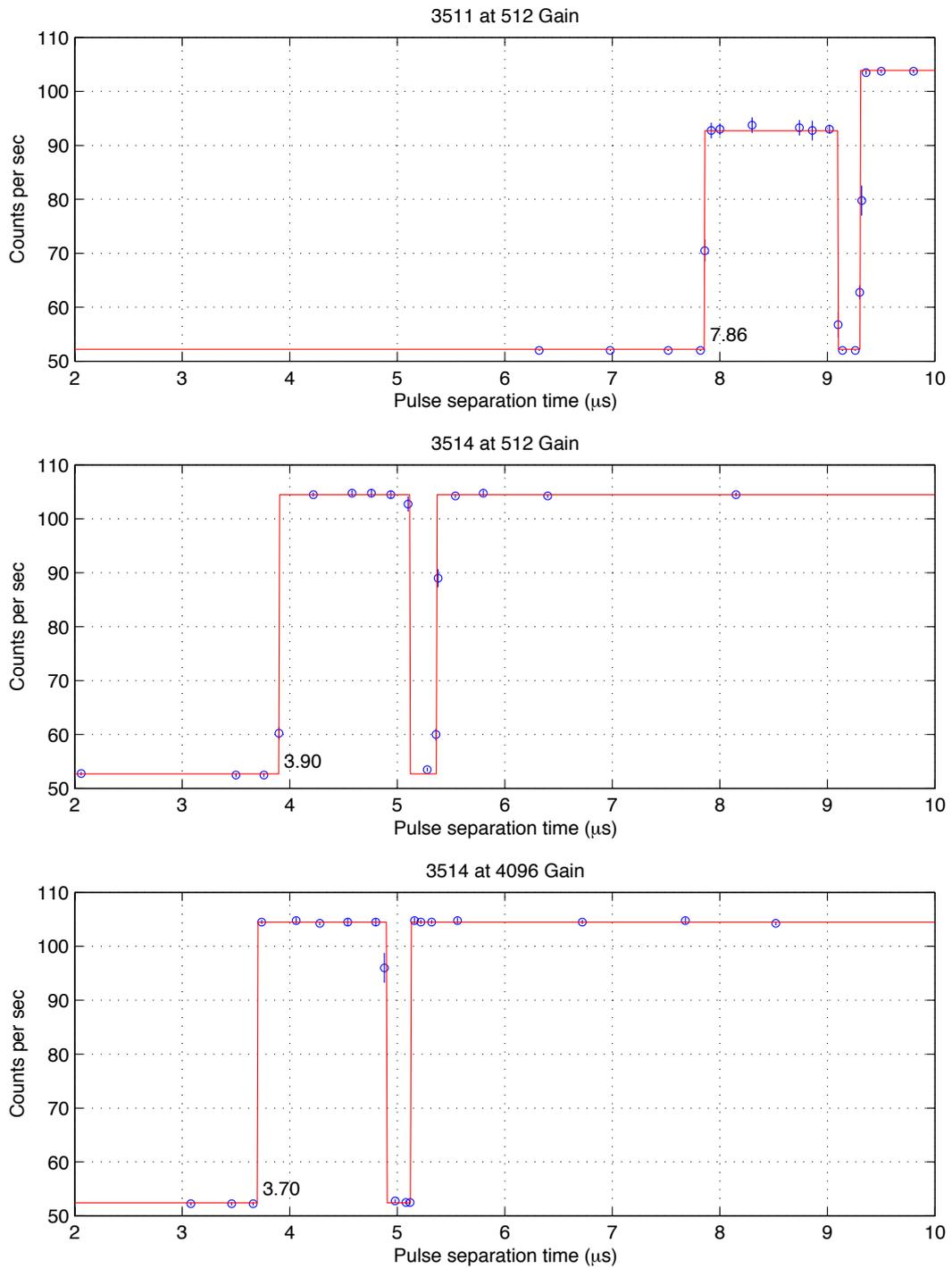

Figure 2.10: The observed output count rates from the ADCs for varying double-pulse separation times

### 2.8.2  HexScaler

The role of the HexScaler in the DAQ system is to store the current time value from the start of the Beam Off time which is typically incremented every second, while simultaneously allowing the value to be read by the CMC100 at rates of up to around 10 kHz. A problem with this mode of operation, called "the ripple error", was found by Barnett[2].

**The Ripple Error**

The Kinetic Systems 3615 HexScaler uses what are called asynchronous or ripple counting chips. An increment of the value stored in a binary representation on the memory is done using a chain of "*add one - carry over*" operations starting with the lowest order bit. For instance, if the binary pattern `1011` is to be incremented to `1100`, the state of the memory on the chip goes through the sequence: `1011`→`1010`→`1000`→`1100` (that is, in decimals: `11`→`10`→`8`→`12`).

It was found that if the counter is read during the above "ripple" process, the value of the immediate state on the chip is returned. The numbers containing ripple errors are always $2, 4, 8, \ldots 2^n$ less than the expected value. This could pose a serious problem in a high precision half-life measurement where the Event Mode Data must be known accurately. It was decided that a test was required to determine to what extent the Ripple errors would affect the results should be undertaken.

The previous test performed by Ref.[2] was initially repeated using the new system. The counter was set to increment every $10\mu s$. The differences between consecutive counter values, read at intervals of $1.7\mu s$, were calculated. It was found that a Ripple error occurred (that is if the difference observed was $\geq 2$) with a probability of 1 in 700 reads. Although this test demonstrates the Ripple Error effects, it does not use proper running conditions.

A new set of tests was done using the events from a $^{137}$Cs $\gamma$-source incident on a detector to trigger the reading of the counter. This simulates real experimental conditions where readings of the Hexcounter occur randomly in time (as opposed to at fixed intervals which may contain aliasing effects). Realistic time increments of 1ms - 1s were also used. Testing was done in several 40s or 80s blocks, the uncertainty quoted is $1\sigma$.

| Time Inc. (ms) | Nominal rate (s$^{-1}$) | Error probability (read$^{-1}$) |
|:---:|:---:|:---:|
| 1 | 10,000 | $(1.3 \pm 0.7) \times 10^{-5}$ |
| 10 | 10,000 | $(2 \pm 1) \times 10^{-6}$ |
| 1 | 5,000 | $(1.2 \pm 0.5) \times 10^{-5}$ |

Table 2.1: The Ripple error probabilities.

The results of these tests indeed showed the trend one would expect if the time taken for the rippling process was roughly constant. That is, the ripple error probability is expected to be proportional to the read rate and inversely proportional to the time increment. From these results the error probability for the typical time increment of 1s could be extrapolated, nonetheless, it was decided that the actual test should be performed to remove any doubt of what may be occuring.

With the time increment set at 1s and the read rate at 10,000 s$^{-1}$ nominally, two tests were done for a duration 19 and 59 hours. The longer test cleared the hexcounter every 400s, thus further resembling the conditions of a real experiment. The first test observed 10 ripple errors, while the second observed 33 ripple errors, both giving the same error probability of $2 \times 10^{-8}$ read$^{-1}$.

Initially it was thought that an increased ripple error probability should be observed in the first test since the counter used much higher values. Intuitively, this is because the higher counter values mean an increased length of the binary presentation. This should result in, on average, an increased required number of ripples per increment. Although this was correct, the order of the effect was calculated to be only a minuscule increase of 0.25% in the ripple probability which explains why this effect was not observed.

It is possible to device a method for removing the effects of ripple errors. This requires having two consecutive reads of the Hexcounter performed approximately 1$\mu$s apart by the List Sequencer (this is called "Double Hexcounter read" in E-Lifetime). Due to the short spacing between the two reads and the long interval between time increments it is guaranteed that, at most, only one of the two values returned will contain a ripple error.

If we arbitrarily define the time value to be at the occurrence of the first read, the recognition and handling of the four possible scenarios (of the Double reads shown in figure 2.12) will be required as follows to remove the effect of the error:

- If the second value returned is equal to or less than the first, then either Scenario (i) or (ii) has occurred. In both these scenarios, the time value returned by the first read is chosen.

- If the first value returned is less than that of the second, then either Scenario (iii) or (iv) has occurred. In both these scenarios, the time value of the second read decremented by one is chosen.

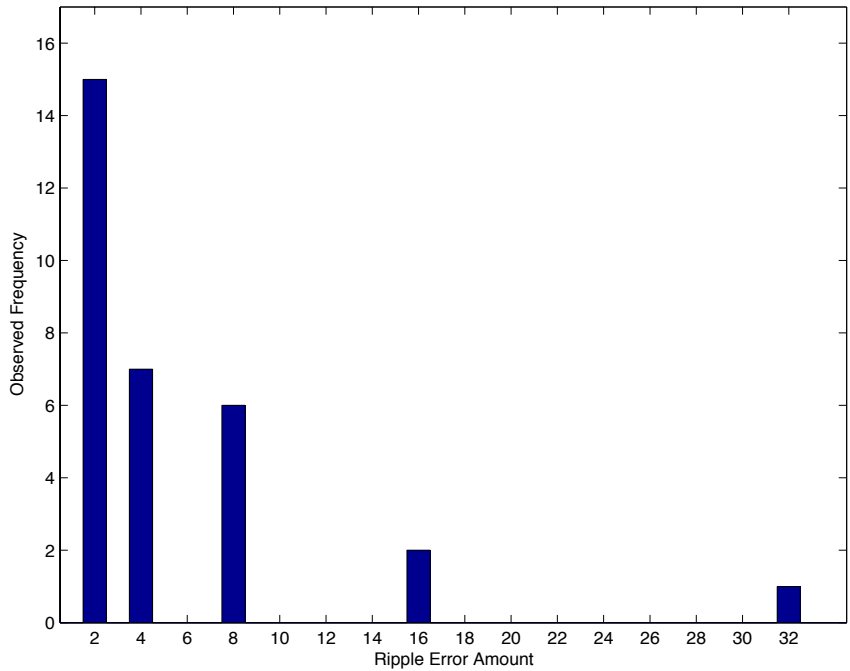

Figure 2.11: The distribution of the ripple errors observed

The implementation of the above procedure removes the effects of Ripple Errors completely. This occurs at the expense of an additional ∼1μs dead time per event collected, as well as the reduction in the possible events held in the Response Buffer by one third.

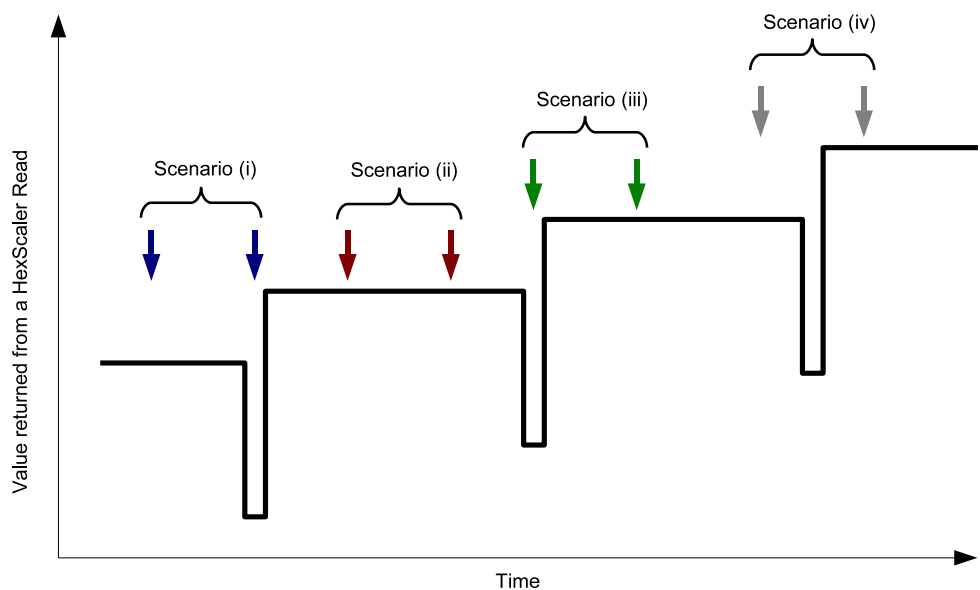

Figure 2.12: The possible scenarios taking place from a Double HexScaler read

**Quantifying the Effect from Ripple Errors**

Assuming that the ripple probability is proportional to the count rate, the expected number of ripple errors occurring in a pass is given by:

$$\langle \text{ripple errors/pass} \rangle = \int\limits_{t=0}^{t_{end}} \left( P_{N_0} e^{-\lambda t} \right) \left( N_0 e^{-\lambda t} \right) dt = \frac{P_{N_0} N_0}{2\lambda} \left( 1 - e^{-2\lambda t_{end}} \right)$$

(2.1)

where $N_0$ is the initial count rate, $P_{N_0}$ is the probability of a ripple error occurring per read at the initial count rate, $\lambda$ is the probability of the radioactive decay and $t_{end}$ is the duration of data collection.

The initial count rate is not expected to exceed $10{,}000\text{s}^{-1}$, using $P_{10{,}000\text{s}^{-1}} = 2 \times 10^{-8}\text{read}^{-1}$ as measured previously, $\lambda = ln(2)/20s = 0.03\text{s}^{-1}$ and $t_{end} = 300s$, typical parameters of the $^{10}$C half-life measurement, 0.003 number of ripple errors are expected to occur per pass; that is, one ripple error per 300 passes.

A single count shifting from one time bin to another every 300 passes is deemed to be a negligible effect compared to that of Poisson fluctuations intrinsic to radioactive counting results. From this it was decided that the
Double HexScaler Read method would not be needed for the remainder of the
experiment.

### 2.8.3   Homemade Sequencer & the BNC Output Box

The Homemade Sequencer when connected to the BNC Output Box allows
voltage levels to be produced from within Kmax. The voltage on the six
channels of the BNC Output Box is determined by the bit pattern on the
Control Register; bit $x$ controls the voltage on channel $x - 15$.

A NAF write command is used to alter the pattern on the Control Register
from within Kmax. A logic pulse is produced by the toggling of the value on
the Control Register using two successive Kmax commands. The duration of
a pulse produced this way is limited by the cycle time of commands in Kmax
to around 1 ms.

Voltages along separate channels should behave independently, unfortu-
nately this was not the case. It was found, using an oscilloscope, that when a
voltage level was activated on channel 2 ($0\rightarrow3$V) a pulse of -500mV amplitude
lasting 100ns was seen on the unused channel 7 (when the voltage level was de-
activated ($3$V$\rightarrow0$) a pulse of +700mV amplitude of 200ns duration was seen).
There appears to be some differential coupling between the two supposedly
independent channels.

Although surprising, this problem is not difficult to overcome. Since the
outputs carrying pulses control the target motor movement go through Single
Channel Analysers before being sent down the long cables to the Beam Room,
the increase of the lower discrimination level on these devices will filter out
the strange unwanted pulses. As for the channel that the DAQ gate exists on,
activity on the other channels only exists during the period when the gate is
off. This causes no problems as data collection is also disabled by a CAMAC
inhibit in the E-Lifetime program.

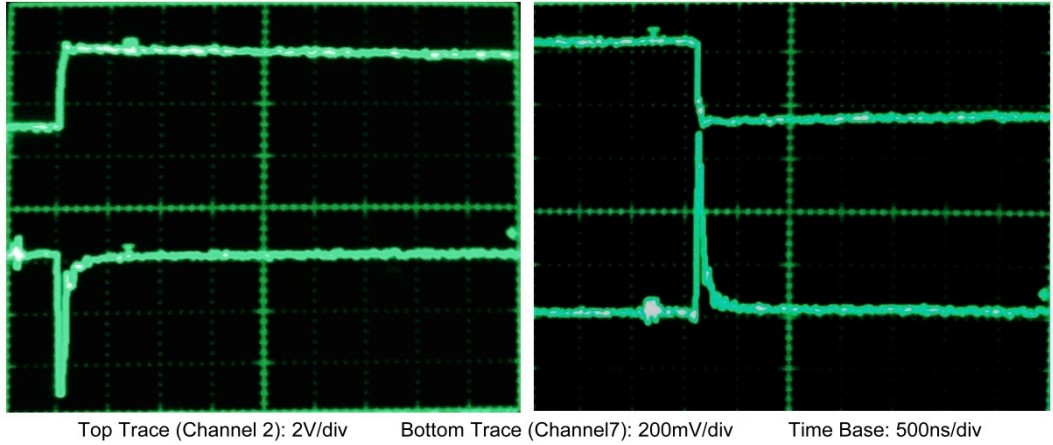

Top Trace (Channel 2): 2V/div    Bottom Trace (Channel7): 200mV/div    Time Base: 500ns/div

Figure 2.13: Images from the scope showing the coupling of the BNC Output Box channels

## 2.8.4 Time Phasing

The time histograms of the passes may be summed before fitting to reduce Poisson error. This is possible even if the absolute time phases between different passes are slightly out of phase (in this case by at most 1s) by discarding the data from the first and last time bins. This is because any time phase (and initial count rate) difference, $\Delta t_i$, between exponential decays when summed can be separated as a single constant of the fit:

$$\sum_i N_i e^{-\lambda(t+\Delta t_i)} = \left( \sum_i N_i e^{-\lambda \Delta t_i} \right) e^{-\lambda t} = N' e^{-\lambda t} \qquad (2.2)$$

The same thing can not be done for variations of the time scale during or between passes. This is because, in $\sum_i N_i e^{-\lambda(c_i t)}$, the time scale variation factor $c_i$ can not be separated from the decaying exponential.

The delay between the detection of an event by the system and the actual reading of the time corresponding to this event is due to the time required for the ADC to digitalise the pulse and set up the LAM which then triggers the time reading. The existence of this delay merely shifts the time phasing of all events by a constant amount, provided the delay is constant. If it is not constant, there will be a time scale fluctuations during the data collection process which is highly unwanted.

I then set out to measure the time between the strobe pulse arriving at the ADC and the reading of the HexScaler. The arrival of the strobe pulse going into the ADC can be observed from an output from the Gate and Delay Generator. Observing when the HexScaler was read proved to be more elusive.

The N LED on the front of the HexScaler indicated the time when the unit was accessed, that is, the occurrence of the time reading. But using the scope, it was found that the voltage across this LED was stretched to ∼1ms long by an IC circuit of some kind. This proved to be useless as I wanted to be sure that the delay remained constant for high data rates. Since the N line directly connects the HexScaler and the CMC100, modifications deep inside the circuitry of either would be required to measure the voltage on this line. This was to be avoided as both are crucial to the execution of the experiment.

The solution, instead, was to measure the voltage on the common Busy (B) bus line, by a modification of the Blinking Lights Box unit. A Busy signal is present whenever a N signal is present, thus the Busy line indicates when the HexScaler (as well as all other devices) is accessed. The Blinking Lights Box is a passive unit that sits on the CAMAC Bus, indicating when any of the lines are accessed via the LEDs on its front panel. The modification made was by soldering two wires onto the IC circuit chip (across a $100\Omega$ resistor to prevent damage to the chip in case shorting) which was connected to the Busy line pin on the back of the unit. The two wires were brought to outside the the unit so an oscilloscope could be connected easily.

What was observed on the oscilloscope is shown in figure 2.14. At first, the voltage on the Busy line looked very strange with logic pulses having been expected. But after some thought, it was decided that differential of the voltage on the Busy Line was actually being observed. This is most likely due to a misinterpretation of the circuitry (since no diagram existed) resulting in the voltage being measured across some kind of capacitive network. The confirmation that this is exactly what was happening comes by comparing the pattern produced when two and three NFA commands are present in the list sequencer.

The observations showed that the time between the strobe pulse and the

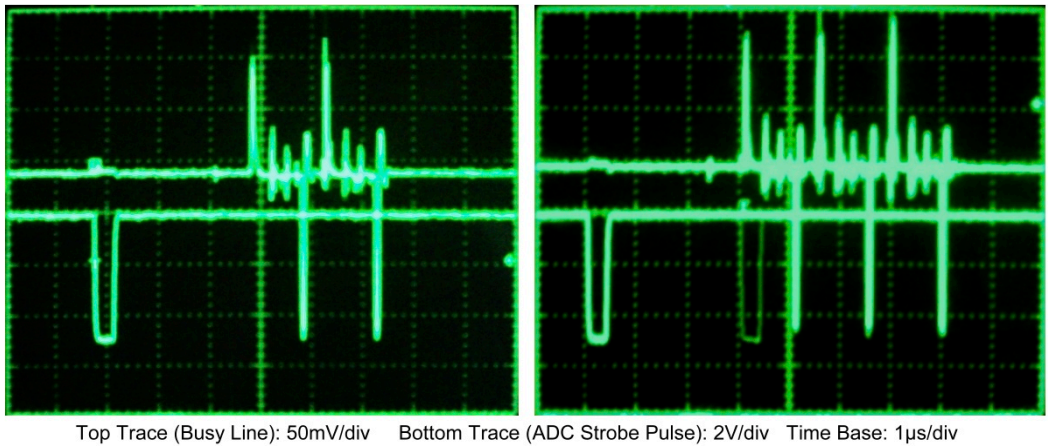

Top Trace (Busy Line): 50mV/div    Bottom Trace (ADC Strobe Pulse): 2V/div    Time Base: 1µs/div

Figure 2.14: Measurements of the ADC strobe pulse and the Busy Line voltage with a digital oscilloscope with two (left) and three (right) NAF commands executed by the list sequencer.

HexScaler Read command was constant to an accuracy (determined by the time resolution and thickness of the trace) of around 100ns. This test, along with the ones previously described, justifies the use of the current data acquisition in a high precision nuclear experiment.

# Chapter 3

# The Detection System

## 3.1 Introduction

The detection system consists of four plastic scintillation detectors as well as the associated electronics required in producing appropriate signals required by the DAQ system described in the previous chapter.

A scintillation detector consists of a scintillator, a light guide and a Photo-Multiplier Tube (PMT). Radiation produces optical light in the scintillator which, with the aid of the light guide, is collected by the PMT which converts the optical light into a measurable electrical pulse. Pulses of interest are selected by various NIM electronics and then collected by the DAQ system.

The construction and considerations required in attaining a scintillation detector of a standard suitable for our experiment are described in section 3.2. The detection principle and electronics system required are described in sections 3.3 and 3.4.

### 3.1.1 Scintillators in General

Energy from radiation such as charged particles or high energy photons (x-rays and gamma-rays) is deposited in a scintillator through atomic or molecular excitations. De-excitation results in the emission of scintillated light of a characteristic spectrum in the optical region in a process called fluorescence.

The two types of scintillators discussed here are: inorganic crystals such as sodium iodide (doped with thallium) NaI(Tl), and organic plastics, a solution of aromatic hydrocarbon crystals in a solid transparent plastic solvent. The two types undergo different scintillation processes, the former involving transitions of electrons in the crystal band structure, and the latter transitions of free valence electrons in molecular orbitals. Owing to the differing nature of the scintillation processes, in general, a different set of characteristics exists for each of the two types of scintillator.

### 3.1.2   Plastic Scintillators

Scintillated light pulses originating from plastic scintillators are very short in duration. This is because the characteristic decay time of the molecular de-excitations is only around 2ns; an order of 100 times shorter than that from sodium iodide detectors. This allows them to be used at much higher count rates without pile-up or gain shift problems.

Other practical advantages of plastic scintillators are that they are rugged, flexible and relatively inexpensive. This means that blocks of plastic scintillator can be obtained and used for the construction of large sized detectors with relative ease.

The major disadvantage of the use of plastic scintillators is poor energy resolution, that is the inability to resolve the energy of the incident radiation from analysis of the pulse height spectrum. For high energy photons (gamma energy region) this is primarily due to the combination of two properties: the relatively low light output and the lack of a distinguishing peak in the spectrum.

The light output of a scintillator is the conversion efficiency of deposited energy to optical light. This process follows Poisson statistics so that the energy resolution scales as the square root of the light output. Plastic scintillators produce around 10,000 optical photons per MeV of deposited energy, a factor of four less than sodium iodide crystals, thus the intrinsic energy resolution due to the difference in light output is a factor of 2 poorer.

In the energies we are interested in, the two primary interactions allowing a photon to deposit energy in a scintillator are the photoelectric effect and

| | NaI(Tl) | BGO | CsI(Tl) | Pure CsI | BaF₂ | GSO: Ce | Plastic | LSO: Ce | PWO | YAP: Ce |
|---|---|---|---|---|---|---|---|---|---|---|
| Density (g/cm³) | 3.67 | 7.13 | 4.51 | 4.51 | 4.88 | 6.71 | 1.03 | 7.35 | 8.28 | 5.55 |
| L$_{rad}$ (cm) | 2.59 | 1.12 | 1.85 | 1.85 | 2.10 | 1.38 | 40 | 0.88 | 0.87 | 2.70 |
| Refractive Index | 1.85 | 2.15 | 1.80 | 1.80 | 1.58 | 1.85 | 1.58 | 1.82 | 2.16 | 1.97 |
| Hygroscopic | Yes | No | Slightly | Slightly | Slightly | No | No | No | No | No |
| Luminescence | 410 | 480 | 530 | 310 | 325 | 430 | 400 | 420 | 470 | 380 |
| (nm) | | | | | 220 | 310 | | | | |
| Decay time | 230 | 300 | 1000 | 10 | 630 | 30 | 2.0 | 40 | 15 | 30 |
| (nsec) | | | | | 0.9 | | | | | |
| Relative Light Output | 100 | 15 | 45 to 50 | <10 | 20 | 20 | 25 | 70 | 0.7 | 40 |

Figure 3.1: Common scintillator characteristics (from Hamamatsu[25]). The two main properties of sodium iodide and plastic scintillators discussed are circled.

Compton scattering. The former is an absorption process where all of the incoming photon energy is deposited thus producing a sharp peak (called "the photo-peak") in the energy spectrum. The latter is a scattering process where only a fraction of the incoming photon energy is deposited depending on the scattering angle. This produces a continuous distribution (see section 3.2.7) from zero energy to the Compton Edge energy, the maximum possible energy deposited from 180° back scattering. The energy of the photo-peak is always slightly higher than that of the Compton edge.

The cross-section for the photoelectric effect depends roughly on $Z^5$ (where Z is the atomic number of the atoms involved). While the cross-section for Compton scattering depends only linearly on Z[24]. Hence interactions in plastic scintillators is predominantly of the Compton scattering type. This results in an energy spectrum containing just the continuous Compton distribution with no photo-peak. This makes plastic scintillators almost impossible to use for most spectroscopy purposes.

## 3.2 Detector Construction

Four plastic scintillation detectors were constructed by me at the University of Auckland's Physics Department. The process included optical and structural coupling of the main components of a detector (plastic scintillator, light guide and PMT) using optically transparent epoxy glue, reducing the amount of scintillated light loss by painting the external optical surfaces with reflective paint and sealing it off from ambient light by wrapping it in reflective aluminium foil.

In this section the construction process of the detectors is described in detail. This is done to provide a reference for the four completed detectors as well as to serve as a guide for interested detector builders.

### 3.2.1   The Plastic Scintillator Blocks

The BC400 Premium Plastic Scintillator (NE102 equivalent) from Saint Gobain Crystals[26, 27] was used. This is a "general purpose" organic scintillator dissolved in a polyvinyltoluene plastic base. It has a light output of 30% relative to NaI(Tl), a peak emission wavelength of 423nm (see figure 3.3 for emission spectrum), a decay constant of 2.4ns, a bulk light attenuation length of 250cm and a refractive index of 1.58.

A large rectangular block of dimensions $50 \times 100 \times 430$mm was obtained. The large block was cut into four smaller blocks of $50 \times 100 \times 100$mm by the Physics Department's workshop at the University of Auckland. All the scintillator surfaces have to be highly polished to encourage total internal reflection. It was found that the polishing done by the manufacturers were of a higher quality than that of the workshop's which had visible"lathe" marks. With hindsight, due to the soft nature of the scintillator material, extra care should have been made when submitting the polishing specifications to the workshop.

The thickness of the original block from the manufacturers is only approximate; the actual thickness was found to vary between 50-51 mm along its length. This meant that the dimensions of the scintillator did not always exactly match those of the light guide. This was estimated to cause a light loss of around 1-2% which is negligible compared to other light loss factors.

### 3.2.2   The Photomultiplier Tubes

A PMT is a standard device used for the conversion of scintillated light to electrical pulses. It consists of an input window, a photocathode, focusing electrodes, an electron multiplier (dynodes) and an anode all housed in a vacuum glass tube (see figure 3.2). Scintillated light strikes the photocathode material causing the emission of photoelectrons by the photoelectric effect. The emitted photoelectrons are directed by the focusing electrodes toward the electron multiplier part of the tube. The photoelectrons strike the first dynode of the electron multiplier causing the emission of secondary electrons. The secondary electrons emitted from the first dynode are accelerated toward the second dynode causing the emission of more electrons. This cascading process continues for each of the remaining dynodes, amplifying the flow of charge at each stage. The charge is finally collected at the anode where it is integrated.

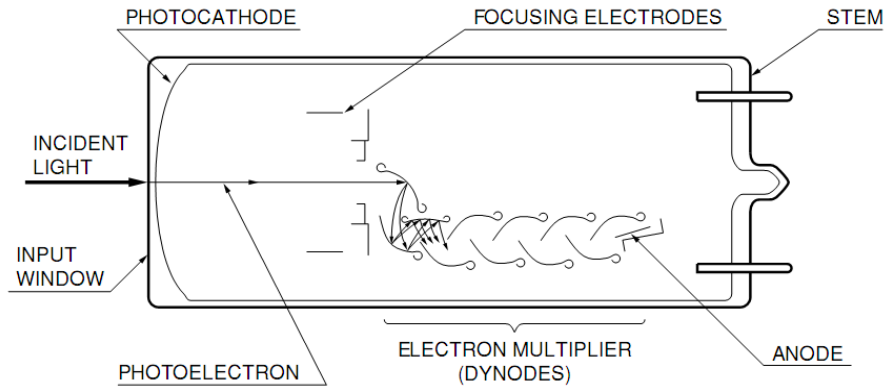

Figure 3.2: Diagram of the parts of a linear focussed PMT

The PMTs used are the R329-02 model from Hamamatsu Photonics. This model has a tube diameter of 51mm[cathode diameter?] with a linear focusing dynode structure (as shown in figure 3.2) consisting of 12 dynodes. The main advantage of this PMT design is the fast time response. Typical characteristics given by the manufacturers[25] for this PMT are: 2.6 ns rise time (the time taken for the anode output pulse produced from a delta-function light pulse to rise from 10% to 90% of the peak amplitude), a 48ns transit time (the time interval between the arrival of a delta-function light pulse at the photocathode and resultant pulse at the anode output) and a 1.1 ns transit time spread (FWHM of the frequency distribution of transit times).

To maximise efficiency of the detector it is important to match the spectral response of the photocathode to the characteristic spectrum of the light emit-ted from the scintillator. The spectral response is the photocathode material's sensitivity for the production of photoelectrons through the photoelectric effect as a function of incident photon energy. It is usually expressed in terms of the quantum efficiency, defined as the ratio of the number of emitted photoelectrons per incident photon. The spectral response of the bialkali photocathode used in the R329-02 PMTs and characteristic emission spectrum of the BC400 plastic scintillators used are well matched (see figure 3.3).

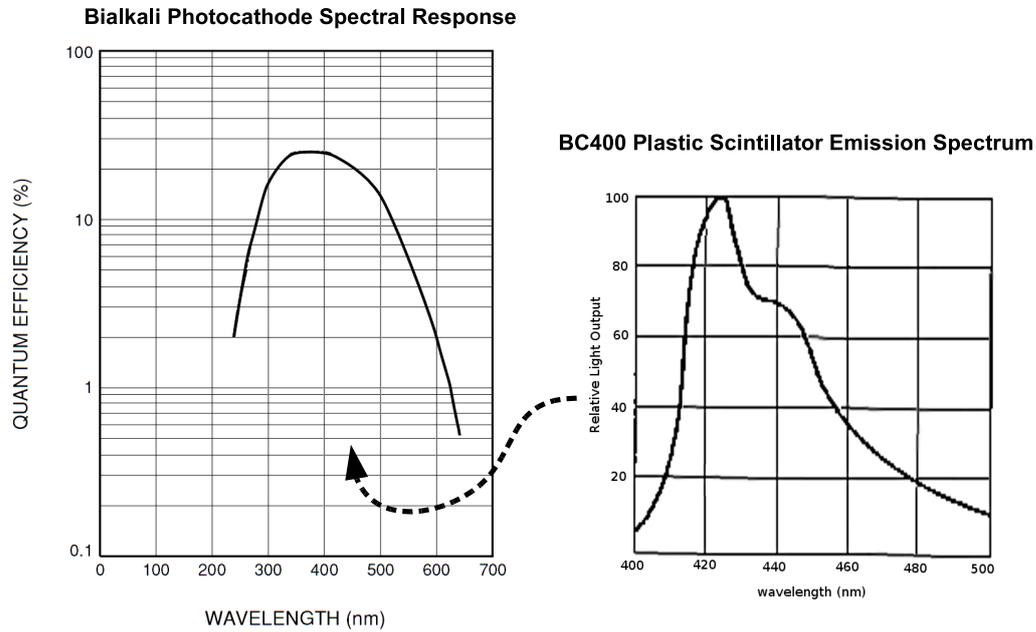

Figure 3.3: The spectral response of the bialkali photocathode and emission spectrum of the BC400 plastic scintillator (from Hamamatsu[25] and Saint-Gobain[26] respectively).

## The Bases

The voltage distribution across the dynodes in the electron multiplier section is established by a voltage divider network housed in a section attached to the end of the PMT called "the base". A base consists of resistors and capacitors that divide a single high voltage input (typically on the order of 1000V) to the distribution required at the dynodes for the electron multiplication process. It also contains a capacitive circuit that is connected to the anode so that the anode integrated charge produces a voltage difference across the external output.

As one can guess from their names, the anode is required to be at a higher potential than the cathode in order for the electrons to be accelerated towards it. Two schemes are thus possible: the anode grounded and the photocathode at a high negative potential or the photocathode grounded and the anode at a high positive potential.

In practice the latter scheme, and the one used by our E5859-03 bases (which are also from Hamamatsu), is usually applied in scintillation detectors. That is because the scintillator (and light guide) is usually grounded by an external metal case or by aluminium foil. If the photocathode is at a high potential, a small discharge current may flow through the PMT glass window creating noise in the photocurrent from the cathode before the amplification process[25, 24].

The trade off in having the anode at a high potential is the requirement of a coupling capacitor in order prevent the DC level signal produced by the potential from appearing at the output. This has the effect of reshaping the output pulses causing a loss in time information[24].

### 3.2.3 The Light Guides

Funnel shaped perspex (polymethyl methacrylate or acrylic glass) light guides are used to optically couple the $50\times100$ mm rectangular faces of the plastic scintillators to the 51 mm diameter circular input windows of the PMTs. The light guides were produced by the University of Auckland's Physics Department workshop. See figure 3.4 for the design.

The dimensions of the two end faces of the light guide are obviously set to match the optical surfaces to be coupled. The length, however, determines the effects of several factors.

A longer light guide better approximates what is known as an adiabatic light guide; a light guide of constant cross-sectional area and slowly changing geometry. In fact it can be shown[28] that the maximum amount of light transferable by total internal reflection occurs for the case of an adiabatic light guide. Theoretically this amount is given by the ratio of outgoing area to the incoming area of the light guide, $A_{out}/A_{in}$ (which in our case is approximately 0.4).

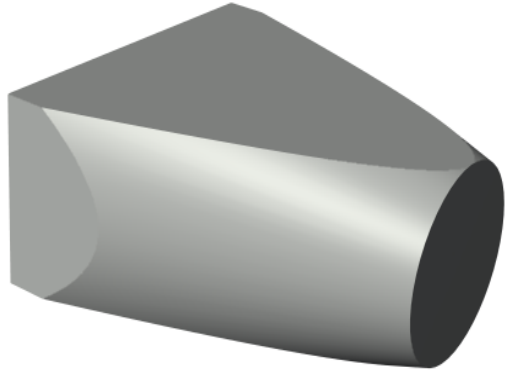

Figure 3.4: 3D CAD drawing of the light guides used

Another advantage of having a longer light guide is that a scintillation event close to the PMT results in greater light collection than that from an event further away. A long light guide reduces this variation giving an improvement in the overall detector response homogeneity[29].

In the end, a structurally robust detector is the most desired property because regardless of light collection efficiency in a detector, it serves no purpose if it is broken! This is obviously achieved with a shorter light guide. A length of around 10cm was chosen. With this modest length, it becomes necessary to use a reflector on the surfaces of the light guide (see section 3.2.5) since total internal reflection can not be solely relied upon.

### 3.2.4   The Gluing Process

Selleys Ultra Clear Araldite[30], a two part epoxy glue with a 5 minute curing time at room temperature, was used for both the scintillator to light guide and light guide to PMT optical joints.

The material providing the optical interface should have an index of refraction close to that of the two materials to be coupled in order for transmissive losses to be minimised. The glue used has a refractive index of n = 1.54, which satisfies the criterion very well for the plastic scintillator and perspex light guide, which have refractive indices of 1.58 and 1.49 respectively.

Typically, transparent silicone grease (n = 1.47) is used for the light guide to borosilicate glass window of the PMT interface[24, 27] (both having n = 1.49). The grease, however, does not provide structural support other than from surface tension effects, hence, mechanical support is usually provided for by a padded or spring loaded metal enclosure attached to the PMT. This design was decided against as the scintillators are required to be as close together as possible in the experimental setup (see section 3.3).

From Fresnel's equations[31], assuming the scintillated light is randomly polarised, it can be calculated that the increase in the transmissive loss due to reflection across the optical interface when the glue is used instead of the silicon grease is only 0.02% for normal incidence and 0.1% for 45° incidence. Thus the use of the glue for the perspex light guide and PMT glass window is well justified.

**Methodology**

The main difficulty in the gluing process is providing good mechanical support while the epoxy glue is left to set over several hours. Practice runs were done on several miss-cut light guides (with similar dimensions to the real ones) and a malfunctioning PMT. This provided valuable lessons on learning the techniques required to produce scintillator to light guide and light guide to PMT glue joints of a high quality.

Two different glue joints are required to be made for each detector. Instead of building the four detectors one by one, it was decided that the same type of gluing were to be grouped together and done separately. The scintillator to light guide gluing was chosen to be done first primarily because there were 5 scintillators and 5 light guides but only 4 PMTs so that if the scintillators and light guides were glued together first, we would have the option of selecting the best four out of five done to then be glued to the PMTs.

A tried and tested procedure used to perform the gluing was as follows:

1) The surfaces to be glued were cleaned. This was done by washing with mild detergent mixed with warm water, followed by rinsing with filtered water, then finally drying thoroughly with lens tissue. The use of alcohol for cleaning and standard cloth for drying was avoided as it was found to leave streak marks or scratches on the highly polished surfaces.

2) A metallic plate, heated to around 70-80°C with hot water, was used to mix the glue on. This heated the glue giving it a more "runny" texture (lower viscosity) which allowed the glue to be spread more easily hence reducing the amount of air bubbles formed. The trade off was that the higher temperature increased the curing time of the glue by a few minutes.

3) Equal proportions of the two parts of the epoxy glue were used. Around 5-10 ml of glue was needed for the smaller PMT surface and around 10-15 ml for the larger plastic scintillator surface.

4) The glue was mixed gently to minimise the number of trapped air bubbles. Any large air bubbles that may have formed in the mixture were removed by pricking them with the corner of a plastic rod.

5) The glue was slowly poured on to the centre of the surface to be glued. A circular blob for the round PMT surface and a dumb-bell shape blob for the rectangular scintillator surface was found to work best.

6) The two surfaces were slowly pressed together to carefully spread the glue across the interface. Lapping the surfaces together in smooth circular motions was then done to increase uniformity of the glue across the interface and also to remove air bubbles that may have been trapped in the glue by bringing the pocket of air out to the edge of the interface. This technique was found to work very effectively at removing unwanted air bubbles.

7) If a satisfactory glue joint was not achieved within 5 minutes of when the glue was first mixed, the two surfaces were immediately pulled apart and glue cleaned off with lens tissues and a relatively weak solvent such as ethanol or iso-propanol. Any glue that remained dried on was later scraped off with lens tissue soaked in solvent wrapped around a plastic ruler to prevent scratching the polished surfaces.

8) The components, once glued, were then placed in a mechanical structure to provide pressure on the joint and keep good alignment of the pieces. The structures were designed so that the glue joints were horizontal in order to avoid non-uniformity in the thickness of glue due to gravitational effects. Usually the glue was left overnight to reach its maximum bond strength. Any unwanted glue that may have leaked out of the joint and dried on to the external optical surfaces was removed with the scraping technique as described previously.

*Scintillator to light guide*

9) Duct tape was wrapped around the glue joint to reduce the sliding of the surfaces and also to provide a temporary surface for the clamps used for keep alignment. The tape, chosen for its low stretchability, offered reasonable support but minor movements of the surfaces were still possible. It was found that the adhesive from the duct tape seeped into the glue joint due to the slight relative movements before applying the clamps. Thus before using the duct tape, plastic insulation tape, with the non-adhesive side facing outwards, was taped on to the duct tape to prevent contact of any adhesive directly with the edge of the glue joint.

10) The scintillator and light guide assembly was then placed in the mechanical structure as shown in figure 3.5. A large clamp maintained downward pressure on the glue joint, while smaller clamps (two for the longer edge and one for the shorter edge) were used to keep alignment of the pieces. Care was taken not to overly-tighten the clamps as this might have caused mechanical deformation, particularly in the softer plastic scintillator, such that when the glue had dried and the pressure removed, mechanical relaxation might have cracked the glue. Also the clamp tightnesses were not re-adjusted after the glue had had sufficient time to harden (around 10mins) as minor movements at this stage might have caused the appearance of stress features in the glue.

*PMT to Light guide*

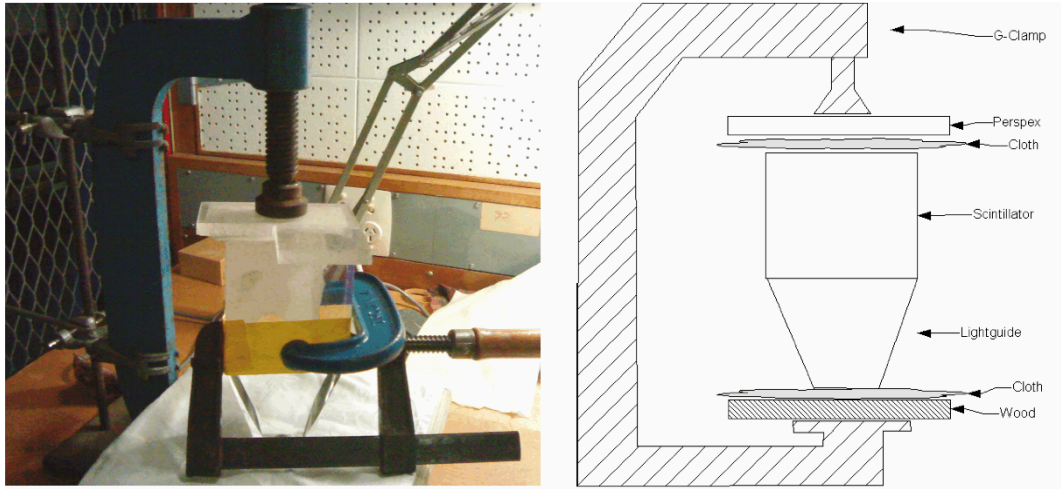

Figure 3.5: A photograph and diagram of the mechanical structure used for the gluing of the light guide to scintillator optical joint.

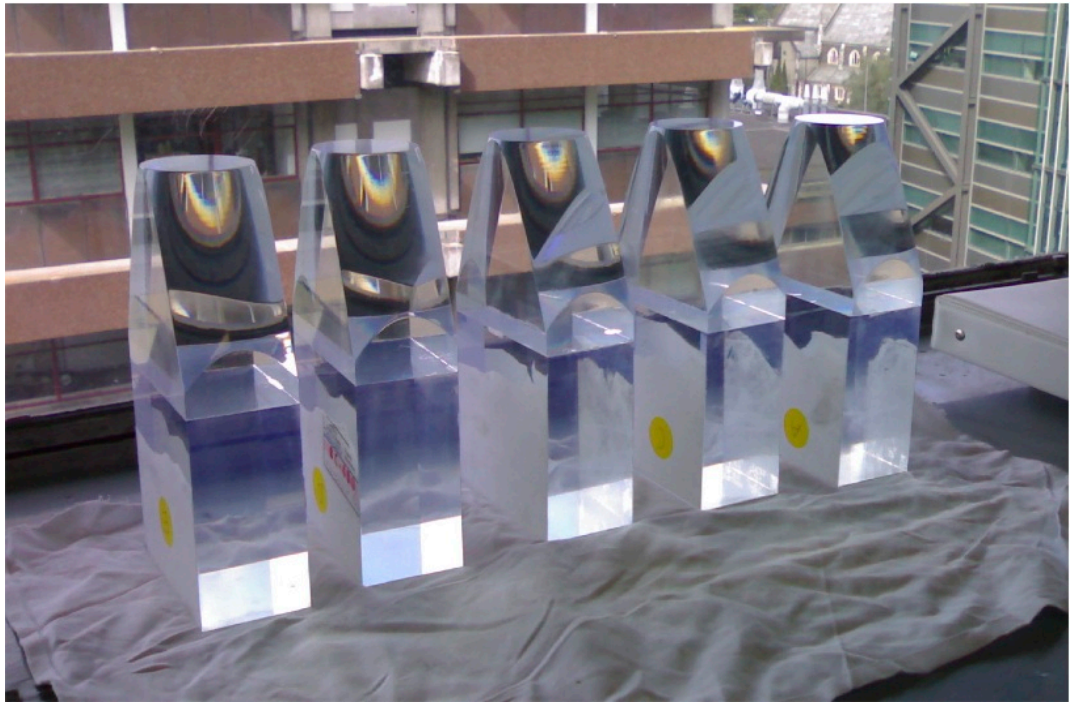

Figure 3.6: A photograph of the five scintillator to light guide assemblies optically coupled together with epoxy glue.

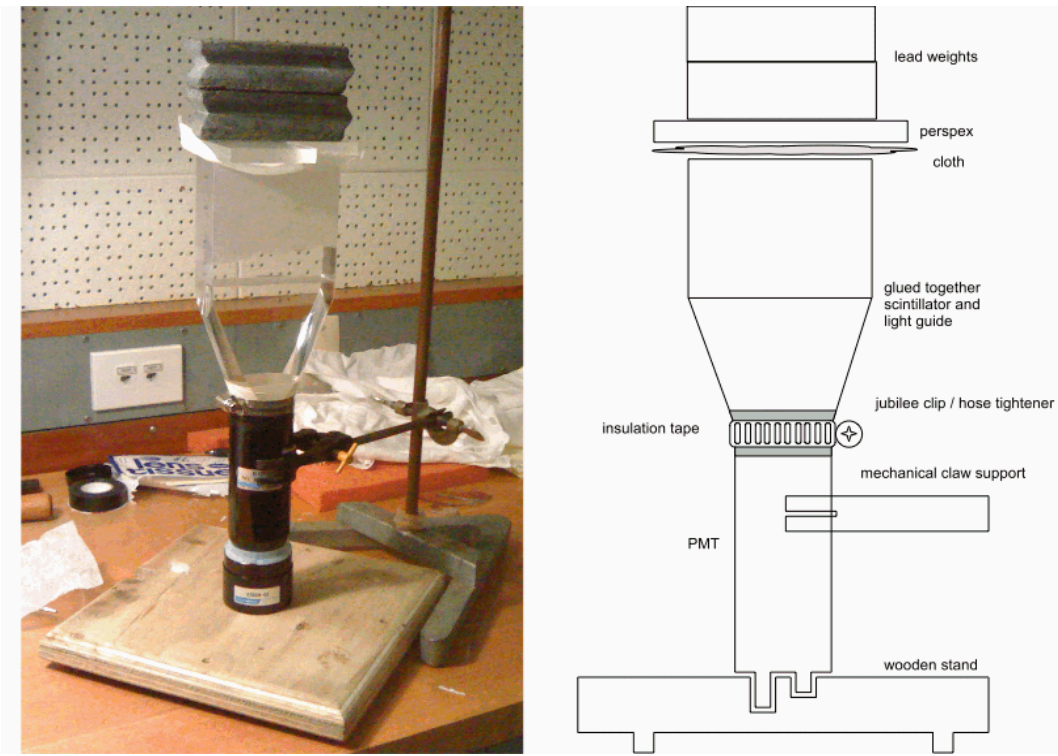

Figure 3.7: A photograph and schematic of the mechanical structure used for supporting the scintillator to light guide assembly and the PMT during the gluing process.

11) Downward pressure on the joint was provided by lead blocks while alignment was kept using a jubilee clip, see figure 3.7. Note the use of insulation tape wrapped around the light guide and PMT, applied before the start of the gluing, to prevent the steel jubilee clip from scratching the surfaces.

12) Due to the tapered edge of the light guide, care had to be taken when positioning and tightening the jubilee clip otherwise the light guide could be lifted off the PMT surface. The successful procedure found was to tighten the clip until it was just free to slide up and down (done before the gluing). Then while the light guide was held in alignment with respect to the PMT, the clip was raised slowly until it gently touched the light guide evenly all around. The clip was then only tightened by a very small amount just enough for the it to be fixed.

The gluing was completed in under two weeks and, overall, the process was quite successful. By mix-and-matching the ordering of the two different glue joints of a detector, four detectors having similar optical properties were produced. The optical quality was judged from simply looking down the axis of the detector.

### 3.2.5   The Internal Reflector

Our scintillators and light guides combined do not have a large length to width ratio. This means that there will be a high light loss if total internal reflection (critical angle of a perspex or scintillator to air interface is around 40°) was solely relied upon to keep the scintillated light trapped in the detector. The use of reflecting material on all the external surfaces of the scintillator and light guide was thus required.

The two types of internal reflectors commonly used are aluminium and titanium dioxide ($TiO_2$). Aluminium is almost always used in a foil form. It is a specular reflector, that is it follows the law of (mirror-like) reflection which states that the angle of the reflected light is equal to the angle of incident light. Titanium dioxide is used as an amorphous powder or in a paint form. It is a diffuse reflector, that is the angle of reflection is independent of the angle of incidence and follows Lambert's cosine law:

$$dI/d\theta \propto cos\theta \qquad (3.1)$$

where $I$ is the intensity of the reflected light and $\theta$ is the angle of reflection with respect to the normal.

There are two ways of using reflectors: direct coupling onto the optical surfaces or indirect coupling across an air gap. In the former case the reflector is in direct contact with the optical surface, essentially removing the possibility of total internal reflections because the reflector usually has a refractive index equal to or higher than the plastic light guide and scintillator.

Despite many theoretical arguments for the use of a particular reflector and coupling method over another in a particular circumstance[29] the best demonstration, in the author's opinion, should come from experimental stud-ies. Two such studies[32, 33] have shown little difference between the use of diffuse or specular reflectors. Although the dimensions of the scintillators and light guides involved in the studies are not identical to ours, it is still believed that the results obtained would still be applicable. In the range of our length to width ratios, both studies have shown around a 10-15% increase in light yield when indirect coupling of the reflector was used.

This slight increase in collection efficiency (when compared to the few percent of scintillated light that produce a photoelectron in the photocathode that will actually reach the first dynode of the PMT) is not significant enough to justify going through the added difficulties in obtaining the materials for indirect coupling and the consequent drop in final yield in not being able to have the detectors closer together.

**Methodology**

The paint used is an acrylic based titanium dioxide rutile paint from Golden Colors. It has an opacity rating of 2 (highest possible is 1) and has excellent lightfastness and permanency properties[34]; that is the colour of the paint remains unchanged with both prolonged light exposure and age.

Five thick coats of paint were directly applied onto the scintillator and light guide surfaces with a minimum of 3 hours of drying time in between coats. Increased thickness of the paint gives an increase in the opacity of the surface, see figure 3.8, which indicates more reflection at the surface. There is no real upper limit to the thickness of paint since the detectors are to be used for the detection of gamma rays, as opposed to charged particles, and so attenuation is negligible.

## 3.2.6 Other Details

**Unwanted Ambient Light**

The amount of scintillated light produced by a gamma ray in our detectors is on the order of $10^4$ photons. If ambient light were to leak into a detector, a small amount of equates to a very number (easily $>10^{10}$) of photons. This would produce a high count of unwanted low amplitude pulses in the PMT. Although these are removed at an early stage in the electronics, high count rates should still be avoided. Exposure to large amounts of ambient light may also result in instability effects in the PMT or even destroy it entirely[24].

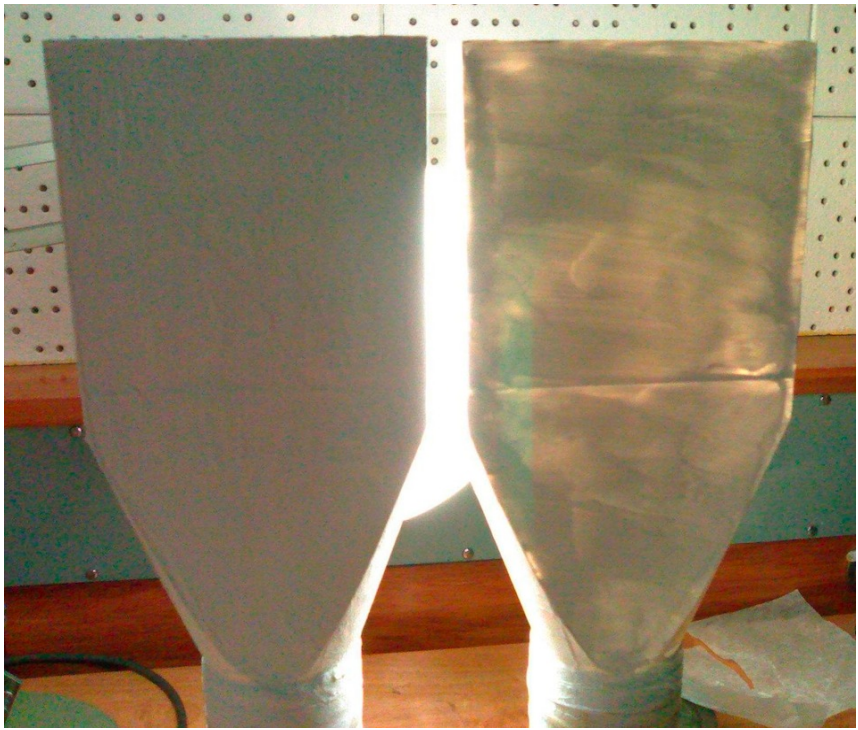

Figure 3.8: Photograph taken with a bright light source behind detectors with 3 layers (left) and 1 layer (right) of titanium oxide paint.

Catering grade aluminium foil[35] from Ullrich Aluminium was wrapped around the detectors to seal them off from ambient light. This high quality foil has a nominal thickness of 15 microns and was chosen to be void of tiny pin-sized holes that may be present in low grade aluminium foil (12 microns thick). The added thickness made the foil a lot less susceptible to tearing and ripping from the folding and wrapping of the foil around sharp edges and corners of the detectors. All detectors were wrapped so that one of the side faces only had a single thickness of foil. This was done so that the detectors could be placed closer together and attenuation reduced for emitted radiation in the plane of the aluminium foil surface. The front face was also made to be a single thickness of foil to minimise attenuation of head on radiation.

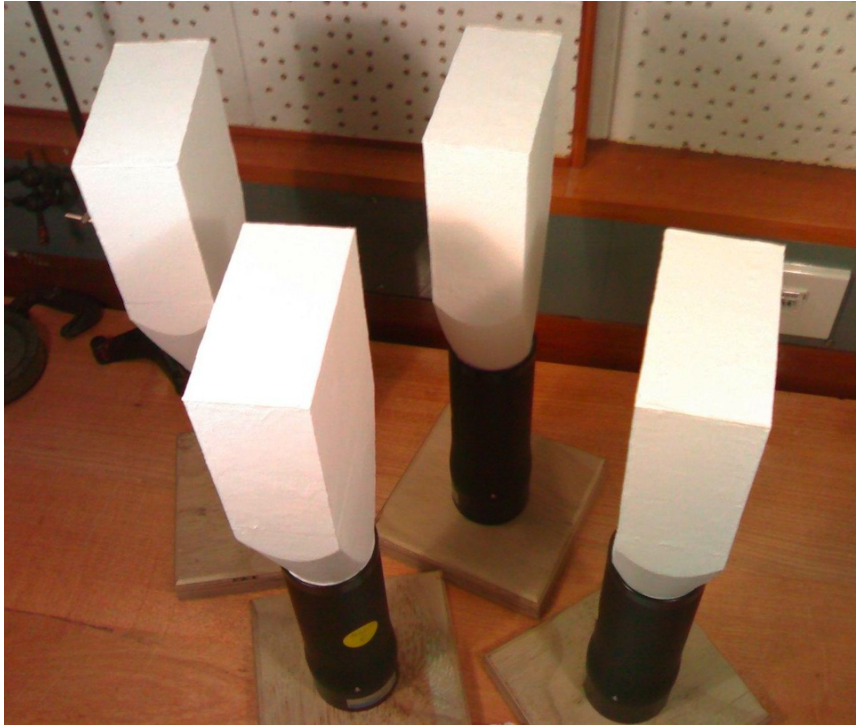

Figure 3.9: Photograph of the four detectors painted and with the its magnetic shielding and bases attached.

To further reduce ambient light and to cover up tiny holes in the foil which appeared after the shipping of the detectors from Auckland to Canberra, several of them were completely wrapped in black electrical tape.

**Magnetic Shielding**

Magnetic fields can alter the path of the electrons in the PMT assembly causing them to deviate from their designed trajectory resulting in a reduction of the gain. The most susceptible part of the PMT to magnetic fields is the electron focusing section. Any unwanted deflections here might prevent the electrons from the photocathode ever reaching the first dynode.

The PMT is enclosed in a mild steel cylinder to shield it from magnetic fields. Ideally the shielding should extend past the front face of the PMT by at least the radius of the tube[25]. In our case, due to the tapering of the light guide, the shield can only extend several millimeters past the front face. This is sufficient shielding for operation in the presence of low magnetic field strengths[24].

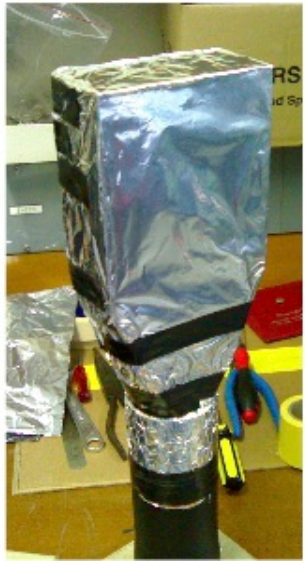

Figure 3.10: A photograph of a detector wrapped in the catering grade aluminium foil.

### 3.2.7   Review of Detectors

**Pulse Shapes**

The pulses caused by gamma rays in the detectors and observed in an oscilloscope are shown in figure 3.11. Ability to distinguish consecutive events is determined by the width of the pulses in time which is around 25ns in our detectors.

The shape of the voltage pulses from a scintillation detector is determined by many factors such as, the decay time of the scintillator, the time spread in the light collection process, transit time spread in the PMT and the time constant of the PMT's anode circuit. To a good approximation[36] the shapes of the voltage pulses coming from the detectors V(t) can be described by (with

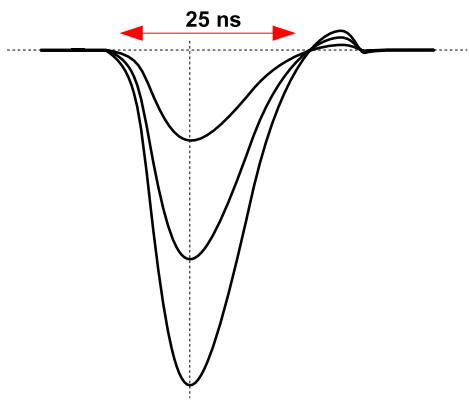

Figure 3.11: Drawing of the pulses from our plastic scintillator detectors as observed on an oscilloscope.

initial condition V(0)=0:

$$V\left(t\right) = \frac{1}{\lambda - \theta} \frac{\lambda Q}{C} \left( e^{-\theta t} - e^{-\lambda t} \right) \qquad (3.2)$$

where $\lambda$ is the reciprocal of the scintillator's decay time constant, $\theta$ is the reciprocal of the anode circuit time constant, $Q$ is the total charge collected by the anode over the entire pulse and $C$ is the capacitance of the anode circuit.

The amount of scintillated light produced in the scintillator is proportional to the energy deposited by the incident gamma photon. Since the PMT is a linear, the amount of charge collected will also be proportional to the energy deposited.

For differing amounts of charge $Q$ collected at the anode of our detector, only the amplitude and not the shape of the pulses should vary. This is indeed what was observed in our detectors.

**Pulse Height Spectra**

A typical pulse-height spectrum obtained in E-Lifetime of a $^{137}$Cs (0.662MeV) gamma source with the PMT operated at 1500V nominally is shown in figure 3.12. The spectrum is characterised by a continuous Compton distribution and the "Compton edge peak". The Compton edge corresponds to 180° back-scattering of the incoming gamma ray in the scintillator and lies at an energy

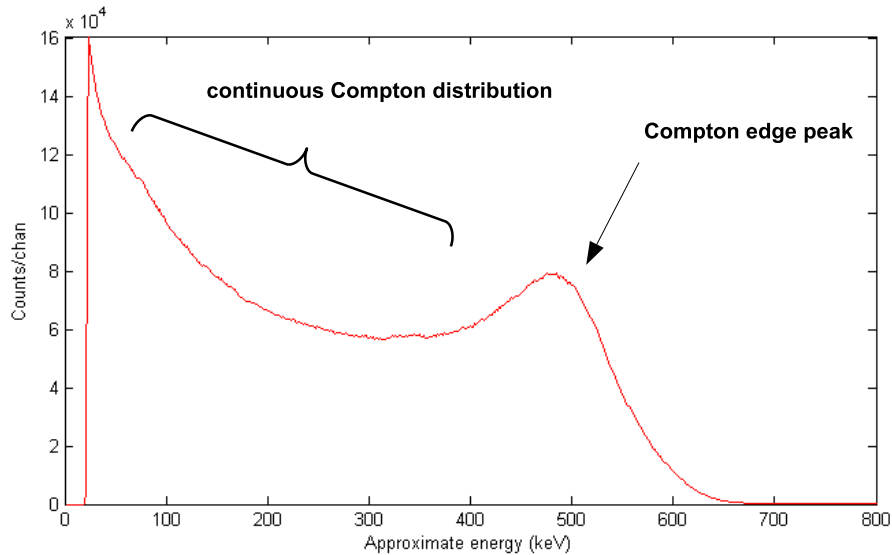

Figure 3.12: Typical pulse-height spectrum of a $^{137}$Cs (0.662MeV) gamma source with the PMT operated at 1500V.

$\Delta E_C$ less than the incident gamma-energy given by:

$$\Delta E_C = \frac{h\nu}{1 + 2h\nu/m_e c^2} \tag{3.3}$$

where $h\nu$ is the energy of the incoming gamma and $m_e c^2$ is the rest mass energy of the electron. This expression is only true for Compton scattering with free or unbounded electrons. The effects of the binding energy of the electron is usually small and masked by other energy resolution factors in our detectors.

The origin of the Compton edge appearing as a peak in plastic scintillator spectra is often misinterpreted[37]. Why is there a peak when the cross-section for a 180° Compton back scatter is a minimum? The reason is because the pulse-height spectrum is essentially a plot of the number of electrons per unit recoiling energy interval (not per unit scattering angle). The maximum in the recoiling electron's kinetic energy occurs for 180° back scattering. Combined with smearing effects described below, a peak is produced in the pulse-height spectrum.

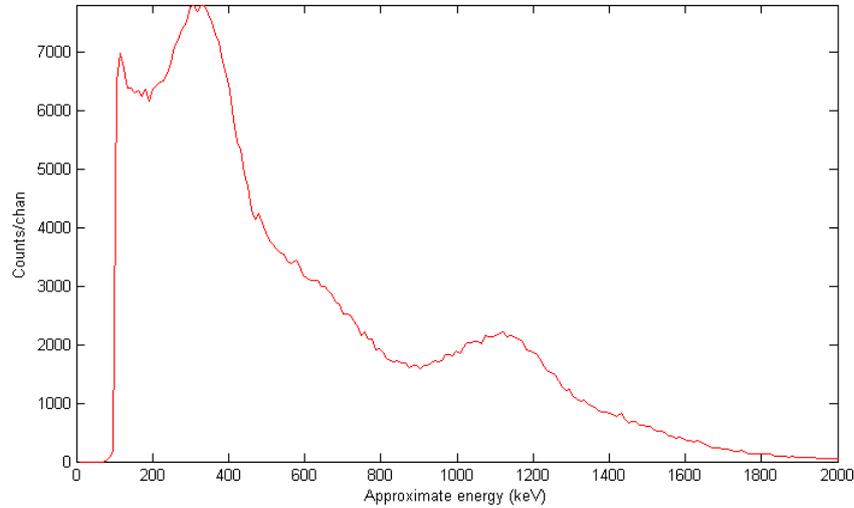

Figure 3.13: A spectrum of a $^{22}$Na positron source

**Energy Resolution**

The main effects that can smear the Compton edge peak feature in a plastic scintillator spectrum[38] are: variation in the light output of the scintillator, the occurrence of one or more Compton interactions within the scintillator and variation of light collection efficiency with the position of the event in the scintillator. The first is intrinsic to the scintillating material while the latter two are determined by the scintillator size and geometry.

A typical spectrum of a $^{22}$Na source (which emits 2 ×0.511MeV & 1.28MeV gammas for each decay) taken with our detectors is shown in figure 3.13. As one can see, it is possible to resolve the two Compton edge peaks that exist at 0.341MeV and 1.067MeV respectively in our detectors.

**Detector Gains**

Many factors determine the gain of a detector. Factors from the construction process such as optical glue joint quality and internal reflector opacity alter the scintillated light collection efficiency and may result in small changes of the gain of a detector. But by far the dominant factor comes from the early stages of the PMT. The quantum efficiency of the conversion of scintillated photons to photoelectrons is on the order of 10-20%. Of these, only a fraction (a further 10-20%) actually reach the first dynode to undergo the electon multiplication process.

The previously described effects alter the gain of a detector by a fixed factor. The gain of a detector can be varied by adjusting the bias voltage applied across the electron multiplier section of the PMT. A higher bias voltage results in a higher potential difference between the dynodes resulting in an increase in secondary emissions at each stage of the multiplication process. The gain $G$ of the PMT assembly is then related to the applied biased voltage V by:

$$G \propto V^{\alpha n} \tag{3.4}$$

where $n$ is the number of dynodes ($n = 12$ in our case) and $\alpha$ is a coefficient determined by the dynode material and geometric structure (usually $\alpha = 0.7$ to 0.8). Since four detectors of equal gain are desired in this experiment, the gain variations resulting from the manufacturing process are compensated for by altering the applied voltage.

To measure the required voltages to match the gains, the detectors were individually powered by four separate high voltage power supplies. The voltages were adjusted until the $^{137}$Cs Compton edge peak from each of the detectors lay in the same channel on the spectra. The voltages were then measured using a high voltage probe across the second parallel output of each of the high voltage power supplies. A set of voltages that allowed the gains to be matched is shown in table 3.2.7. The uncertainties are estimated from effects such as the shifting of the peak channel caused by varying the power supply voltage.

| Detector A | Detector B | Detector C | Detector D |
|------------|------------|------------|------------|
| 1506 $\pm$ 3 V | 1491 $\pm$ 3 V | 1371 $\pm$ 3 V | 1493 $\pm$ 3 V |

Table 3.1: A set of detector bias voltages that enabled the gains of the four detectors to be matched.

As can be seen, three of the detectors have very similar intrinsic gains except for the fourth not. Due to the high dependence of a detector's gain on the PMT's internal structure, the difference observed is probably a result of the different characteristics of each PMT.

## 3.3 The "3 out of 4" Coincidence Method

### 3.3.1 Description

$^{10}$C decays via positron (or $\beta^+$) emission into the excited states of $^{10}$B. The decay to the first excited state of $^{10}$B is the dominant branch with a probability of 98.53%. In this process, the emission of the positron is followed by the immediate (within 1ns) emission of a 718keV $\gamma$-photon from the de-excitation of the excited nucleus. The decay to the second excited state of $^{10}$B occurs with a probability of 1.46%. In this process, the emission of the positron is followed immediately by the emission of 1022keV and 718keV $\gamma$-photons from two cascading de-excitations of the excited nucleus. The emitted positron in both cases annihilates almost immediately in nearby material, emitting two 511keV photons in opposite directions.

The setup of the "3 out of 4" coincidence method, suggested by Barker[39], is to have four detectors in the symmetric arrangement shown in figure 3.15. In this setup, the following can occur upon the decay of a $^{10}$C nucleus. The two opposite out going 511keV photons can be detected by two opposite detectors, while the 718keV-$\gamma$ photon (or the 1022keV photon in the 1.46% branch) can be detected in one of the remaining two detectors. If this detection signature of a triple coincidence is unique to $^{10}$C decays and the probability of it occuring is constant then its rate can be used to identify the $^{10}$C decay rate and half-life.

### 3.3.2 Why the need?

There are three sources of unwanted events present: contaminant activity, low-level noise and high energy cosmic rays. Due to the poor resolution of plastic scintillation detectors, it is almost impossible to apply spectroscopy techniques to remove the unwanted counts (with maybe the exception of the last). How the "3 out of 4" coincidence method deals with the above activities is described below.

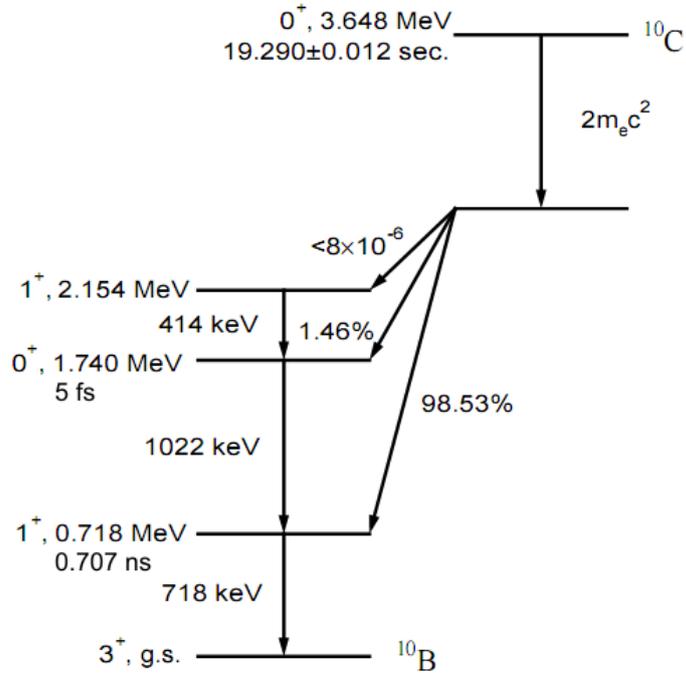

Figure 3.14: Decay scheme of $^{10}$C to the excited states of $^{10}$B

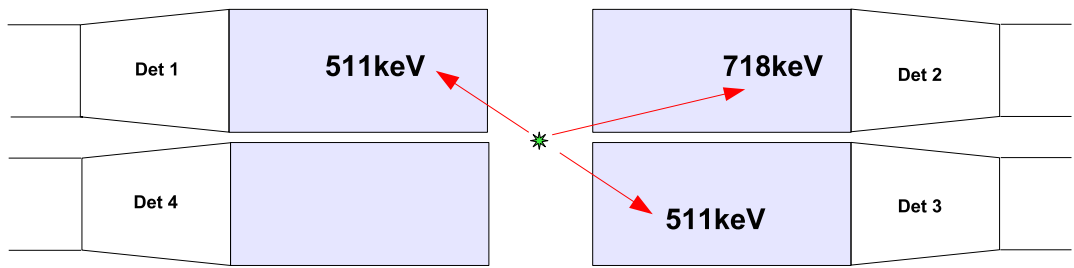

Figure 3.15: Diagram of the "3 out of 4" coincidence setup used to detect the decay of $^{10}$C.

**Contaminants**

The main contaminant by far is $^{11}$C, a positron emitter with half-life of 20.38 minutes, produced from impurities in the Boron target. Another contaminant is $^{13}$N, positron emitter with a 10 minute half-life, which can be produced from $^{13}$C found in oil deposits that may be present on the surface of the target. The deposits may have been from oil vapours from the vacuum pumps or contact with hands or fingers during the handling of the target. Although discussion will be restricted to $^{11}$C, the same points can be made for $^{13}$N.

$^{11}$C decays into $^{11}$B by the emission of a positron, like $^{10}$C, the positron annihilates and emits two oppositely out-going 511 keV photons. But unlike $^{10}$C there is no correlated $\gamma$-photon, so in principle the "3 out of 4" coincidence technique should not detect $^{11}$C decays. However, it was discovered that the signature of the $^{11}$C decay can actually masquerade a triple coincidence event from $^{10}$C decay (see section 3.3.3). This produces a long-lived decaying component in our time histogram which has to be corrected for.

**Noise**

In all detectors, there is a continuous noise current which may be the result of ambient light leaks, the PMT's anode dark current effects[25] or general noise in the electronics. The standard technique in the removal of noise is by setting a threshold at an early stage in the electronics below which signals are ignored.

The occurrence of noise events is purely random in time and thus uncorrelated in each of the detectors. This makes the "3 out of 4" coincidence technique very efficiency in removing the noise.

**Cosmic Rays**

There is a presence of background cosmic rays, in particular high energy muons, in our detector spectra. Figure 3.16 shows the background spectrum from a single detector. The peak at the very high channels is due to very high energy events from the cosmic rays exceeding the linearity range of the electronics.

When the "3 out of 4" coincidence mode was first turned on with no sources present a count rate of around 1 s$^{-1}$ was observed. The pulse-height spectrum of the summed pulses from all four detectors is shown in figure 3.17. Because of the high energy distribution of the events, it was guessed they were "3 out of 4" coincidences coming from cosmic rays. For this to be occurring, the counts in separate detectors must be correlated. Thus the following must be occurring, either: a) the cosmic rays were coming in showers such that different particles were incident on different detectors simultaneously, or b) a single high energy particle was being detected in multiple detectors within the resolving time of the detector so that the events appeared simultaneous. This prompted the investigation of the cause of this.

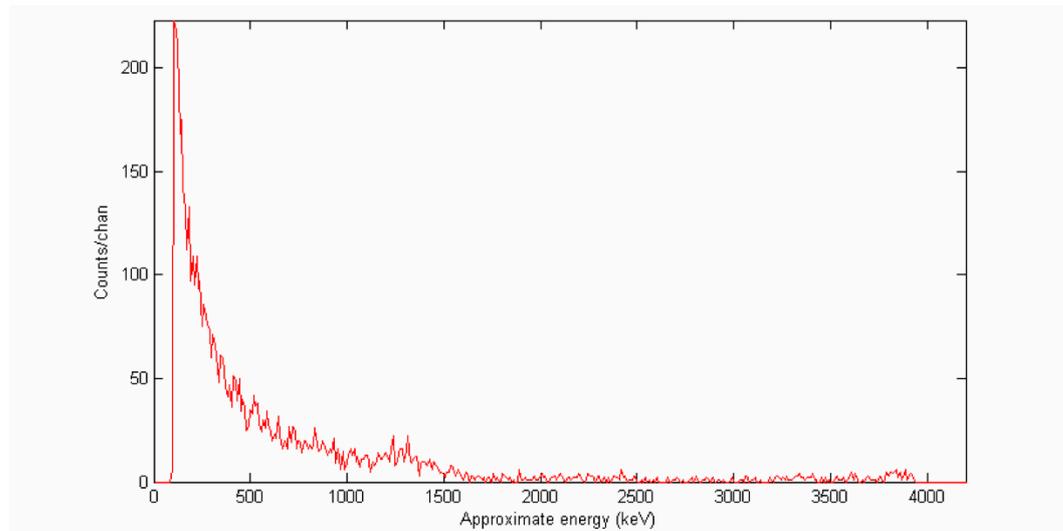

Figure 3.16: Detector background spectrum with detector bias at 1400V and an ADC conversion gain of 512.

It is worth noting that with the detectors setup in the proper experimental configuration in the Beam Room (recall this is where the target and detectors are placed during the experiment) the background of "3 out of 4" coincidences was around 1-2 every 5s.

### 3.3.3  False Triple Coincidence

Since the primary interaction of photons in plastic scintillator is the Compton effect, a detected photon is not absorbed but instead scattered. The scattered photon can then enter another detector and be detected again. In fact, if the scattering angles are just right, a single photon can be detected in multiple detectors. The time involved in the Compton scattering process is so small that the time separation between successive detection events is just the time taken for the photon to travel from one detector to the next. Thus the events occur within nano-seconds of each other making them appear simultaneous for a system with a time resolution on the order of 30ns. This is what we shall call a "false coincidence".

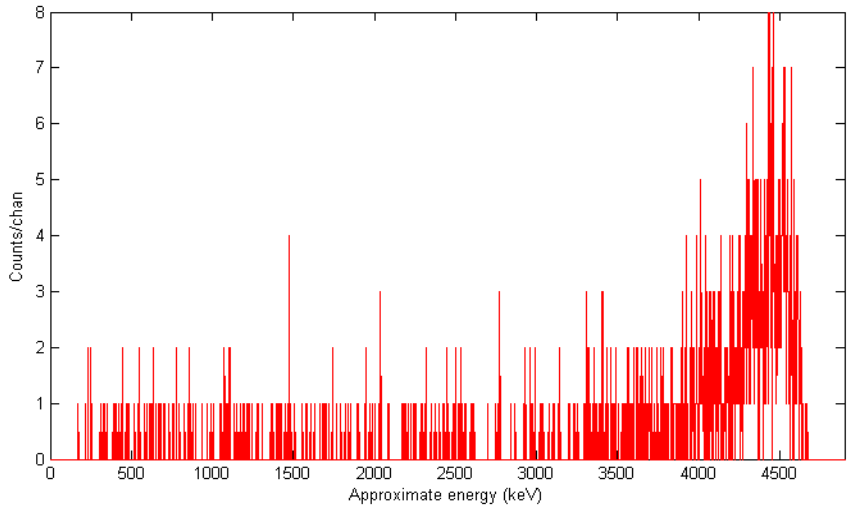

Figure 3.17: The "3 out of 4" coincidence spectrum with no sources present using the electronics setup described in section 3.4. Detector biased voltages of 1500V were used along with an ADC conversion gain of 2048.

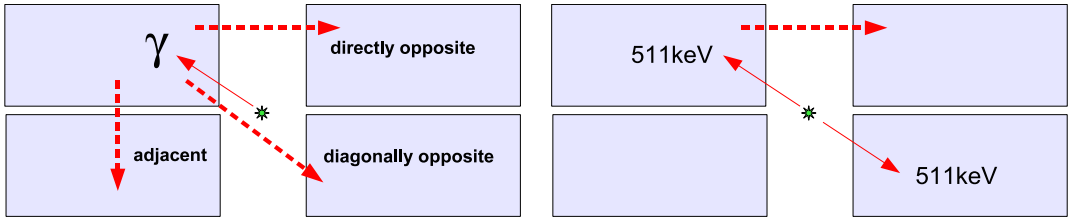

Figure 3.18: Left: The possible false coincidences from a single photon. Right: The most probable way that a false "3 out of 4" coincidence may be caused by $^{11}$C.

The geometry of the detectors is such that the occurrence of false coincidences is favored for large scattering angles. The differential cross-section for Compton scattering, which is a function of scattering angle and incoming photon energy, is described by the Klein-Nishina formula. The ratio of the cross-sections for large to small scattering angles is greatly reduced with increasing photon energy. Thus the probability of false coincidences varies as a function of photon energy. In fact it has been suggested that this may be used to increase the energy resolution of systems involving multiple plastic scintillators[40].

The effect of false coincidences in our system is tested using uncorrelated 662keV $\gamma$-photons from $^{137}$Cs since they have a similar energy to what we are interested in. The $^{137}$Cs source was enclosed in at least 6mm of perspex to prevent any $\beta^-$ radiation from escaping. Initially it was found that the probability of false coincidences was highest for adjacent pairs of detectors followed by directly opposite pairs and then diagonally opposite pairs (see figure 3.18 for explanation of these terms). This was exactly as expected from scattering angle cross-section and solid angle considerations.

It was found that placing a layer of lead between adjacent sets of detectors reduced the probability of false triple coincidences for both adjacent and diagonally opposite sets of detectors but had very little effect on directly opposite sets. Increasing the thickness of lead meant that the detectors were moved further away from the source which caused a reduction in the count rate. A nominal thickness of 11mm of lead was found to serve the purpose well. A summary of the results from the tests is shown in table 3.3.3. The "exp. setup" column contains the results obtained with the detectors in their final experimental setup positions. In this setup, the lower probability of false coincidences in the diagonally opposite and directly opposite sets of detectors is due to attenuation in the stainless steel "wheel" which is part of the beamline vacuum system.

## 3.4   The Beam Room Electronics

The circuit diagram on page 84 shows the setup of the NIM electronics located in the beam room (recall the beam room is also where the detectors and target are located). The function of the setup is to gather pulses from the four detectors, determine when a "3 out of 4" coincidence has occurred, and trigger the DAQ system to acquire the analogue pulse corresponding to this event (the DAQ system is described in Chapter 2).

| *(values in $s^{-1}$)* | **no Pb** | **3.5mm Pb** | **11mm Pb** | **exp. setup** |
|---|---|---|---|---|
| single detector | $8,094 \pm 6$ | $7,301 \pm 6$ | $6,525 \pm 6$ | $10,020 \pm 9$ |
| adjacent dets | $1,303 \pm 4$ | $157 \pm 1$ | $23 \pm 1$ | $25 \pm 1$ |
| diagonally opp. | $249 \pm 2$ | $114 \pm 1$ | $84 \pm 1$ | $24 \pm 1$ |
| directly opp. | $305 \pm 2$ | $259 \pm 2$ | $240 \pm 2$ | $129 \pm 2$ |
| any 3 of 4 | $104 \pm 1$ | $9.6 \pm 0.4$ | $5.4 \pm 0.4$ | $0.2 \pm 0.1$ |

Table 3.2: Summary of the false "3 out of 4" coincidence rates from a $^{137}$Cs source with increasing thickness of lead in between adjacent detectors.

### 3.4.1 Description

- The negative analogue pulses from each of the four detectors are split by a Lecroy 428F Quad Fan-in/Fan-out unit. The unit is used in the Normal (non-inverting) mode. In this mode the unit can handle pulses with rise and fall times down to 3ns and 4ns and rates of up to 100MHz[41].

- The first set of analogue pulses from the fan-out are summed using a channel on a Phillips 744 Quad Linear Gate Fan-In/Fan-out unit. The unit can handle rise times of 1 ns and rates of up to 250Mhz[42]. The output is quoted to be linear ($\pm$ 0.4%) for upto $\pm$2V across two 50 ohm loads. Since such high linearity is not required in this setup, the range used in the experiment was determined by a separate linearity test (see section 3.4.2). The summed analogue pulses from a "3 out of 4" coincidence are around 30ns wide (compared to the 25ns wide pulses from a single detector) due to not being able to find cables of equal lengths for the signal from the four detector.

- The summed analogue pulse is passed through a channel of the Phillips 744 used as a linear gate. The gate is open only when a "3 out of 4" coincidence has occurred (see below) so that only relevant summed pulses are passed through to the slow linear amplifier.

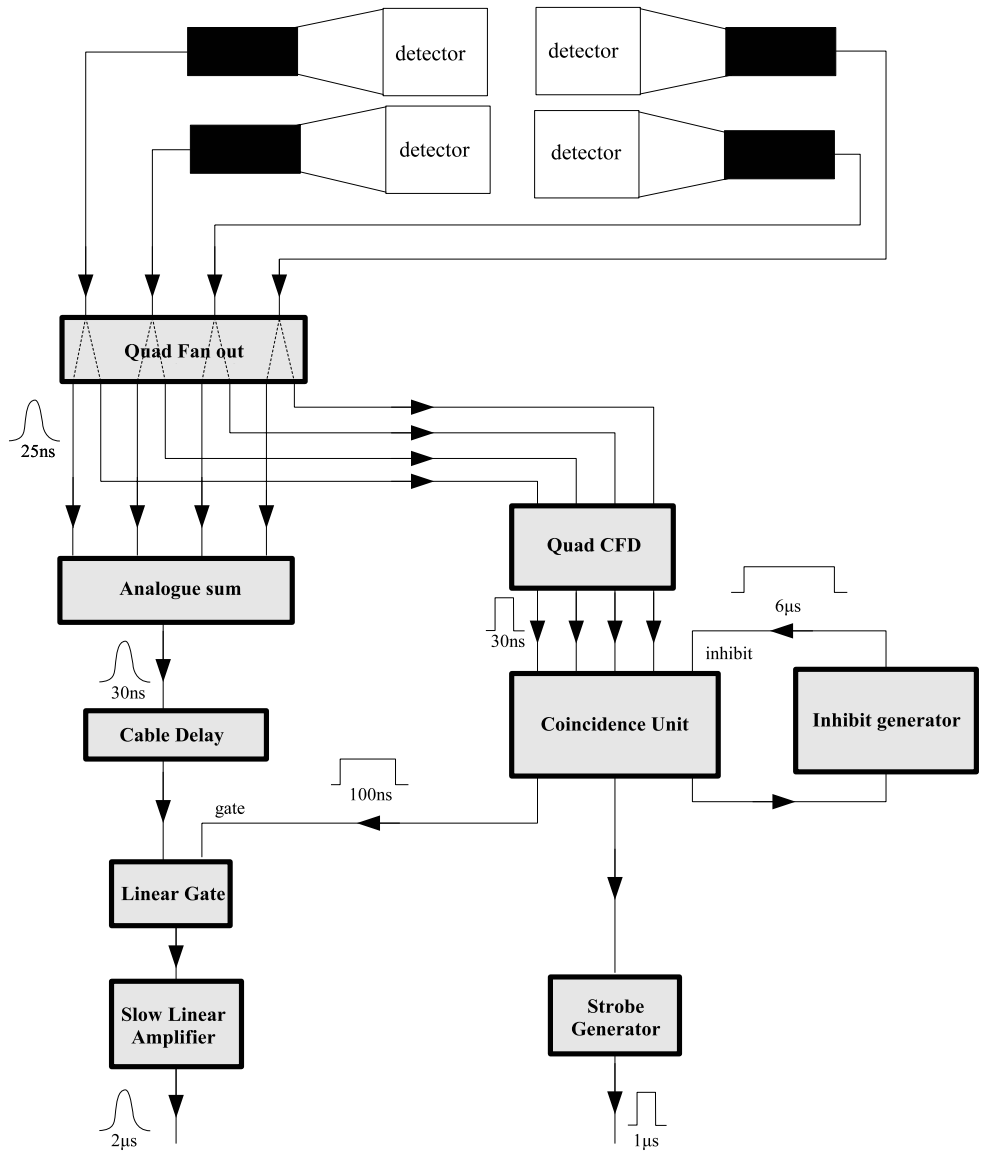

Figure 3.19: Circuit diagram of the electronics in the beam room

- The slow amplifier used is an Ortec 440A Selectable Active Filter Amplifier. It is a pole-zeroed Gaussian shaping amplifier, offering good signal-to-noise characteristics, widely used for spectroscopic purposes. The gain is set such that output amplitude does not exceed +8V (the range required by the ADC). Positive unipolar pulses with a shaping time of $0.25\mu s$, the shortest available on the amplifier, were used. This was done to ensure that the trailing tails of a pulse did not overlap with the next pulse.

- The pulses from the amplifier are sent down around 50m of cable to the control room. The integrated pulses are much more slowly varying than pulses directly from the detectors (1 $\mu s$ in duration compared to 30ns) and so suffer much less distortion in shape in the journey down the long cables.

- The second set of analogue detector pulses from the LeCroy 428F fan-out is converted to 30ns wide logic pulses by a Tennelec TC 455 Quad CFD (constant fraction discriminator). The logic pulses contain the time information of the pulses from each of the detectors. A CFD is used to reduce the jitter in the timing of the generation of the logic pulses since the shapes of the analogue pulses remain even for different amplitudes.

- The "any 3 out of 4" condition is tested for by a LRS 165 5-fold logic unit. Coincidence is defined as an overlap in time of the logic pulses from the Quad CFD. When a "3 out of 4" coincidence event occurs, a 100ns wide logic pulse is produced across the unit's outputs. Three of such outputs are used.

- The first of the outputs is used to generate a linear gate to exclude signals not lying within it. The time taken for the gate to open and close is less than 2ns. Because of the added electronic procedures required in producing the gate, a low-loss cable of appropriate length is required to delay the summed analogue pulse so that it is in time coincidence with the gate at the linear gate unit.

- The second of the outputs from the coincidence unit is used by a strobe generator unit to generate the strobe pulses required to trigger the DAQ system (described in Chapter 2). These slower TTL logic pulses are a lot less susceptible to distortion in shape when sent down the long cables to the control room. Adjustments of the timing of the strobe pulse to match the amplified analogue pulse peak is done in the Control Room.

- The third of the coincidence unit outputs is used to generate a system inhibit. This is done by using the inhibit generator, a second LRS 165 used in single coincidence mode, to produce nominally $6\mu s$ long logic pulses which are then passed back into the original coincidence unit's inhibit input. This creates the effect that for $6\mu s$ after a "3 out of 4" event has been detected, all events will be ignored by the coincidence unit and no further pulses will pass the linear gate and on to the slow amplifier. This manually set inhibit time reduces pile-up and enforces a constant system dead time per event which is essential for eventual dead time loss correction.

### 3.4.2   System Tests

**Fan-in Linearity**

The linearity range of the Phillips 744 Fan-in's output (quoted as $\pm 0.4\%$ for $0 \rightarrow \pm 2V$) determines the detector gains used. The PMT biase voltages were set such that the sum of the three pulses from a "3 out of 4" coincidence (assumed to be $2 \times 511\text{keV} + 718\text{keV}$) did not exceed the linearity range of the fan-in's output. This was done by making sure the heights of the largest pulses resulting from a 662keV $\gamma$-source were less than 1/3 of the full linearity range of the fan-in's output.

Since high resolution spectroscopy is not required, an actual test of the linearity of the fan-in unit was performed to find a tolerable linearity range. The larger this range the higher the detector gains can be set to increase resolution and more important the count rate. The test was done using a Tektronix AFG 3252 Dual Channel Arbitrary/Function Generator set to produce pulses of the same shape as that from our detectors. A single output from this unit was split using a LeCroy 428F Fan-out with three of the outputs then summed by the Phillips 744 Fan-in unit. Care was taken to use cables of roughly equal length between the fan-out and fan-in units to keep the pulse peaks matched in time. The height of the summed pulse was observed on a digital oscilloscope. A plot of the input pulse heights against the fan-in output's pulse height is shown in figure 3.21. From this it can be inferred that output is linear for three input pulses of up to 1.1V amplitude.

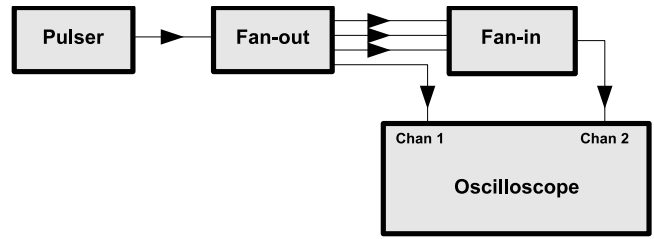

Figure 3.20: Circuit used to find a tolerable linearity range of the Phillips 744 fan-in.

The advantage of testing with pulses of the same shape as real operating conditions is the possibility of testing for and observing effects that may not have been considered. Ideally the linearity of the whole system, the position of the peak in E-lifetime measured as a function of the input pulse amplitudes, should be tested for. Unfortunately when we had access to the Tektronix pulser (the only available pulser capable of matching the shapes of the real pulses) in Auckland, the Ortec 440A amplifier was not functioning properly (see later).

## CFD Timing

It was found that the Tennelec TC 455 Quad CFD had different "fraction discriminator chips" for differing channels. The chips are the circuitry that determined at what fraction of the total peak height to initiate the triggering of the logic pulse. Channels 1 and 3 had the same discrimination value of 0.2 while channel 2 had a value of 0.3 and channel 4 had a value of 0.5.

The term "amplitude walk" refers to the variation in the timing of the logic pulses derived from analogue pulses of differing amplitudes while the term "time jitter" refers to the variation in the timing of pulses from the same amplitude. The former effect is small when a CFD is used for pulses whose shape does vary with amplitude while the latter effect is determined by the noise level in the pulses. The timing for pulses of the same amplitude and zero noise will be different if the value of the constant fraction discriminator is different (larger fractions will result in a later timing). This effect shall be called the "discrimination walk".

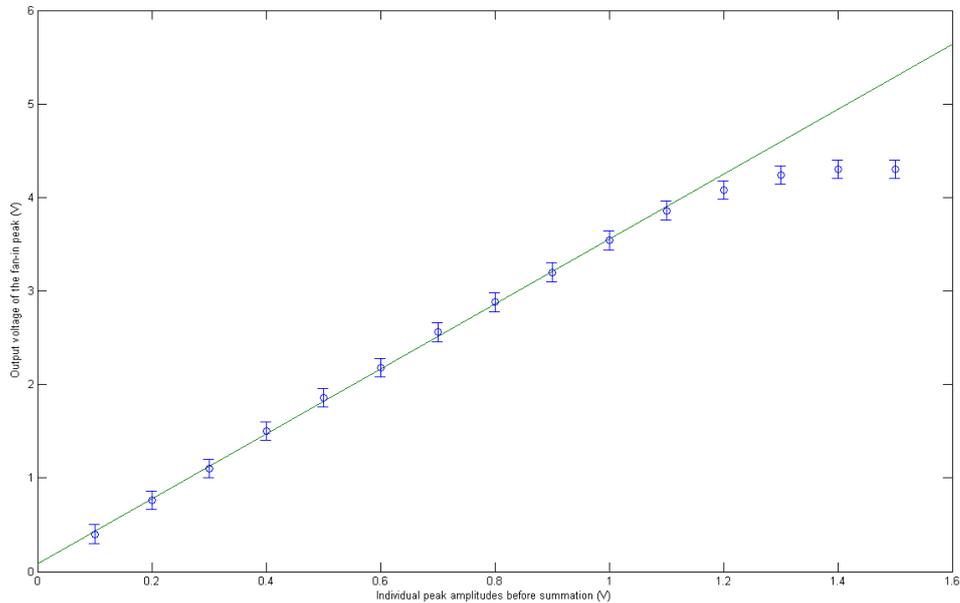

Figure 3.21: Plot of the Phillips 744 Fan-in output versus the one of the three input voltage peaks.

In order to measure the discrimination walk, a TAC (Time Amplitude Converter) unit was used in the setup shown in figure 3.22. A TAC unit outputs an analogue pulse with a height proportional to the time difference between the start and stop input pulses so that the width of the pulse-spectrum formed will allow jitter and walk effects to be determined. A fixed delay was used to ensure that the stop pulse comes after the start pulse by a comfortable amount for the TAC device to measure. The TAC was used on the 100ns full scale mode and DAQ at 512 channels. The scale was calibrated by observing the shift caused by a cable of a known delay (measured with an oscilloscope). Three TAC spectra were recorded, one where the timing difference between two 0.2 fraction discriminators were used, another between the 0.2 and the 0.3, and the last between the 0.2 and 0.5. Comparison of the latter two with the timing of the first allows the discrimination walk effect to be measured.

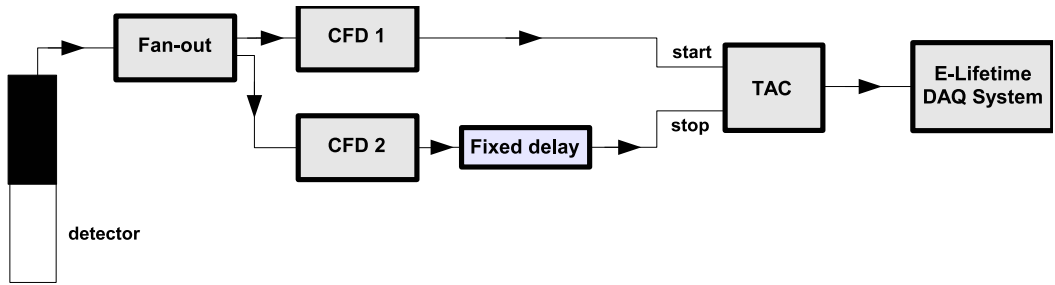

Figure 3.22: The circuit used in testing the discriminator walk effect of the CFD.

The TAC spectra are shown in figure 3.23 and the results in table 3.4.2. It is worth noting a smaller plastic scintillator than the ones described previously was used for this test so that the pulses were only 18ns wide. Because of this, the discrimination walk effect for our 25ns wide pulses are maybe 1.5 times larger than the $0.4 \pm 0.6$ ns and $1.1 \pm 0.6$ ns values quoted. Another interesting thing to note from the result is the variation of time jitter (quantified as the FWHM of the peak in the TAC spectrum) for differing fraction discriminators. Our results agree well with other empirical tests which find that fractions of 0.1-0.2 to be the best in minimising time jitter effects[36].

| Disc. fractions | Timing difference (ns) | Disc. walk (ns) |
|---|---|---|
| 0.2 & 0.2 | $16.8 \pm 0.4$ | - |
| 0.2 & 0.3 | $17.2 \pm 0.4$ | $0.4 \pm 0.6$ |
| 0.2 & 0.5 | $17.9 \pm 0.6$ | $1.1 \pm 0.7$ |

Table 3.3: Results from the discrimination walk test made using a TAC to measure the timing difference between using the different discriminators on the Tennelec 455 Quad CFD.

Since a coincidence is defined as an overlap of 30ns wide pulses, the several nano-seconds of discrimination walk is negligible. Therefore the use of the Tennelec TC 455 Quad CFD with different fraction discriminators is justified.

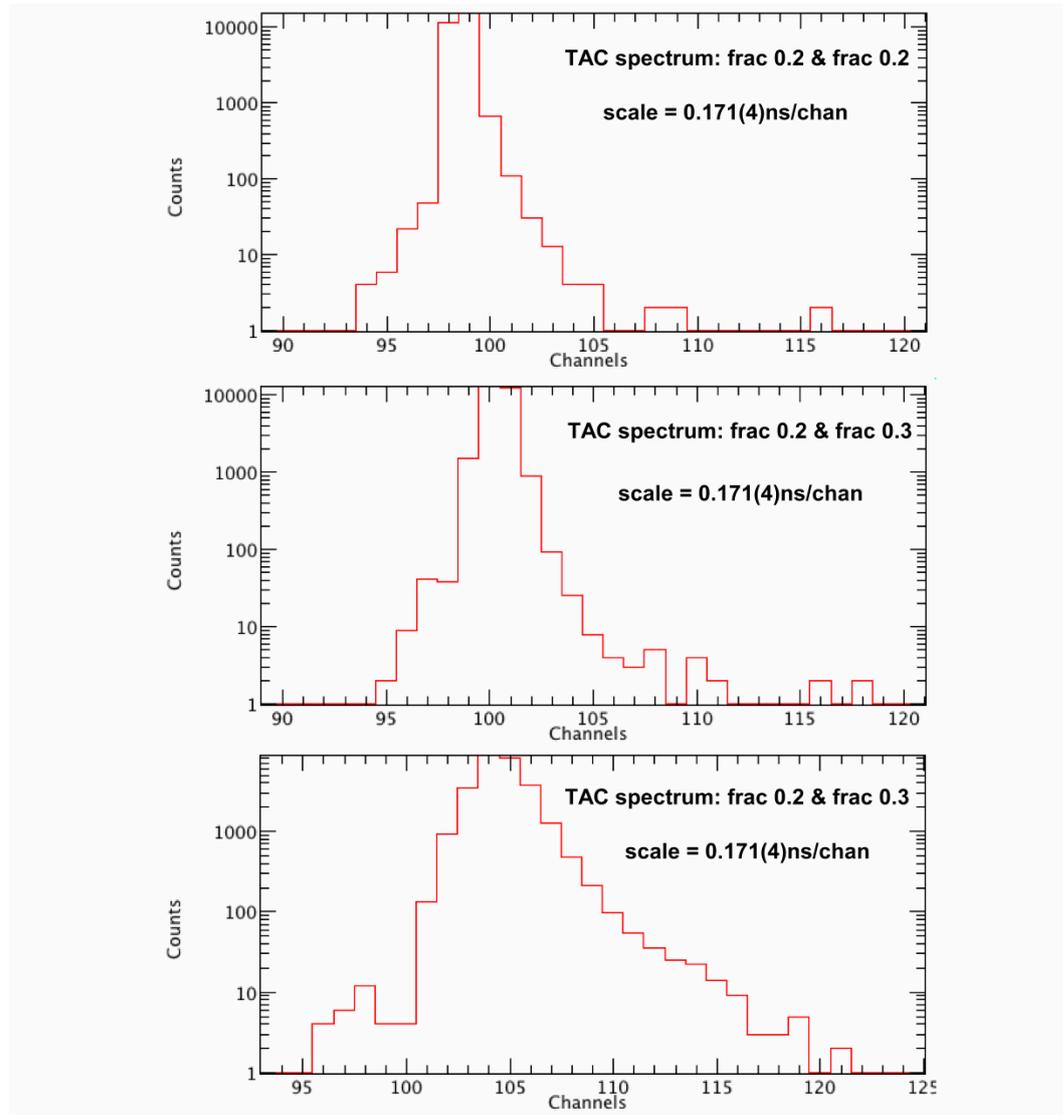

Figure 3.23: The TAC spectra of time difference between the logic pulses produced from different fraction discriminators.

**Stability**

There had been problems with the gain stability of the Ortec 440A amplifier used in Auckland. Large gain walks in periods of around 5-10 minutes were observed. The actual experimental setup at ANU thus used a different Ortec 440A amplifier unit for which the stability had been measured for over 10 hours. Stability problems were also encountered with a high voltage supply before a suitable one was found. In the end an Ortec 556 High-voltage Power Supply was used to power all four detectors. This unit is capable of delivering up to 10mA (2.5mA was actually used) with stability rated at better than 0.01% per hour and 0.03% per 24 hours[43]. The different voltages across the detectors were achieved by using a passive voltage distributor unit.

A test of the stability of the system several days out from the experiment was performed. The channel location of the $^{137}$Cs Compton edge peak was used to determine the gain of the system. Spectra from one of the detectors were taken for 10s every 4 minutes for 10 hours. The peak channel plotted as a function of time is shown in figure 3.24 along with the event rate in the detector. The system showed excellent gain stability over the duration of the test ($< 1\%$ fluctuation over 10 hours).

**Dead Time**

A measurement of the intrinsic dead time per pulse of the entire system, without the use of the inhibit generator, was performed. Recall that the dead time is the time after a collected event for which the system is busy and ignores any subsequent pulses. The output from Tektronix Pulser, set to produce double analogue pulses of variable separation, were fed into the LeCroy fan-out unit. The separation times between the pulses observed on a digital oscilloscope were plotted against the count rates measured in E-Lifetime (see figure 3.25).

The intrinsic dead time per pulse of the system is determined to be $5.4 \pm 0.1 \, \mu s$. By far, the largest contribution to this is from the use of CAMAC and the CMC100 device in the DAQ system (see section 2.8.4 on page 51). The inhibit generator is required to produce a inhibit time larger than this to ensure that it is this time that determines the overall dead time of the system. This allows a constant dead time to be set and accurate dead time loss corrections to be calculated. The inhibit time of the system used in the experiment was measured at the end of all data taking and was found to be $6.43 \pm 0.01$ $\mu$s. The plot of the values used in determining this was found to have the same feature as that in figure 3.25 and therefore was decided to be omitted.

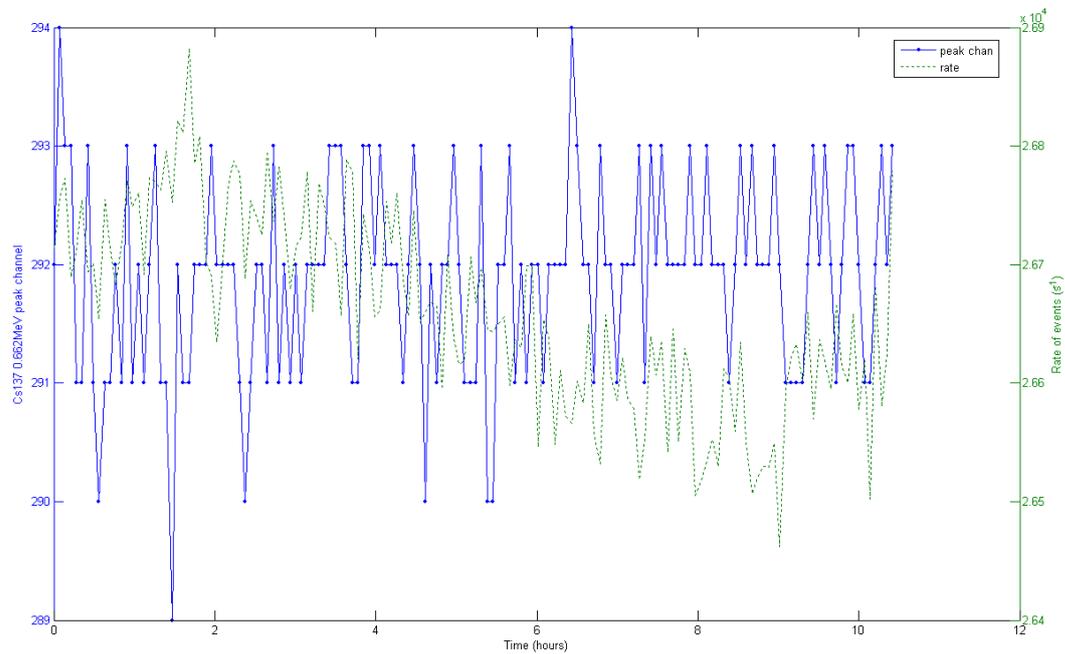

Figure 3.24: A plot of the Compton edge peak channel and the rate of a $^{137}$Cs source versus time used to determine the long term stability of the system.

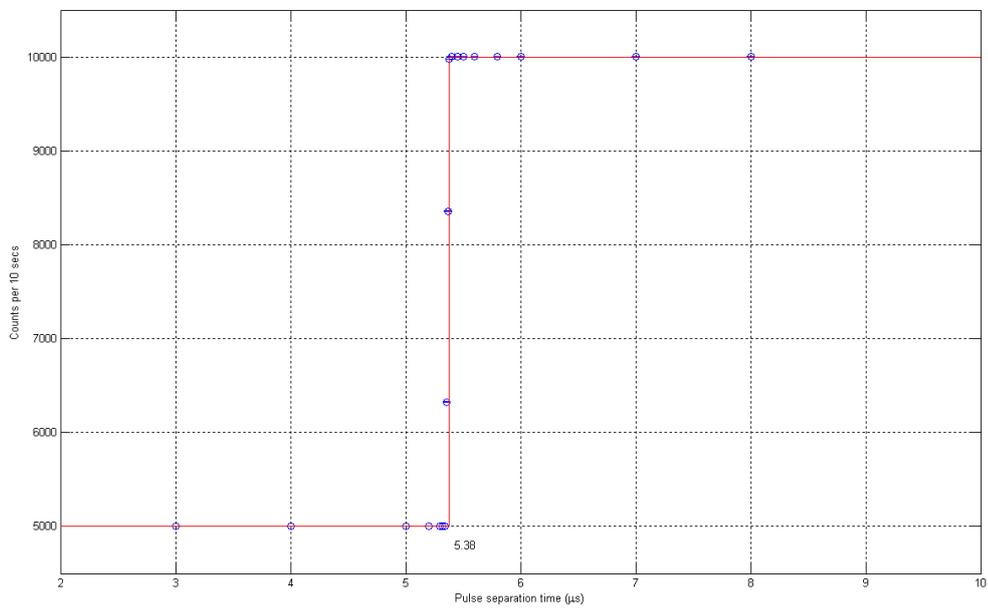

Figure 3.25: Plot of count rate versus separation time of the two double pulses used to determine the dead time of the system without the inhibit generator.

# Chapter 4

# Experimental Method

## 4.1 Apparatus

The experiment was performed at the Department of Nuclear Physics of the Research School of Physical Sciences and Engineering at Australian National University (ANU), Canberra. The beam of high energy protons was produced by the 14UD Pelletron, a tandem Van de Graaff accelerator capable of producing a proton beam with energy upto 34MeV per proton. The energy range used in this experiment (around 8MeV per proton) is therefore easily accessible. In fact, modifications to the standard operating setup were required to produce such a low energy proton beam.

The $^{10}$B$(p, n)^{10}$C reaction is used to produce $^{10}$C. The minimum energy required for the incident protons is $\sim$5MeV for this experiment. Data was taken using two targets of differing purity at differing beam energies and intensities. This was done in the hope of observing and reducing systematic effects resulting from contaminant activity. Details of the targets and beam conditions used can be found in this section.

## 4.1.1 The 14UD Pelletron Accelerator

A high energy proton beam is produced by the 14UD Pelletron (depicted in figure 4.1) in the following way: a low energy beam of singly-negatively charged hydrogen ions (H$^-$) is produced at the ion source by the bombardment of a titanium hydride target with a beam of accelerated cesium atoms. The negatively charged hydrogen ions are accelerated along a positive electric field gradient towards the high voltage terminal. The beam then passes through an equi-potential region of containing a low pressure stripping gas, which strips the hydrogen ions of both their electrons, leaving behind hydrogen nuclei or protons (H$^+$). The positively charged protons are accelerated away from the high voltage terminal by a negative field gradient. The beam is focused by quadrupole magnets and collimated by two narrow slits before and after a 90° bending magnet, after which it is directed down the appropriate beam line by a switching magnet and then collimated once more before reaching the target.

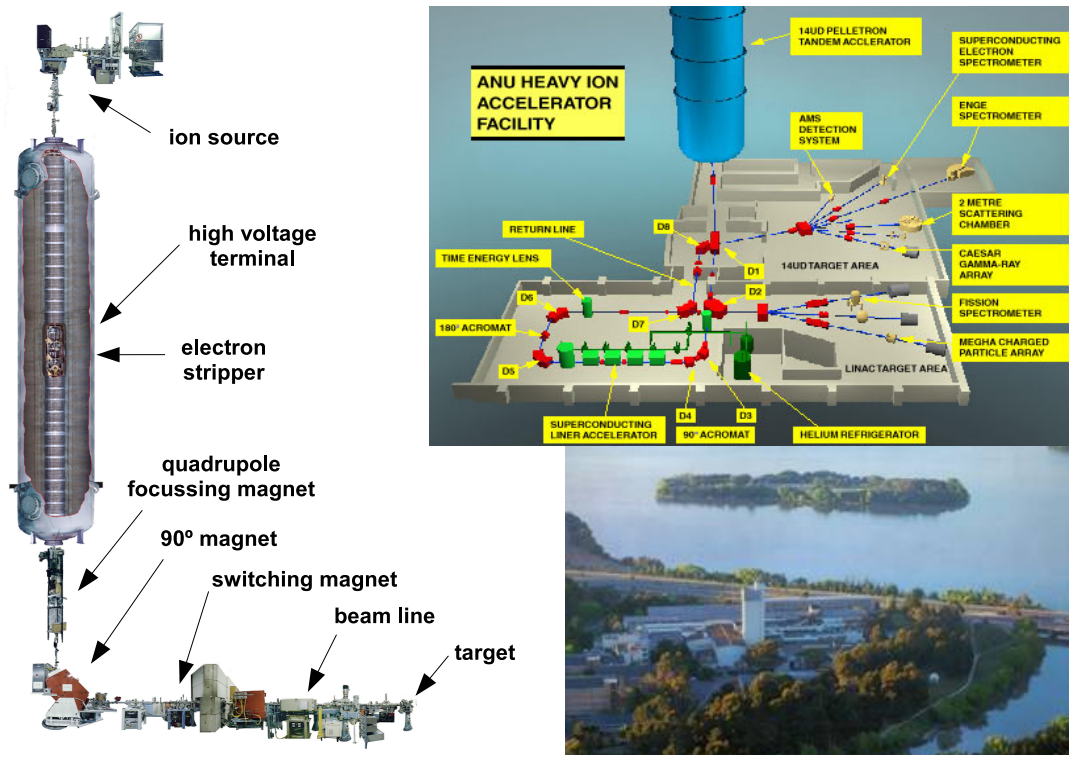

Figure 4.1: The 14UD Pelletron accelerator and facilities. Photographs and diagrams courtesy of the Department of Nuclear Physics, ANU, Canberra.

The 90° bending magnet and the slits ensure that only protons of a narrow energy range (∼1keV) pass through to the switching magnet and then the target. The energy of the protons is approximately equal to the small initial energy of the negative hydrogen ion beam (∼0.2MeV) plus the energy gained in the tandem acceleration process ($2 \times V_{accelerating}$ in eV). An accurate measurement of the beam energy is made by measuring the 90° bending magnet's field strength with a NMR probe.

When the beam is switched off for the data collection process (see chapter 2) it is done by an electrostatic deflector at the ion source. Stopping the beam early on in the accelerator has many advantages, one of which is the reduction of the amount of background activity coming as a result of the beam production process described above (see section 4.1.4 for details).

## 4.1.2 The Targets Used

The target consists a thin sample of $^{10}$B deposited on a very pure thin gold backing (or in "sandwich" structure). The main contaminant in the $^{10}$B target is $^{11}$B which has been shown to be problematic (see section 3.3.3). Therefore the target is required to be as pure as possible. The gold, which provides structural support for the thin film of boron, sits directly in the path of the beam.

Material with large atomic numbers $Z$ such as gold ($Z$=79) and tantalum ($Z$=73) can be used in the path of the proton beam without the worry of contaminant product. That is because the positively charged protons can not penetrate the large Coulomb barrier so that interactions with their nuclei are very rare. This implies that material used must have very high purities.

The two targets used in the experimental runs shall be called the "thin" and the "thick" targets. They differ in the following way:

**The thin target** is the less pure target. The boron used has a composition of approximately 93% $^{10}$B (main contaminant being $^{11}$B) and is around 0.2mg/cm$^2$ ($\sim$10keV to the beam) thick. It sits on a 99.999% pure gold backing of around 12mg/cm$^2$ thick.

**The thick target** is the more pure target. The boron sample used is 99.49% $^{10}$B and is around 23mg/cm$^2$ ($\sim$1.2 MeV) thick. It is sandwiched between two layers of 99.999% pure gold each with a thickness of 14mg/cm$^2$ ($\sim$0.3MeV to the beam).

### 4.1.3   Beam Energy and Current

The beam energy was chosen to increase the yield of $^{10}$C while reducing that of
$^{11}$C contaminant. $^{11}$C is produced through a similar nuclear process as $^{10}$C but
with a lower threshold energy; $^{11}$B$(p,n)^{11}$C occurs with a threshold of around
3MeV whereas $^{10}$B$(p,n)^{10}$C has a threshold of around 5MeV. The yield of $^{11}$C
is thus much higher because the energy will be further above threshold and
also because of the lower spin difference between target and product states
($3/2^+ \rightarrow 3/2^+$ transition for $^{11}$C where as a $3^+ \rightarrow 0^+$ transition for $^{10}$C). All
of this means that it is almost impossible to completely avoid the production
of the $^{11}$C contaminant.

A plot of the ratio of the $^{10}$B$(p,n)^{10}$C to $^{11}$B$(p,n)^{11}$C cross-sections versus
proton energy was made by combining data from Refs.[44, 45, 46, 47, 48] and is
shown in figure [4.2]. Since the $^{11}$C production cross-section is essentially flat
for the 5-10MeV energy range, the features seen in the plot are predominantly
from that of the $^{10}$C production cross-section. The proton energy was chosen
to be as large as possible to maximise the cross-section ratio while avoiding
the dip starting at around 8.5MeV.

The beam energy used on the thin target was 7.7MeV with beam intensities
that varied between 200-400nA. Two energies were used on the thick target,
they were 7.7MeV and 8.3MeV with beam intensities that varied between 80-
200nA. For the thick target the protons lose around 0.3MeV of energy before
reaching the front of the boron due to the front layer of gold. Furthermore,
due to the non-negligible thickness of the boron, there is an energy difference
of around 1.2MeV between the protons at the front and back of it. The effec-
tive energy of the protons for the reactions are $\sim$7.4MeV and $\sim$6.7MeV from
integration of the cross-section curve.

### 4.1.4   The Beam Line Setup

The experiment was performed on the end of beam line 4 of the 14UD Pel-
letron. The setup is depicted in figure 4.3. The fragile target, along with its
tantalum target holder, is attached to a rotating aluminium arm situated in-
side a circular stainless steel vacuum chamber (known as "the wheel"). The target arm is moved 180° between the beam on (bombarding) and beam off (counting) positions by a magnetically coupled motor.

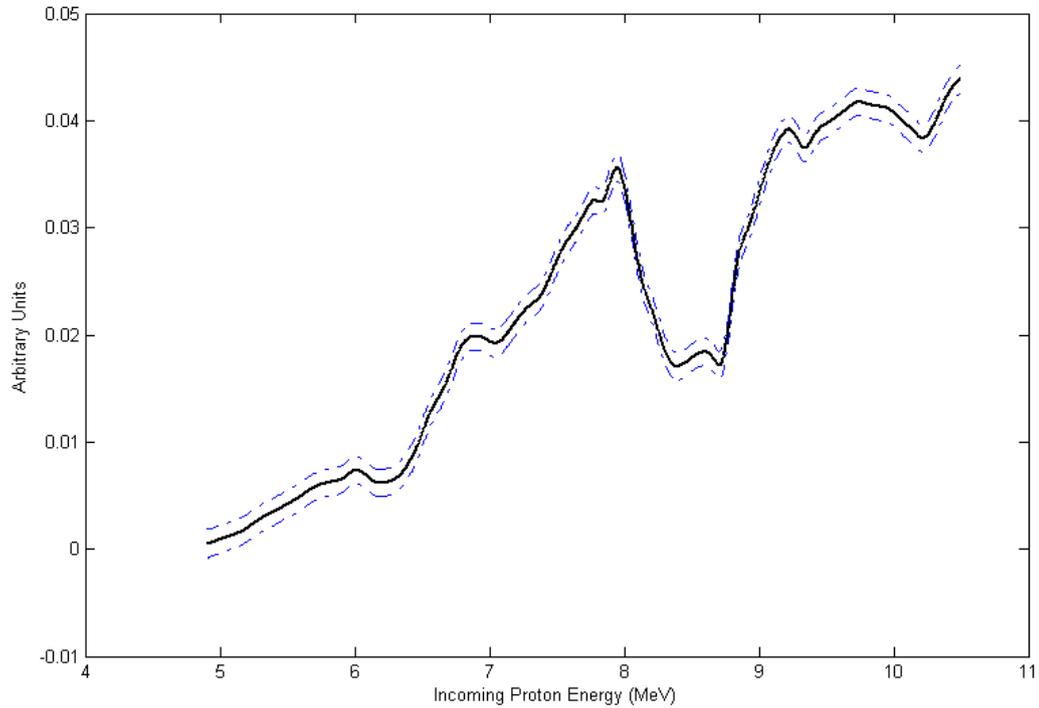

Figure 4.2: Ratio of the $^{10}$B$(p, n)^{10}$C to $^{11}$B$(p, n)^{11}$C cross-sections (in arbitary units) for different proton energies. The dotted lines represent the $1\sigma$ uncertainty of the curve.

The entire beam line is housed in a high vacuum system ($\sim$ a few $\mu$Torr). A turbomolecular pump is used in the section close to the target. Although significantly better than diffusion pumps, turbomolecular pumps may still produce oil vapours in the vacuum which can deposit onto the target, resulting in the production of contaminants (see Chapter 2).

Situated in the path of the beam behind the target is a thick 99.995% pure gold beam stop. This absorbs the high energy protons, terminating the beam at the end of the beam line. The gold beam stop is electrically isolated and the beam intensity is measured as the current flowing from it. There is a slight possibility of protons back-scattering from the beam stop and producing contaminants. Therefore, efforts were made to ensure that all non-high Z material exposed to the beam stop on the back of the target and target holder were covered by tantalum (such as the screws attaching the target to its holder).

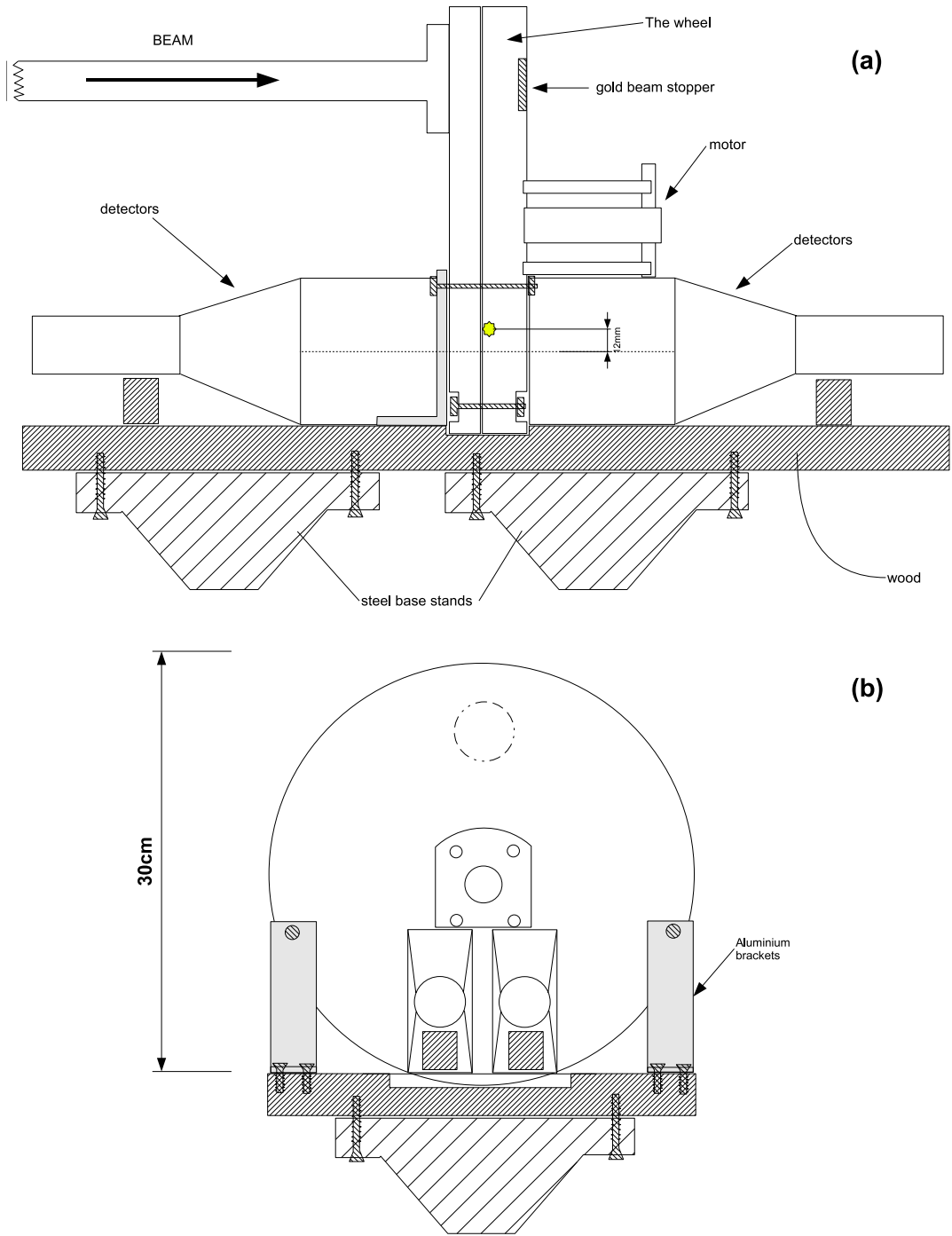

Figure 4.3: Diagram of the beam line setup, (a) side view (b) view looking into the beam.

There may be a flux of unwanted radiation reaching the detectors in the form of bremsstrahlung radiation due to accelerating charges at the high voltage tower and contaminant activity produced at the collimator approximately 1m upstream from our detectors. To reduce the effects of these, thick lead blocks were used to shield the outside of the detectors so that the external unwanted radiation is attenuated by a minimum of 10cm of lead.

## 4.2 The Data

All data was collected in runs of 5 passes each of which had a nominal 60s beam on time, 300s beam off time and 1s time step. Recall that a pass is a single data collection cycle. It consists of the target being bombarded by the proton beam for the duration of the beam on time and then data collected for the duration of the beam off time. Data was collected as event mode data; that is, the time and energy (in this case the summed energy from all four detectors) is recorded for each event. Event mode data can be projected onto either the time or energy axis which allows: a) the decay curve from only events that are within a specified energy range or, b) the energy spectrum from counts only within a specified time period to be used for analysis.

The $^{10}$C produced at the start may have decayed away, due to its short half-life, before the end of a long beam on time has been reached. This restricts the maximum resultant (or final) yield of $^{10}$C produced in a beam on period. This however, does not affect the $^{11}$C contaminant as significantly, for the same beam on duration, due to its much longer half-life. The final yield of both follow an exponential build-up to their respective saturation ($A_\infty$) of the form: $A_\infty(1 - e^{-\lambda t})$, where $ln(2)/\lambda$ is the half-life of $^{10}$C or $^{11}$C. A beam on time of around 3 $^{10}$C half-lives was chosen as it produces an ample proportion (90%) of the maximum saturation yield of $^{10}$C. Any length of time longer than this would act to produce a lot $^{11}$C for very little $^{10}$C (for example a beam on time of 4 half-lives would increase the former by 30% while the latter by only 6%).

The beam off time, which spans approximately 16 half-lives of $^{10}$C, causes the activity from $^{10}$C to drop to 1/30,000th of its original strength. This was chosen to ensure that by the end of the beam off period essentially no activity from $^{10}$C would be observed so that any activity from long-lived contaminants or background could be studied.

The time step defines the width of each time channel used for the histograms of the counts. A short time step compared to the half-life of the existing activity should be chosen to provide good time resolution of the variation in counts while long enough to ensure that each time channel contains sufficiently large number of counts. A time step, thought to be around 1s, thus ensures a good statistical fit of the discrete time channels to a continuous decay curve.

A typical data set collected from a run (that is the data from the 5 passes are combined) using the thick target at 6.7MeV effective beam energy and 100nA beam intensity is shown in figures 4.4 and 4.5.

### 4.2.1 Time Projection

The uncertainties in the time projected data of the entire energy range (shown in figure 4.4) are calculated from the square root of the counts in each channel (the Gaussian approximation to the Poisson distributed uncertainty). A full discussion of the uncertainties is found in section 4.3. The drop of counts in the first time channel is caused by the phase difference between the high precision oscillator used for data timing and the start of the software-controlled data collection procedure and are thus discarded.

From looking at the time projected data in the logarithmic scale it is immediately obvious that not only is the decay of $^{10}$C observed but that a prominent component of a long-lived contaminant is also present. One can also see that even 300s of beam off time, the decay of $^{10}$C counts persist for much of the curve. In hindsight, the beam off time should have been increased by 50-100s to improve the statistics of fitting for the long-lived contaminant which would turn reduce the uncertainty in the fitted half-life of $^{10}$C.

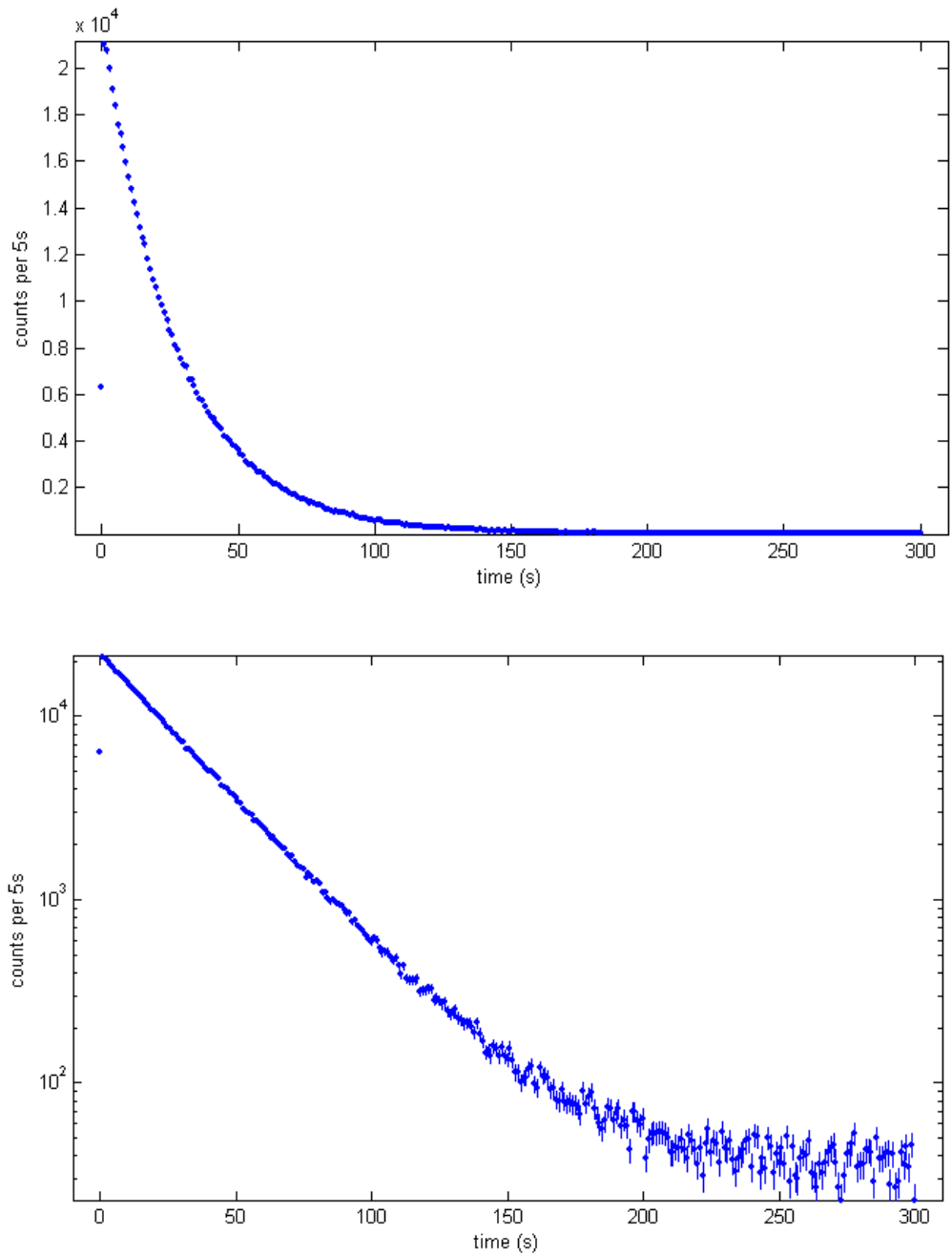

Figure 4.4: The time projected data from a typical run of 5 passes in linear (top) and logarithmic (bottom) scales. The thick target was used at 6.7MeV effective beam energy and 100nA beam intensity.

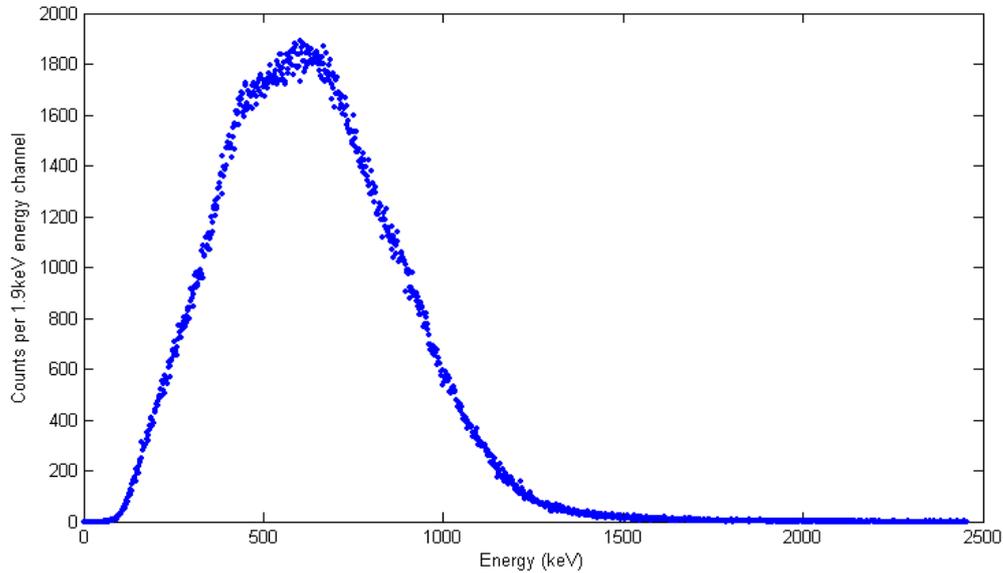

Figure 4.5: The energy projected data of a typical run of 5 passes with the thick target at 6.7MeV effective beam energy and 100nA beam intensity.

### 4.2.2   Energy Projection

The energy projection of the data of all the time channels is shown in figure 4.5. The probability density function (PDF) of the sum of independent random variable is given by the convolution of the PDFs of the individual variables. The shape of the observed energy spectra indeed matches the expected spectra, calculated from simulated individual photon spectra (see figure 4.6), of the energy deposited by two 511keV photons and a 718keV $\gamma$ photon simultaneously incident in our detection system. This is a good demonstration of the central limit theorem which states that: "*the probability density of the sum of N well-behaved independent random variables tends to a Gaussian distribution...*"[49].

### 4.2.3   Removing Contaminant Activity

With event mode data, it may be possible to make suitable energy cuts in the spectra during post-experiment analysis to remove contaminant counts. Due to the prominent continuous Compton distribution, this is only possible in plastic scintillator spectra, if parts of the interested spectrum lie above that of the contaminant spectra. In our experiment the energy per [10]C count comes from the sum of the energy deposited by two 511keV photons and a 718keV $\gamma$ photon. This is expected to be statistically higher than the energy per [11]C which comes from the sum of only two 511keV photons.

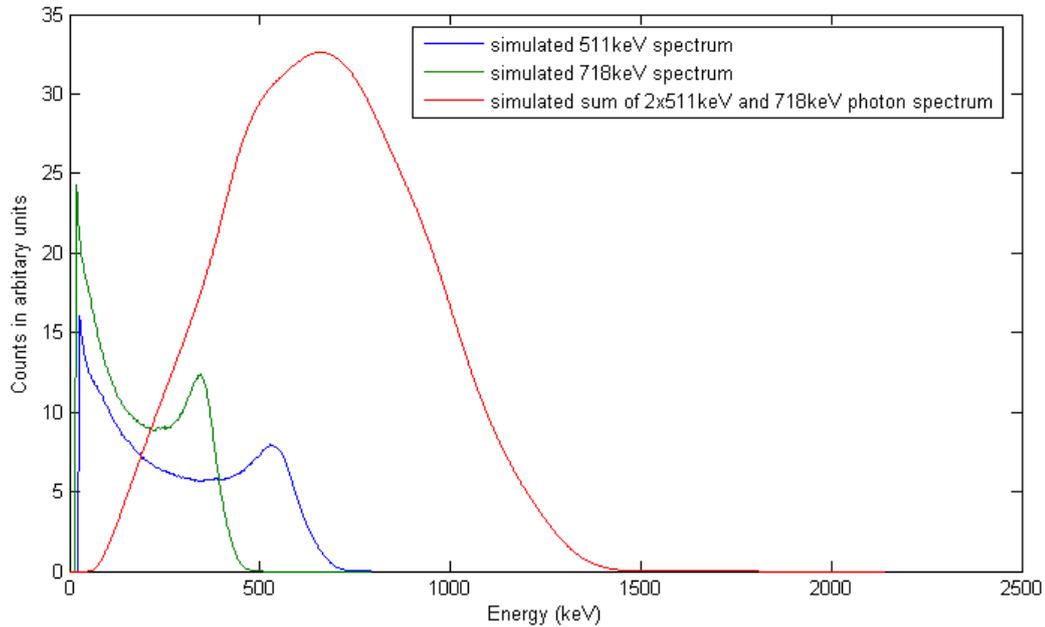

Figure 4.6: Calculated spectra from a 511keV photon and 718keV photon plotted along with the expected spectrum(on a different vertical scale) from 2×511keV and a 718keV photons simultaneous in our detectors.

A method of determining the energy cut is to compare the spectra produced using beam energies above and below the threshold for the production of the interested nucleus[9]. The energy cut can then be made at an energy above which there is no contaminant activity. This technique assumes that the contaminant spectrum does not change significantly between the two energies. The energy range used in the current experiment (up to ∼8MeV) is far from the threshold (∼5MeV) so that this assumption cannot be made. An alternative method (also employed by Ref.[9]) can be used if the half-life of the contaminant is far greater than that of the interested activity. This involves observing the time projected energy spectrum after several half-lives of the interested activity, to find the contaminant spectrum, and then determining a suitable energy cut from this. This method was used because in the current experiment the half-life of the $^{11}$C contaminant is approximately 60 times larger than that of $^{10}$C.

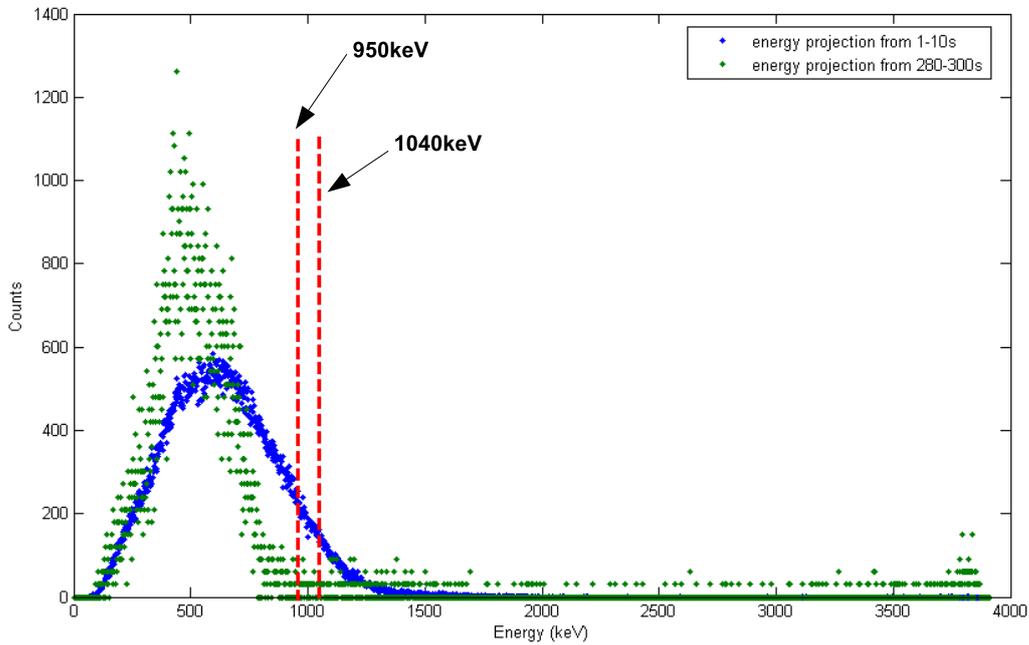

Figure 4.7: A plot of the energy projected spectrum for the 1-10s and 280-300s intervals. The latter is data combined from 8 runs and has been rescaled vertically.

The energy projected data was taken for the time intervals of 1-10s and 280-300s. A plot comparing the two spectra is shown in figure 4.7. An energy cut made at 950keV results in a 90% loss of total counts while a more conservative cut made at 1040keV results in a 95% loss of total counts. Such energy cuts were not made in our data. The reason for which is that the need to fit for contaminants is by far not the main error contributor to our analysis of the data.

### 4.2.4 Dead Time Loss Corrections

The dead time behavior of our system is classified as *nonparalyzable*[36]. That is, due to the inhibit generator, events occurring after a constant time period during the detection of a true event are lost and have no effect on the behavior of the system. Systematic errors will be present in our fitting process if dead time losses are not corrected for. The dead time loss correction for such a system is given by:

$$n = \frac{m}{1 - m\tau_{dead}} \qquad (4.1)$$

where $n$ is the corrected count rate, $m$ is the observed count rate and $\tau_{dead}$ is the constant dead time of the system, set to $6.43 \pm 0.01$ $\mu$s in our case. The amount of dead time loss increases rapidly with the count rate. For the high initial count rates in our system ($\sim$10,000s$^{-1}$) this loss can be as large as 5%.

The correction described in equation 4.1 assumes a constant count rate $m$. Our count rates are in reality dropping over the 1s time interval so that the quoted rate is in fact averaged over the measured interval. The use of the count loss correction equation is valid given the measurement time interval is short compared to the half-life of the decaying source.

The time projected data from the individual passes in a run are usually summed and analysed. In this analysis, the counts in each channel are treated as if they were a single count obtained from the sum of the counts from the separate time intervals. That is, if a run containing $a$ passes was performed then the corrected counts in a time channel $N$ is given by:

$$N = \frac{aM}{a - M\tau_{dead}} \qquad (4.2)$$

where $M$ is the uncorrected counts. In this treatment the variations in the counts in the same time channel between passes are ignored. A proper treatment taking into account this effect is to add the corrected counts calculated individually:

$$N = \sum_{i=1}^{a} \frac{m_i}{1 - m_i\tau_{dead}} \qquad (4.3)$$

where $m_i$ are the individual counts from a pass within a run.

The improper treat of the dead time loss can introduce systematic variations in the corrected counts. Variations of mostly around 20-30% (with one instance up to 50%), thought to be a result of beam current fluctuations, was found. The corrected counts from equation 4.3 are systematically lower than that from equation 4.3 by 0.1% to 0.4% in the first few time channels of most runs. It was decided that the size of this effect is not negligible and therefore dead time losses were calculated for individual passes.

## 4.3   Data Fitting Procedure

The count, $d_i$, in the i'th time channel of a decay curve obeys Poisson statistics, i.e. the probability of observing $d_i$ number of counts is given by:

$$p_{\mu_i}(d_i) = \frac{\mu_i^{d_i} e^{-\mu_i}}{d_i!} \tag{4.4}$$

where $\mu_i$ is the "true" mean of the counts in the channel with its best estimate given by $d_i$.

The uncertainty or standard deviation in $d_i$ determines the weight in the least squares fitting procedure. For large values of $d_i$, the Poisson distribution tends towards a Gaussian distribution and the uncertainty is given by $\sqrt{d_i}$. For small values of $d_i$, the uncertainties calculated in this way causes a bias in the least squares fitting procedure.

The bias problem can be illustrated in the following example: suppose two measurements made over 1s of the activity from a source of constant strength gave a result of 1 count and 2 counts respectively. Intuitively the best estimate of the true count rate is approximately $1.5s^{-1}$ but if a least squares fit (with a flat function) of the data is made with the uncertainty given as $\sqrt{d_i}$, the uncertainties in each of the two values will be 1 and $\sqrt{2}$ respectively. This gives an estimate of the true count rate from the source to be $1.3s^{-1}$. There is clearly a bias towards the values of lower counts. The size of this bias effect is a problem for high precision half-life measurements [50].

To avoid the bias problem, either the maximum likelihood (ML) or the recursive least-squares (RLS) fitting procedure is required. ML overcomes the bias problem by incorporating the details of the Poisson distribution (i.e. equation 4.4) and is the more intuitive procedure. RLS overcomes the bias problem by tending the uncertainties for low count rates towards that coming from a Gaussian distribution. This is the more robust procedure and is therefore preferred in the current analysis.

The discussion will be restricted to the fitting function of the form:

$$N = A_1 e^{-\lambda_1 t} + A_2 e^{-\lambda_2 t} \tag{4.5}$$

where $A_1$ and $ln(2)/\lambda_1$ are the initial counts and the half-life of a short-lived activity of interest and $A_2$ and $ln(2)/\lambda_2$ are the initial counts and the half-life of a long-lived contaminant activity, where $\lambda_2$ is assumed to be known. For this fitting function, the RLS procedure has already been tested to give fitted parameters that match those from ML to a very high precision ($< 10^{-5}$) [51]. The significance of the work described here is therefore to test whether estimated uncertainties in the fitted parameters also match between the two procedures.

## 4.3.1 Maximum Likelihood

The maximum-likelihood (ML) data fitting procedure is based on Bayes' Theorem which relates the parameters $\boldsymbol{x}$ to the observed data $\boldsymbol{d}$. It can be expressed as:

$$p(\boldsymbol{x}|\boldsymbol{d}) = \frac{1}{p(\boldsymbol{d})} p(\boldsymbol{d}|\boldsymbol{x}) p(\boldsymbol{x}) \tag{4.6}$$

where $p(\boldsymbol{x}|\boldsymbol{d})$ is called the "posterior probability". It is the probability of the parameters being $\boldsymbol{x}$ given that the data $\boldsymbol{d}$ has been observed. $p(\boldsymbol{d}|\boldsymbol{x})$ is called the "forward probability" or when viewed as a function of $\boldsymbol{x}$, the "likelihood function". This is the probability that the data observed is $\boldsymbol{d}$ given the set of parameters is $\boldsymbol{x}$. $p(\boldsymbol{x})$ is the "prior probability" and represents assumed knowledge that a given set of parameters $\boldsymbol{x}$ is more probable than any other. Finally, the factor $1/p(\boldsymbol{d})$ is the reciprocal of the probability of obtaining the given data $\boldsymbol{d}$. This is usually calculated by normalising the probabilities.

In the current data fit, where no particular set of parameters is assumed to be more likely than any other, $p(\boldsymbol{x})$ is constant and the posterior probability (the term most relevant to a data fitting problem) becomes:

$$p(\boldsymbol{x}|\boldsymbol{d}) \propto p(\boldsymbol{d}|\boldsymbol{x}) \tag{4.7}$$

therefore the task of ML is to find the set of parameters $\hat{\boldsymbol{x}}$ that maximises the likelihood function $p(\boldsymbol{d}|\boldsymbol{x})$. Since this in turns maximises the posterior probability $p(\boldsymbol{x}|\boldsymbol{d})$, the set of parameters found this way is the most probable given the observed data.

In the case of the current fitting function (equation 4.5), the set of parameters is given by $\boldsymbol{x} = (A_1, \lambda_1, A_2)$ and the data is given by $\boldsymbol{d} = (d_1, d_2, \ldots, d_n)$. If the counts in the separate time channels are assumed to be independent, then the likelihood function, $L(A_1, \lambda_1, A_2)$ is written as:

$$L(A_1, \lambda_1, A_2) = \prod_{i=1}^{n} p(d_i|A_1, \lambda_1, A_2) \tag{4.8}$$

Incorporating the knowledge that the counts $d_i$ are distributed according to the Poisson distribution (equation 4.4), the maximum likelihood function becomes:

$$L(A_1, \lambda_1, A_2) = \prod_{i=1}^{n} \frac{\mu_i^{d_i} e^{-\mu_i}}{d_i!} \tag{4.9}$$

where,

$$\mu_i = A_1 e^{-\lambda_1 t_i} + A_2 e^{-\lambda_2 t_i} \tag{4.10}$$

The "log-likelihood" function, defined as $g(A_1, \lambda_1, A_2) \equiv -ln(L(A_1, \lambda_1, A_2))$, turns out to be far easier to minimise computationally. It can be written as:

$$\begin{aligned} g(A_1, \lambda_1, A_2) \ \ &= -\ln\left(\prod_{i=1}^{n} \frac{\mu_i^{d_i} e^{-\mu_i}}{d_i!}\right) \\ &= -\sum_{i=1}^{n} (d_i \ln \mu_i - \mu_i) + const. \end{aligned} \tag{4.11}$$

The Nelder-Mead simplex (direct search) algorithm[52] is used to optimise the log-likelihood function in order to find the parameters of best fit $\hat{\boldsymbol{x}}$.

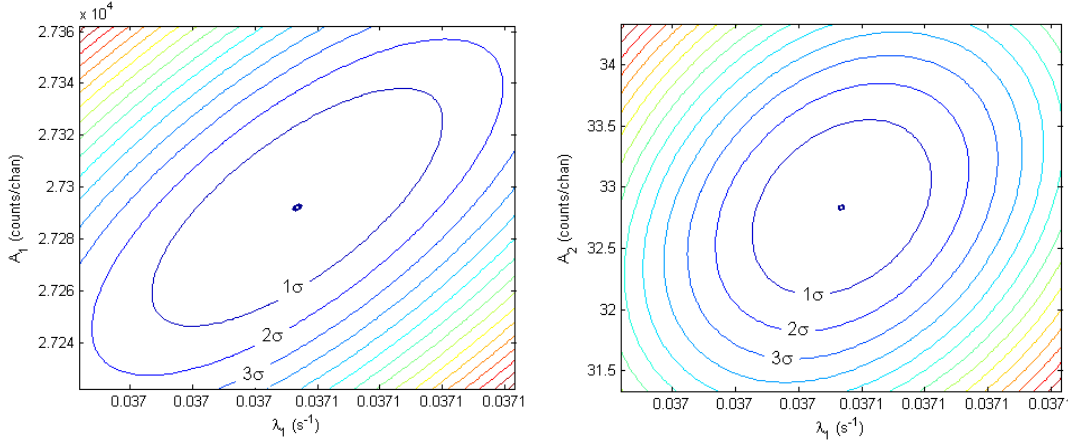

Figure 4.8: $1\sigma$, $2\sigma$ and $3\sigma$ contours of the likelihood function in the $A_1$ -vs- $\lambda_1$ and $A_2$-vs-$\lambda_1$ parameter space representation.

**Estimation of the Parameter Uncertainties**

As well as finding the best parameters from the fitting procedure, one would also like to know the uncertainty associated with each parameter. For most well-behaved fitting problems, the likelihood function can be approximated by a multivariant Gaussian peak in parameter space. The uncertainties in the parameters can then be estimated by the projection of the $1\sigma$ contour of the Gaussian onto the respective parameter axis. Mathematically[49], the approximating Gaussian is given by:

$$L(\boldsymbol{x}) \approx \mathcal{N} \exp\left(-\frac{1}{2}(\boldsymbol{x} - \hat{\boldsymbol{x}})^t \mathbf{Q}(\boldsymbol{x} - \hat{\boldsymbol{x}})\right) \quad (4.12)$$

where $\mathcal{N}$ is a normalisation factor and $\mathbf{Q}$ is the inverse covariance matrix where the elements are obtained from the negative of the mixed partial second order derivatives of the log-likelihood function $g(\boldsymbol{x})$, i.e,

$$Q_{ij} = -\frac{\partial^2 g(\boldsymbol{x})}{\partial x_i \partial x_j} \quad (4.13)$$

The covariance matrix, $\mathbf{R} = \mathbf{Q}^{-1}$. The $1\sigma$ uncertainty in each of the parameters is then given by the square-root of the diagonal elements of the covariance matrix, i.e. $\sigma_{x_i} = \sqrt{R_{ii}}$.

For the current fitting function where $\boldsymbol{x} = (A_1, \lambda_1, A_2)$ with $g(A_1, \lambda_1, A_2)$ and $\mu_i$ given by equations 4.11 and 4.10 respectively, the inverse covariance $\mathbf{Q}$ matrix can be expressed as:

$$\mathbf{Q} = \begin{pmatrix} \sum\limits_{i=1}^{n} \frac{d_i}{\mu_i^2} e^{-2\lambda_1 t_i} & -\sum\limits_{i=1}^{n} \frac{t_i}{\mu_i^2} \left\{ d_i A_i e^{-2\lambda_1 t_i} - \mu_i (d_i - \mu_i) e^{-\lambda_1 t_i} \right\} & \sum\limits_{i=1}^{n} \frac{d_i}{\mu_i^2} e^{-(\lambda_1 + \lambda_2) t_i} \\ (Q_{21} = Q_{12}) & -\sum\limits_{i=1}^{n} \left\{ A_1 t_i^2 e^{-\lambda_1 t_i} \left( \frac{d_i}{\mu_i} - 1 \right) - \frac{d_i A_1 t_i^2 e^{-2\lambda_1 t_i}}{\mu_i^2} \right\} & -\sum\limits_{i=1}^{n} \frac{d_i t_i}{\mu_i^2} A_1 e^{-(\lambda_1 + \lambda_2) t_i} \\ (Q_{31} = Q_{13}) & (Q_{32} = Q_{23}) & \sum\limits_{i=1}^{n} \frac{d_i}{\mu_i^2} e^{-2\lambda_2 t_i} \end{pmatrix}$$

It is evident that analytically calculating the uncertainties in the parameters is difficult. Even a simple change of the form of the fitting function requires a re-calculation of the $\mathbf{Q}$ matrix which may be cumbersome.

### 4.3.2   Recursive Least Squares

The bias problem occurs because low count rates give a poor estimate of the true mean of the distribution in the channel. This then leads to a poor estimate of the real uncertainty. Information of how the counts vary between adjacent channels, provided by the channels of higher counts, should be used to improve the knowledge of the mean for channels of low counts. This can be done by first making an initial least-squares fit of a decay curve, then using the value for the counts in a channel $\hat{d}_i$ found from this fit as the estimate of the true mean $\mu_i$ for the next fit. Repeating this gives progressively better approximations of $\mu_i$ and hence reduces the effect of the bias problem. This is known as the recursive least-squares (RLS) fitting procedure[50].

A normal least squares fitting procedure can be shown to be equivalent to the ML fitting if the noise in the data is additive, independent and Gaussian distributed (with mean zero)[49]. This is because the likelihood function $L(\boldsymbol{x})$ becomes:

$$L(\boldsymbol{x}) \propto \exp \left[ \sum_{i=1}^{n} \frac{(y_i - \mu_i)^2}{\sigma_i^2} \right] \qquad (4.14)$$

and the log-likelihood function $g(\mathbf{x})$ is then given by:

$$g(\boldsymbol{x}) \propto \sum_{i=1}^{n} \frac{(y_i - \mu_i)^2}{\sigma_i^2} \qquad (4.15)$$

Minimisation of this is therefore equivalent to the procedure of least-squares fitting. The implementation of the least-squares fitting used in this experiment employs a Levenberg-Marquardt function minimisation algorithm [49] and the convergence condition is set to when the relative differences between iterated $\lambda_1$ are less than $10^{-5}$.

The RLS procedure of improving the precision of the true mean of the distribution in a channel is also effectively achieved from obtaining larger counts. Thus RLS can be intuitively thought of as tending the uncertainties in each channel towards a Gaussian distribution. This can be viewed as why the result obtained by RLS is expected to be close to that from ML fitting.

**Estimation of the Parameter Uncertainties**

The theory used for the estimation of the uncertainty in the parameters from RLS fitting used is the same as that described for ML fitting, except that, the likelihood function and the log-likelihood function are given by equations 4.14 and 4.15 respectively. One other difference is that the second order mixed partial derivatives required to calculate the inverse covariance matrix $\mathbf{Q}$ are calculated numerically rather than analytically.

### 4.3.3 Comparison

To decide whether using the parameter uncertainty estimates from the RLS fitting procedure is justified for the current fitting function, the results obtained were compared to that from ML fitting. In general, it was found that the fitted half-life $ln(2)/\lambda_1$ and initial activity $A_1$ from the two methods were in agreement to less than 0.001% with their uncertainties the same to better than 1%. Also the fitted contaminant initial activity $A_2$ and uncertainty agreed to better than 0.01% and 2% respectively. The convergence of the RLS fitting procedure to that of the ML procedure for increasing iterations is shown in

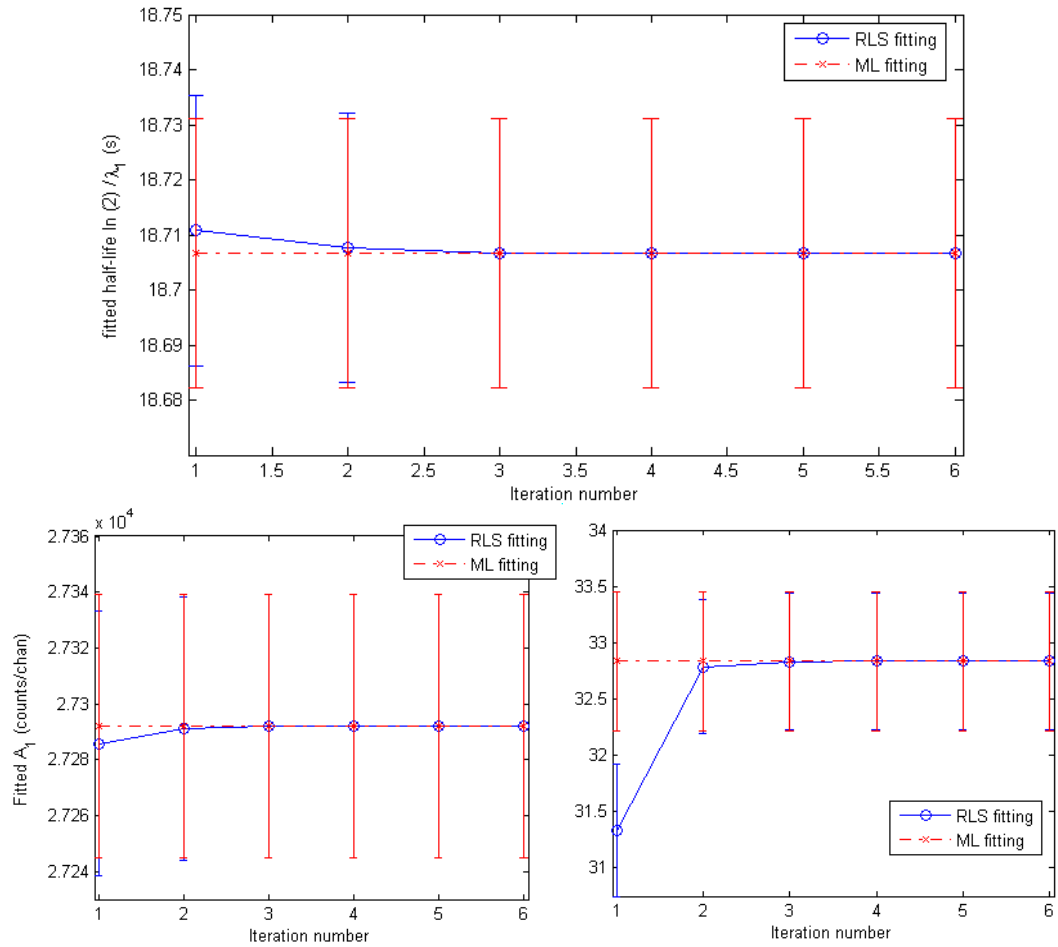

Figure 4.9: The values of the parameters determined by the recursive least-squares (RLS) fitting procedure converging to that from the maximum likelihood (ML) fitting procedure.

figure 4.9. This is the result obtained from the fit of a typical decay curve (run name 'AU011101' described in table 5).

The results obtained from the comparisons show that the use of the RLS procedure in the current fitting problem is well justified.

### 4.3.4 Effect of Contaminants

**Varying Levels of the Long-lived Contaminant**

Contaminant or background activity is unwanted in a high precision half-life measurement. It increases the error of each data channel used in the fit of the interested half-life even if the time distribution of the contaminant is known and fitted for. This can be understood as the following: consider the number of counts in a time channel to be $d$ with an error $\sigma_d$. Correctly fitting for a contaminant activity is essentially subtracting the expected mean of contaminant activity $a$ from the observed activity. Thus the corrected activity is given by $d' = d - a$ and the error in $d'$, $\sigma_{d'} = \sqrt{\sigma_d^2 + \sigma_a^2}$. So even if the precise mean of the contaminant activity in the channel is known and corrected for, the true counts decreases while the error increases so that the fractional error in each data channel used for the half-life fit is drastically increased.

Fitting of simulated decay curves, following the form of equation 4.5, was used to study how the presence of varying amounts of a long-lived contaminant activity affects the value and uncertainty in the fitted half-life $ln(2)/\lambda_1$. Decay curves were generated with $ln(2)/\lambda_1 = 20.00$ s, $A_1 = 10,000$ counts/chan, $ln(2)/\lambda_2 = 20.38$ mins, and $A_2$ at 0.1%, 0.2%, 2% and 10% of $A_1$.

Since this is a statistical effect on the fitted half-life, the curves were generated with Poisson distributed noise added to each time channel. To remove possible correlations in the pseudo-random number generation process, 10 values from the desired Poisson distribution were first generated and then only one value from these was selected using a separately generated random number.

The results of the fitted half-life for varying amounts of contaminant activity are shown in figure 4.10. The plot on the left shows that fitting for increasing amounts of contaminants increases the uncertainty in the fitted half-life but the values are still consistent with the true half-life (indicated by the red line). The plot on the right show the points vertically aligned to emphasis the increase in the uncertainty between the values. An increase of around 50% in the uncertainty of the fitted half-life is observed between the case of 0.1% and 10% contaminants at an initial value of $A_1 of 10,000 \text{s}^{-1}$.

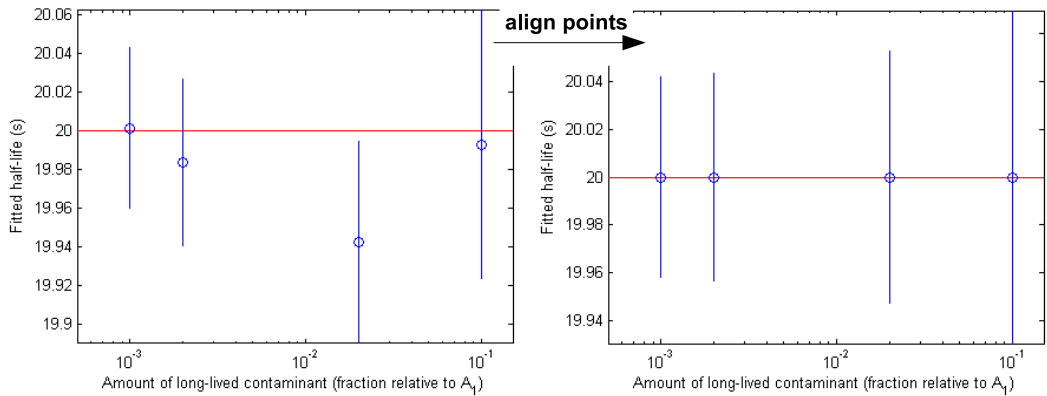

Figure 4.10: The effect of fitting for increasing amounts of contaminant activity on the value of the fitted half-life as well as its uncertainty.

**Use of the Wrong Contaminant Half-life**

The half-life of the long-lived contaminant is assumed to be precisely known in the fitting problem. This might turn out to be false if the contaminant is actually different to what was expected or the half-life of the contaminant had been poorly determined in the past. An investigation of the effect of contaminant half-life error on the fitted half-life of the activity of interest is made. This was done by fitting decay curves with the simulated contaminant half-lives different to that of the half-life used in the fitting function. Since this is a systematic effect, the curves were generated not to include noise so that there is effectively infinite precision in each time channel.

The decay curves were generated with $ln(2)/\lambda_1 = 20.00$ s, $A_1 = 10,000$ counts/chan, $A_2 = 0.2\%$ of $A_1$ and $ln(2)/\lambda_2$ varied from between $\pm 0.1$s to $\pm 1000$s. The fixed value of the half-life used in the fitting function was set at 20.38 mins. The results obtained are shown in figure 4.11. It can be concluded that as long as the half-life of the long-lived contaminant used for the fit is not wrong by more than $\pm 100$ s, the effect of the systematic error on the fitted half-life is negligible.

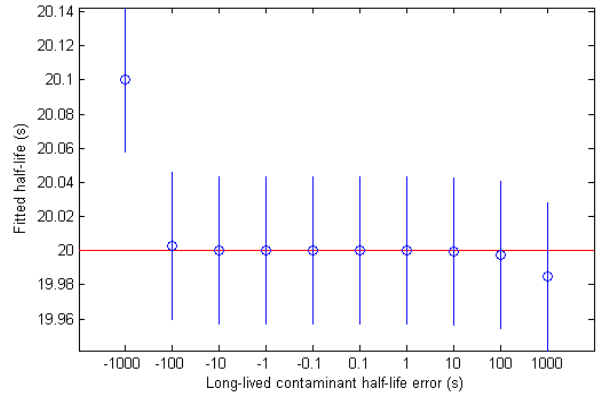

Figure 4.11: The effect on the fitted half-life of using the wrong long-lived contaminant's fixed half-life.

## Presence of an Unaccounted for Contaminant

It is possible for there to be a small amount of some unaccounted for contaminant activity present in the data. If the half-life of this activity is comparable to that of the known long-lived contaminant, then the correction for it would be automatically incorporated. Therefore it is the unknown presence of a contaminant with a short half-life that would be most affect the fitted half-life of the activity of interest the most. Testing of this effect comes from simulated decay curves. Since this is, again, a systematic error in the fitted half-life, no noise will be added to the time channels.

The effect of varying both the unaccounted for contaminant half-life as well as its intensity was investigated. The simulated decay curves have $ln(2)/\lambda_1 = 20.00$ s, $A_1 = 10{,}000$ counts/chan, $A_2 = 0.2\%$ of $A_1$ and $ln(2)/\lambda_2 = 20.38$ mins. The unaccounted for contaminant had its half-life varied between 5s and 30s and the amount varied between 0.1% and 10% of $A_1$. The effects on the fitted half-lives are shown in figure 4.12. It can be concluded that the existence of an amount of unknown short-lived contaminant greater than 1% of the initial counts is enough to cause a significant systematic error in the fitted half-life.

The three tests described previously give an estimate of the possible magnitude of the sorts of statistical and systematic effects that may be present in our data.

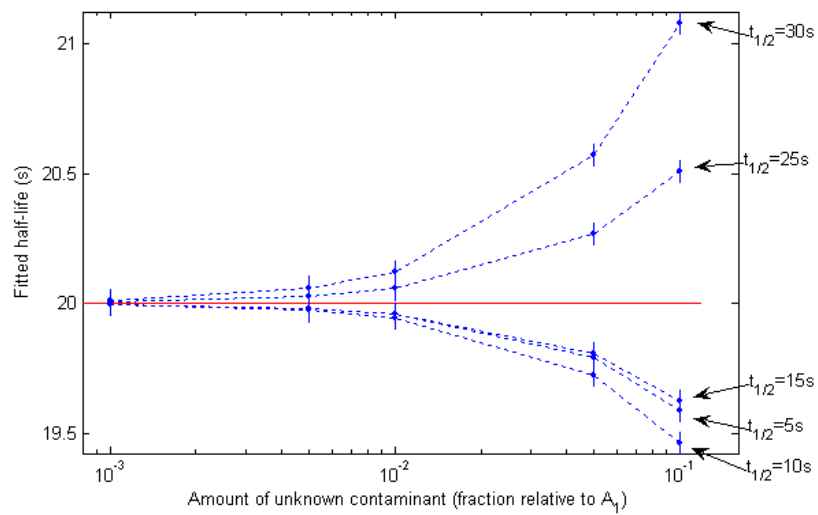

Figure 4.12: The effect of the presence of an unaccounted for short-lived contaminant on the fitted half-life value.

# Chapter 5

# Data Analysis

The numbers of runs (with 5 passes per run) considered acceptable for analysis for the various beam conditions were: 4 with the thin target at 7.7MeV beam energy and 300nA beam intensity nominally (group A), 4 with the thick target at 6.7MeV effective beam energy and 250nA beam intensity (group B), 9 with the thick target at 6.7MeV effective beam energy and 100nA beam intensity (group C) and 8 with the thick target at 7.4MeV effective beam energy and 100nA beam intensity (group D). Details of the runs can be found in Table 5 on p.120.

The sets of data were analysed in `Matlab 7` on a PC with an Athlon XP 1.8 GHz processor and 1 GB of RAM. The event mode data files were imported into the program and then time ot energy projected as required. All procedures, including the cumbersome repeats of large numbers of recursive non-linear least squares fits or dead time loss correction of individual time channels from a pass, took a few minutes to complete.

## 5.1   Initial Analysis

The data analysis was decided to be performed for the separate groups (defined by the target and beam conditions) of data. The recursive non-linear least squares procedure described in the previous chapter was applied. The time

| Run Name | Intensity(nA) | Proton Count | Tot. Counts |
|----------|---------------|--------------|-------------|
| *Group A: Thin target, nominal 7.7MeV energy & 300nA intensity* | | | |
| AU301011 | 330 | 172,614 | 72,990 |
| AU311003 | 300 | 131,598 | 18,363 |
| AU311005 | 340 | 178,898 | 46,311 |
| AU311006 | 390 | 210,741 | 60,614 |
| *Group B: Thick target, nominal 6.7MeV effective energy & 250nA intensity.* | | | |
| AU311007 | 220 | 104,048 | 1,139,773 |
| AU311008 | 270 | 127,412 | 1,325,501 |
| AU311009 | 240 | 107,954 | 988,154 |
| AU311010 | 250 | 110,374 | 922,363 |
| *Group C: Thick target, nominal 6.7MeV effective energy & 100nA intensity.* | | | |
| AU011101 | 140 | 67,959 | 728,983 |
| AU011102 | 105 | 52,243 | 608,043 |
| AU011103 | 90 | 39,582 | 535,387 |
| AU011104 | 80 | 38,200 | 477,585 |
| AU011105 | 105 | 51,193 | 682,040 |
| AU011107 | 100 | 48,564 | 682,852 |
| AU011109 | 80 | 35,677 | 474,646 |
| AU011110 | 70 | 31,641 | 409,417 |
| AU011111 | 100 | 54,734 | 672,244 |
| *Group D: Thick target, nominal 8.3MeV effective energy & 100nA intensity.* | | | |
| AU011113 | 85 | 36,820 | 328,367 |
| AU011114 | 80 | 43,509 | 346,845 |
| AU011116 | 80 | 47,813 | 689,378 |
| AU011116b | 80 | 43,176 | 599,497 |
| AU011117 | 110 | 54,509 | 463,667 |
| AU011118 | 90 | 46,168 | 384,875 |
| AU011119 | 85 | 43,398 | 298,845 |
| AU011120 | 90 | 42,339 | 344,817 |

Table 5.1: The runs collected decided acceptable for data analysis. They are split into four groups determined by the target and beam conditions. Details given are: The name of the run, which contains "AU" followed by the date then run number of this day (e.g. AU311003 was the 3rd run performed on the 31st of October 2007), the proton count given by the sum of the beam on proton counts from E-Lifetime, and the total counts collected in the run.

projected data was fitted in this way by the function:

$$N = A_{10C}e^{-\lambda_{10C}t} + A_{11C}e^{-\lambda_{11C}t} \tag{5.1}$$

where $N$ is the observed counts, $t$ is the time of the channel, $A_{10C}$ and $A_{11C}$ are the counts of $^{10}$C and $^{11}$C at $t = 0$ and $\lambda_{10C}$ and $\lambda_{11C}$ are the decay probabilities of $^{10}$C and $^{11}$C. The decay probability is related to the half-life by $\lambda = \ln 2/t_{1/2}$.

The process of fitting the half-life of an exponential decay curve to observations which are made over only a fraction of the half life is highly prone to error. So if the long half-life of $^{11}$C were to be varied in the fit, the uncertainty associated with doing so would propagate into the uncertainty of the fitted half-life of $^{10}$C. This is undesirable and since the half-life of the $^{11}$C contaminant is well-known, it is be fixed at the standard accepted value [53] of 20.39(2) mins in all fitting procedures. The uncertainty in this value was shown in section 4.3.4 to have a negligible effect on the fitting problem. In hindsight, a run (or even just a pass) performed with a 4-5 hour beam off time would have been useful for checking that the contaminant activity observed was indeed coming solely from $^{11}$C and not other long lived contaminants (such as $^{13}$N with a 600s half-life).

There is also a constant cosmic ray background of around 1 count per second. Attempting to fit the constant background increases error in the whole fitting process by requiring an additional free parameter in equation 5.1. This causes an increase in error in the other fitted parameters. Due to the small magnitude of the background counts and its close resemblance to the shape of the long decaying $^{11}$C component, this correction can be absorbed into the $A_{11C}$ term.

The AU311007 run of Group B will be used as an example to discuss the validity of the data fitting procedure. The fit of the data from this run resulted in a poor normalised chi-squared ($\chi_\nu^2 = 2.2$) with the fitted half-life found to be 18.367 ± 0.019 s. Other fitted parameters for this run were $A_{10C} = 7,394$ ± 11 s$^{-1}$ and $A_{11C} = 17.38 \pm 0.18$ s$^{-1}$.

A plot of the data misfits of the individual counts, given by: $\varepsilon_i = (N_i -$

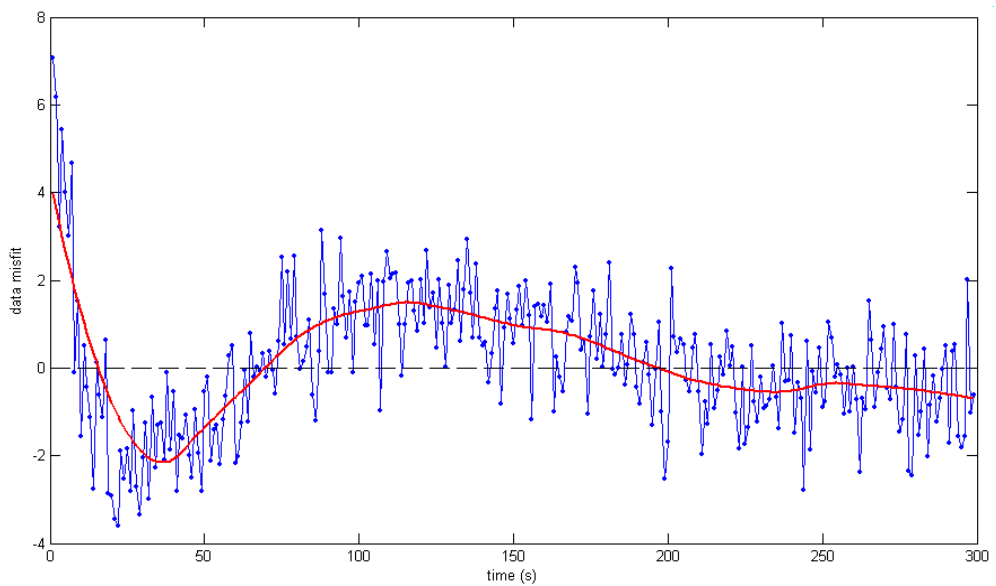

Figure 5.1: A plot of the data misfit ($\varepsilon_i = (N_i - \hat{N}_i)/\sigma_{N_i}$) against time for the AU311007 (Group B). The solid red line is a smoothed curve of the points displayed to guide the eyes of the reader.

$\hat{N}_i)/\sigma_{N_i}$, is shown in figure 5.1. The time distribution of the data misfits is expected to distributed about zero for the values for which a good fit is made. This is not what was observed for the current fit. From the shape of misfits, it can be concluded that the size of the deviation from the expected curve is largest for high count rates. This could indicate the presence of count rate dependent (CRD) effects in our data.

## 5.1.1   The Amount of Contaminant Activity

The ratio of the amount of $^{10}$C activity of the first time channel to that from the $^{11}$C contaminant ($A_{10C}/A_{11C}$) was found to be approximately constant for the individual runs in each group. A ratio of approximately 50:1 was found in group A, 500:1 in group B, 500:1 in group C and 400:1 in group D. These values are roughly as expected from the different target purities and beam conditions used. These amounts of contaminant activity intensity have been shown to increase the statistical uncertainty in the fitted half-life by only a small amount (see secion 4.3.4).

## 5.2 Count Rate Dependence (CRD) of the Fitted Half-life

A method referred to as "channel chopping" is used to observe how the count rate dependence affects the fitted values of the half-life [12]. In this, a single decay curve is analyzed multiple times by shifting the starting point of the fit by approximately half a half-life (10s) for several half-lives. The fitted half-lives obtained from the different starting points are plotted against the initial count rate. To improve statistics, the points obtained in this way from a group were combined into a single plot to obtain fitted half-life versus initial count rate relations for the different beam conditions.

The CRD effects may cause a large increase in the true counts for the high rates. Following on the same arguments as for the dead time loss corrections in section 4.2.4, rate dependent effects of this magnitude may require the analysis of the passes individually. The data are thus analysed in two ways for comparison: channel chopping decay curves from a run and decay curves from individual passes. The poor statistics in the latter, due to the lower counts in each channel, is made up for by more fitted half-life versus initial count rate points.

What initially appeared to be linear trends of the fitted half-life versus initial count rate were fitted with a weighted least-squares linear regression procedure. The weightings used were the uncertainties in the fitted half-lives calculated from the recursive non-linear least squares fitting procedure (plotted as $1\sigma$ error bars). The best calculation of the true half-life of $^{10}$C, free of CRD effects, from this data is to find the extrapolated fitted half-life corresponding to zero initial counts. These values, along with the slope of the linear regression, are shown in table 5.2 for the different groups.

### 5.2.1 Discussion

- The decay curves from Group A have small initial count rates, restricting the number of "channel chops" that can be performed. This makes it difficult to fit a sloped line since the points have a small initial count rate range and a large relative uncertainty (due to the poor statistics). It appears the values of the extrapolation slopes for this group are consistent with zero, so it may be better to fit the data with a flat line (equivalent to taking the weighted mean). In doing so, it is assumed that the count rates are sufficiently low for any CRD effects to be negligible. Whether this is valid for this count rate range will be discussed later.

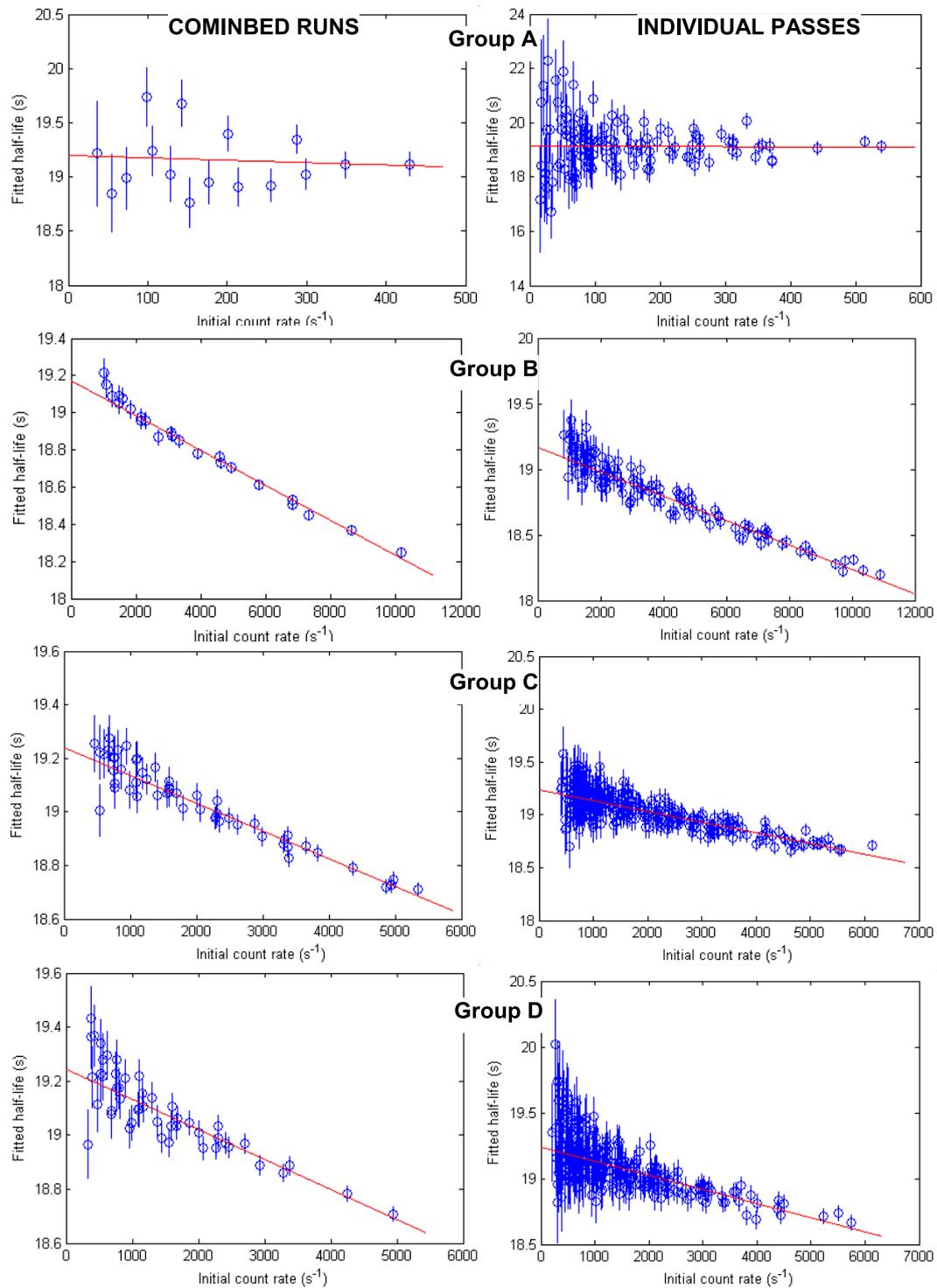

Figure 5.2: Plots of the fitted half-lives versus initial count rates found using the "channel chopping" procedure for the different groups of data. Left hand column are plots of the data analysed from the combined runs; right hand column from individual passes. The different scales and offsets of the axes between the fits should be noted.

| grp | Fitted with combined runs | | Fitted with individual passes | | chops |
|---|---|---|---|---|---|
| | **half-life** (s) | **slope**($10^{-4}s^2$) | **half-life** (s) | **slope**($10^{-4}s^2$) | |
| $A$ | 19.20± 0.12 | -2.2±4.1 | 19.16±0.11 | -6±18 | 4×10s |
| $A$ | 19.14±0.04 | (zero) | 19.126±0.044 | (zero) | 4×10s |
| $B$ | 19.172±0.015 | -0.938±0.024 | 19.169±0.015 | -0.928±0.023 | 6×10s |
| $B$ | 19.187±0.014 | -0.960±0.022 | 19.184±0.014 | -0.950±0.022 | 8×10s |
| $B$ | 19.192±0.013 | -0.967±0.022 | 19.189±0.013 | -0.957±0.021 | 10×10s |
| $C$ | 19.241±0.014 | -1.038±0.042 | 19.236±0.014 | -1.014±0.041 | 6×10s |
| $D$ | 19.243±0.016 | -1.11±0.06 | 19.241±0.015 | -1.08±0.06 | 6×10s |

Table 5.2: Summary of the extrapolated half-lives and the slope of the extrapolated line, from the groups of data using the "channel chopping" procedure. Analyses from combined runs and individual passes were performed for comparison. The "chops" column describes the number of "channel chops" and time interval between each.

- Although in agreement when considering uncertainties, all the values of the half-lives calculated from individual passes were lower than that from combined runs. If the systematic effect (described previously) had not observed then the differences would have been expected to be randomly distributed about zero. This does not appear to have been the case, indicating a small presence of the systematic effect in analysing data in runs as opposed to individual passes.

- The plots of group B shown in figure 5.2 are from "channel chopping" covering approximately 3 half-lives of $^{10}$C (chops at 6×10s). Due to the high initial count rate in group B, more "channel chops" can be made to observe the fitted half-life over a greater initial count rate range. The plots of the "channel chopping" covering 4 and 5 half-lives are shown in figure 5.3. A non-linearity of the data points becomes increasingly visible

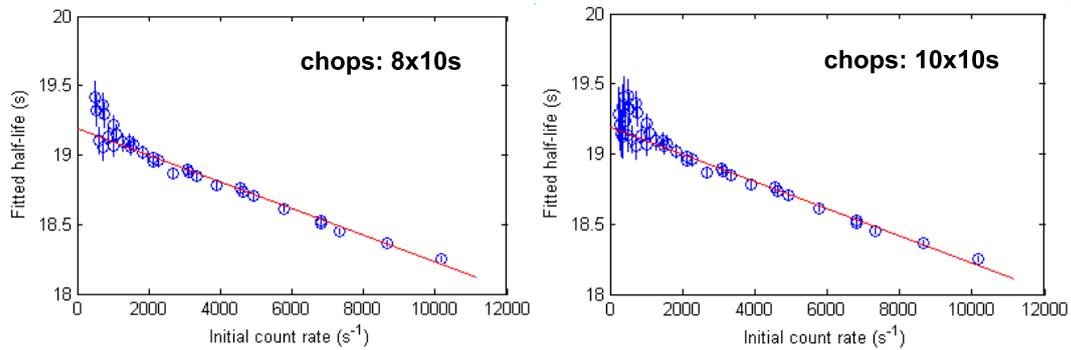

Figure 5.3: Plots of the fitted half-life versus initial count rate of group B with increasing range of "channel chopping".

> as the range of initial count rates observed increases. This can be seen as an increase in both the extrapolated half-life and in the size of the extrapolation slope.

The discovery of the non-linearity is very worrying for the analysis of the results in this way. It might mean that the form of the fitted half-life versus initial count rate dependence may have to be analytically determined and corrected for. Even more worrying is the fact that the non-linearity behaviour is positively concaved. That is, the slope of the points is greatest for the lower initial count rates (the same statement still applies if the fitted $\lambda_{10C}$ is plotted instead). This means that, in our system, even if low initial count rates are used, there will still be some count rate dependence in the fitted half-life.

## 5.3 The Count Rate Dependent Effects

### 5.3.1 The Source of the Count Rate Dependent Effects

So far in discussing the data, all the count rates mentioned were those from "3 out of 4 coincidences" in our system. Measurements made of the rate of events from a single detector when the "3 out of 4" coincidence count rate was around 6,000s$^{-1}$ (in groups B and C) was found to be approximately 300,000s$^{-1}$.

One might think that this is still a rather low rate considering the mean time interval between events is large (3$\mu$s) compared to the resolving time of the system (50-100ns). This is a false assumption. For the Poisson distribution, the distribution function $I_1(t)$ for the intervals between adjacent random events $t$ for a count rate of $r$ is given by:

$$I_1(t) = re^{-rt} \tag{5.2}$$

That is, although the mean interval between events is given by $1/r$, the most probable interval is zero. The rate at which more than one event will appear within a time interval $T$ is then given by:

$$r \int_{t=0}^{T} re^{-rt}dt = r(1 - e^{-rT}) \tag{5.3}$$

In the limit that $rT \ll 1$, the approximation that $e^{-rT} \approx 1 - rT$ can be made so that the rate of more than one event occuring within a time $T$ (equation 5.3) becomes $r^2T$.

**Count-rate dependent effects in our system**

With the dead time of the system already accounted for, there are two other ways, due to the finite time resolution of the system, that can cause the observed count rate to differ from than the expected rate: pile-up and chance coincidence.

In the current system, only a triple coincidence event generates a 100ns wide gate from the logic unit. For low count rates, only the summed analogue pulse corresponding to the coincidence event falls within the gate. As the count rate increases the probability of other pulses, that are not correlated to the coincidence event, falling within this gate increases. When separate analogue pulses in the gate are amplified and integrated by the slow amplifier, they will be shaped into a single peak. The result is a pulse of greater magnitude than just the pulses being amplified separately. This is how the pile-up effect occurs in our system.

The increase in observed counts from pile-up comes only from pulses brought above the system threshold due to their increased amplitudes (see Ref.[36].

This is because all the counts in the pulse-height spectrum are used in our analysis. The probability of counts being close to the threshold is low (see figure 4.5) and only a small shift in the "3 out of 4" coincidence spectrum was observed for increased count rates. Therefore pile-up is believed to contribute only a small amount to the large ($\sim$15%) CRD increase in count rates. To test that pile-up does not play a large effect in our system, one could measure the count rate while varying the width of the gate.

The other way that the observed counts might increase due to CRD effects is from an increased number of coincidence events caused by chance coincidences. This is when any of the events that appeared as a "3 out of 4" coincidence to the system did not originate from the same decaying nucleus. Recall that coincidence is defined as the overlap of the logic pulses from the CFD, hence the width (30ns) of the pulses gives the resolving time of the system for separating true and chance coincidence events. Note that in this description, the false coincidences (see section 3.3.3) are not considered chance coincidences and they do not play a role in count rate dependent increase of the expected counts.

The effect of chance coincidence involves overlapping of logic pulses (i.e. almost infinitely steep edges). This makes determining the size of its true effect more easy than for pile-up where partial overlaps can occur. The rate of chance coincidences in the current system will be lower than for pile-up since the 30ns resolving time is involved and not the 100ns gate time. In reality, the resolving time should be optimised (most likely made smaller) for the detection of true triple coincidence events so that the chance coincidence rate can be reduced.

**Theoretical considerations of the count rate dependence**

The general result of the chance coincidence rate $r_{ch}$ from two uncorrelated Poisson distributed rates $r_1$ and $r_2$ is given by:

$$r_{ch} = 2t_R r_1 r_2 \qquad (5.4)$$

In a simple double coincidence system the total chance coincidence rate is exactly as that from equation 5.4 when $r_1$ and $r_2$ become the rates in each individual detector. If the count rates in the detectors come predominantly from the activity of interest then the true double coincidence rate can be expressed as:

$$r_{2of2} = \alpha_{2of2:r_1} r_1 = \alpha_{2of2:r_2} r_2 \qquad (5.5)$$

where $\alpha_{2of2:r_1}$ and $\alpha_{2of2:r_2}$ are the constant ratios of double coincidence to the single rates $r_1$ and $r_2$. If the activity of interest decays with a probability $\lambda$ then the chance coincidence as a function of time is given by:

$$r_{ch}(t) = A_{ch} e^{-2\lambda t} \qquad (5.6)$$

So the chance coincidence rate manifests itself as a decaying signal of twice the probability (or half the half-life) of the activity of interest.

In a triple coincidence system, a chance coincidence may come from coincidence of 2 correlated events and an uncorrelated event (1st order) or 3 uncorrelated events (2nd order). The rate of the chance coincidence from these two types of events is expected to approximately take the form:

$$r_{ch}^{(1)} \propto r_{3of4}^2 \quad \text{and} \quad r_{ch}^{(2)} \propto r_{3of4}^3 \qquad (5.7)$$

and will appear in the decay curves as:

$$r_{ch}^{(1)}(t) \propto e^{-2\lambda t} \quad \text{and} \quad r_{ch}^{(2)}(t) \propto e^{-3\lambda t} \qquad (5.8)$$

A treatment of the chance coincidence rate in a simple "3 out of 4" coincidence system is given in Ref.[54]. An analytical correction for the chance coincidence may exists if the observed "2 out of 4" coincidence rate is simultaneously measured (perhaps by using a separate coincidence unit). Unfortunately, due to the nature of the $^{11}C$ contaminant in our experiment, the chance coincidence mechanism is far more complex so that an elegant approach such as this cannot be taken.

The rate of chance coincidences depends on the count rate in the individual detectors. In the discussion so far it has been assumed (such as in equation 5.7) that the count rates in the individual detectors come predominantly from the activity of interest. In the current experiment, the amount of activity from

$^{11}$C may be comparable to that from $^{10}$C (it only appears small in our "3 out of 4" coincidence counts due to the relative efficiencies of the system for the detection of the two types of event). This means that for any corrections to be applied, knowledge of the rates of singles counts resulting from both $^{10}$C and $^{11}$C is necessary.

A further complication is the fact that the two 511keV photons from the contaminant $^{11}$C are also correlated in time to each other. This means that there may be a coincidence between 2 correlated events from the $^{11}$C contaminant and an event from $^{10}$C. This is an additional source of 1st order coincidence which is equally as large as the coincidence between 2 correlated events from $^{10}$C and one from $^{11}$C.

Both of these factors combine to make the analysis of our system very complicated. The possibility of finding an analytic correction for the count rate dependence is therefore very slim.

### 5.3.2   Estimating from the data

The difference between the observed count rate and the true triples count rate with no CRD effects will give the magnitude of the count rate dependence. Therefore a procedure of calculating the magnitude of the count rate dependence from the data is to estimate the parameters corresponding to the decay curve with no CRDE. This was precisely the aim of the method of extrapolating the fitted half-lives to zero initial count rate described in the previous section. The remaining two parameters, $A_{10C}$ and $A_{11C}$ (from equation (5.1)), are required to be estimated in a slightly different fashion.

The extrapolation to zero initial count rate of the fitted initial count rates is, of course, zero. Thus, a slight modification of the previous procedure is required. The process of fitting decay curves starting at later initial times is decreasingly affected by CRD effects since the starting time is related to initial count rate. An estimate of $A_{10C}$ and $A_{11C}$ free from CRD effects can then be given by extrapolating the fitted $A_{10C}$ and $A_{11C}$ values to very large starting times. This technique is valid in principle but in practice, there may be large errors involved in the estimated parameters. It is therefore stressed at this stage that this technique is used only to estimate the general form of the count rate dependence and no quantitative analysis will be based on the results.

To gain the best indication of the form of the count rate dependence, decay curves which span a large range of count rates are required. The highest total counts, and therefore also the largest ranges of count rates per channel, are offered by the passes in the 'AU311010' run of group B. As many passes as possible should be combined to improve the statistics of the fitting procedure but if the different passes have varying counts in each time channel then when they are combined the CRD effect may become distorted. The counts in the first channel of the passes from the 'AU311010' run are: 10091, 10962, 9969, 9866 and 9742. The analysis was performed on the combined data from run with the 2nd pass discarded. This results in an initial count rate variation of only 3% (much smaller than the 12% originally).

The number of "channel chops" made on the decay curve is 20 at 5s intervals. The plots of the fitted $A_{10C}$ and $A_{11C}$ versus initial time channel are shown in figure 5.4. Fitting with a function of the form: $Ae^{-\lambda t_{chan}} + C$ was performed, where $C$ was then taken to be the extrapolated value for very large initial times and thus free from CRD effects. Both curves appear to agree well with the fitted function and the parameters estimated to be free from CRD this way were: $A_{10C} = 3.57 \pm 0.04 \times 10^4$ counts/chan and $A_{11C} = 69.9 \pm 0.4$ counts/chan. When combined with an estimated half-life of $19.25 \pm 0.10$ s, these parameters determine the curve with no CRD effects.

The estimated magnitude of the CRD effect as a function of time, calculated from the difference between the observed counts and the estimated counts without CRD, is shown in figure 5.5. It was guessed that a single decaying exponential ( $A_{CRD}e^{-\lambda_{CRD}t}$) approximated the shape of this well ($\chi_\nu^2 = 1.2$). The error in each point was taken to be equal to the error in the observed counts. A conservative estimate of the uncertainty in the determined half-life of the CRD ($ln(2)/\lambda_{CRD}$) was taken to be the range of values given by the possible combinations of the uncertainties in each parameter.

The size of the CRD is around 16% for a count rate of $10,000\text{s}^{-1}$. The

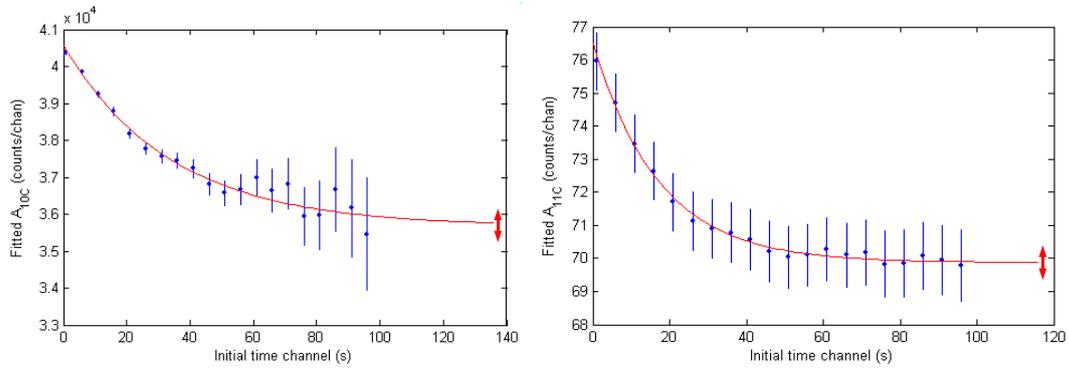

Figure 5.4: Plot of the fitted $A_{10C}$ (left) and $A_{11C}$ (right) versus initial time channel of the "chop" used. The red double-ended arrows on the right hand vertical axes indicate the $1\sigma$ error range of the extrapolated value.

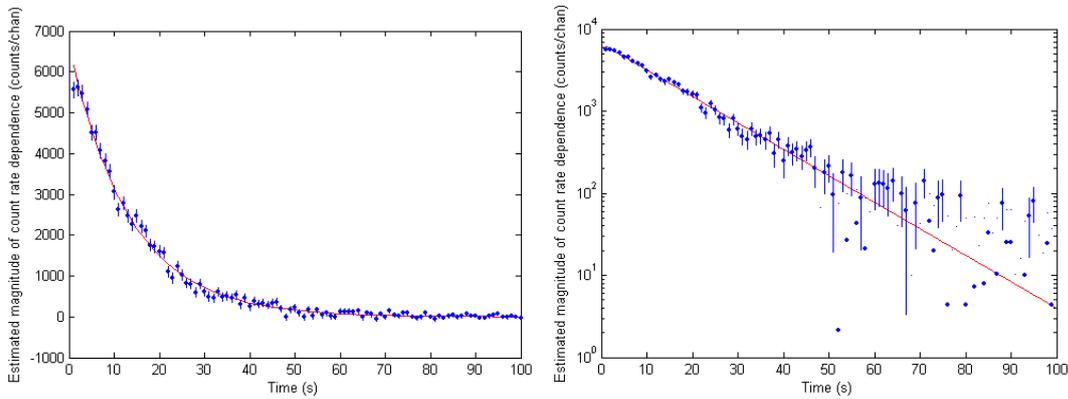

Figure 5.5: Linear (left) and logarithmic (right) plots of the estimated magnitude of the CRD effect versus time. Only the first 100s are plotted since the points beyond this are consistent with zero.

shape of the CRD effect appearing in our decay curves is close to the form:

$$N_{CRD} = A_{CRD}e^{-\lambda_{CRD}t} \qquad (5.9)$$

where the determined "half-life" of the CRD was in this case, $ln(2)/\lambda_{CRD=10,000s^{-1}}$ = $8.3 \pm 1.0$s. The same procedure was performed on data from group B where the initial count rate was $\sim 5,000s^{-1}$. The size of the CRD at this count rate was found to be $\sim 10\%$. The fitted "half-life" of the CRD decay was found to be $ln(2)/\lambda_{CRD=5,000s^{-1}} = 9.3 \pm 1.6$s with a $\chi^2_\nu = 1.1$ for the fit.

The final results represent a good approximation to the CRD effects. There is also a hint of the "half-life" of the CRD effect tending towards that of half the main half-life (signature of the 1st order CRD effect) for low count rates.

### 5.3.3 Simulation of the Fitted Half-life versus Count Rate

From the previous discussion and analysis, the nature of the CRD effects has been shown to be large and very complicated, making analysis with high data rates very difficult. The "channel chopping" technique in section 5.2 revealed a non-linear trend in the fitted half-life versus initial count rate for large count rates but sufficient linearity existed for the lower count rates. By relying on the principle that "all well-behaved functions can be approximated linearly over a sufficiently narrow range"[39], it may be possible for an accurate half-life free from CRD effects to be extrapolated using an appropriate initial count rate range. For this technique to work the linear trend caused by CRD had to continue to the true value of the half-life at zero count rate. In order to study this condition, simulated data with the true parameters known were put through the "channel chopping" procedure.

The model data was generated by drawing the value of the $i$'th time channel from a Poisson distribution with mean $\mu_i$ given by:

$$\mu_i = A_{10C}e^{-\lambda_{10C}t_i} + A_{11C}e^{-\lambda_{11C}t_i} + A_{CRD}e^{-2\lambda_{10C}t_i} \qquad (5.10)$$

where $log(2)/\lambda_{10C} = 19.20$s and $ln(2)/\lambda_{11C} = 23.39$mins. The initial count rate was set to $A_{10C} = 10,000\text{s}^{-1}$ with $A_{11C}/A_{10C} = 1/500$ and $A_{CRD}/A10C = 1/10$ to be similar to actual data. The procedure of generating the randomly distributed Poisson number is that described in section 4.3.4.

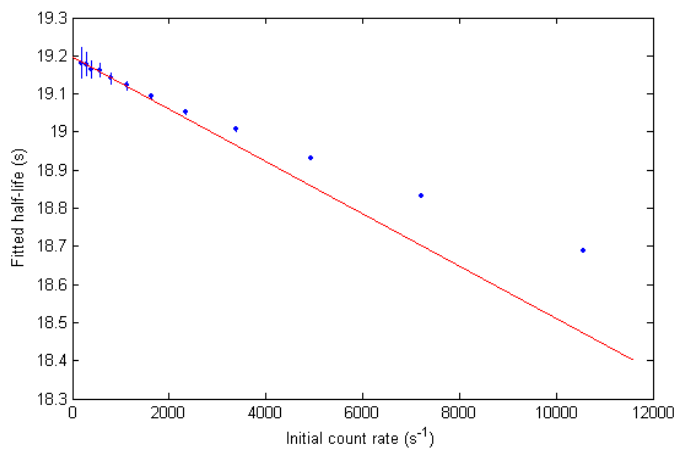

Figure 5.6: Fitted half-life versus initial count rate plots from simulated data containing standard 1st order count rate dependent effects.

Although the model of the count rate dependence is only simulated using the 1st order, it still serves as a good starting point for understanding CRD effects in more complex systems. This is especially the case since only the behaviour at the lower count rates is of concern in the proposed extrapolation technique.

The results from a simulation containing data from 200 passes is shown in figure 5.6. Each point is the weighted mean (from 200 passes) of the fitted half-life value for a specific "channel chop" ($12\times10$s). This was done to improve the overall statistics at each initial count rate. It is encouraging to see that for low count rates the points do tend toward the true simulated half-life (19.20s). The departure from linearity is also evident for count rates greater than $\sim$1,500s$^{-1}$. This occurs even for simple 1st order CRD effects. The line shown that extrapolates accurately to the true half-life value for zero initial count rate is fitted with only the 5 points of lowest count rate ($<$1,000s$^{-1}$). This suggests the existence of a possible count rate range for which the data can be fitted linearly to extrapolate a CRD free half-life value from.

## 5.4 Final Analysis

It has been shown that even a single first order count rate dependent effect (typical of simple 2-fold coincidence systems) in our data causes high non-linearities in the points of fitted half-life versus initial count rates. Due to the added complexities of a 3-fold coincidence system such as ours, attempting to find the form of the non-linearity has been avoided. Even if the form were able to be determined analytically, the poor statistics would make it impossible to fit a curve with a function which would presumably contain many free parameters.

An alternative method to overcome this may be to find a suitably small range of initial count rates from which to extrapolate the half-life free of CRD effects. This approach assumes only that the count rate dependence for low rates approaches that of the 1st order (which had been hinted at from the results in section 5.3.2).

### 5.4.1 Finding an Appropriate Linearity Range

Ideally one would like to have fitted half-lives for count rates but for very low values, the counts are predominately from $^{11}$C resulting in very poor fitting statistics. A nominal count rate of $50s^{-1}$ was fixed as the smallest initial count rate for the analysis of all the groups. At this small rate, the ratio from $^{10}$C to $^{11}$C is estimated to be around 2:1 to 5:1.

The result that suitable linearity exists for rates up to $1,000$-$1,500s^{-1}$ from the simulated fitted half-life versus initial count rate plot in the previous section is only approximate. The actual range should be found in each of the fitted half-life curves. The methodology of doing this is to examine a plot of the extrapolation slope for increasing initial count rate at $100s^{-1}$ increments. As the initial count rate gets large and starts including points of non-linearity, the value of the extrapolation slope should decrease monotonically toward zero. The largest initial count rate possible, before the onset of the monotonic decrease, should be chosen in order to gain the best statistics for the extrapolation slope. The plots of this for the data from group B, C and D (group A does not have good enough statistics for a meaningful result to be extracted from this method) are shown in figures 5.7, 5.8 and 5.9.

The plots of the extrapolation slope turned out to be quite complex (mainly due to the poor statistics) and it was not always clear when the onset of the monotonic increase due to non-linearity began. In order to gain a better understanding of the form of the plots, the following was done: linear trends of fitted half-life versus initial count rate points were generated at the initial count rates of the original data. Gaussian noise was then added to each point with the standard deviation the same as the originally fitted half-life data. In this way fitted half-life versus initial count rate points were generated with the same statistics as the original points but without non-linearity effects. The simulated points generated for each group were then put through the same analysis procedure as the original data.

The plots shown alongside the actual curve in the figures occurred approximately once in every five simulations. It was found that as long as the first few of the points from simulated data were distributed in a similar way to the actual data, the rest of the curve would more or less have the same shape and features. The simulated plots tended towards the value of the slope used to generate them (indicated by the red line), as expected. The important thing to note is that they all flatten off towards the generated slope value at the same portion (indicated by the red arrow) of their curve shape. This was used to determine the initial count rate at which the best extrapolation gradient can been reached. It can be seen that the monotonic increase due to non-linearity occurs shortly after this (only on the actual plot, because the simulated plots do not contain non-linearities in the data). This is indicative of the large CRD effects present in our data.

## 5.4.2   Results

The extrapolated half-life, extrapolation slope and the initial count rate range used for the three data groups are shown in table 5.4.2. Their spread and their weighted mean are shown in figure 5.10.

The combined weighted mean is $19.303 \pm 0.013$ s. It can be seen that the three half-life values from data collected using different beam conditions (the data from the target of lower purity was discarded) agree well with each other. It is also encouraging to see an agreement of the extrapolation slopes between the groups. The significance of the half-life value will be discussed in the next chapter.

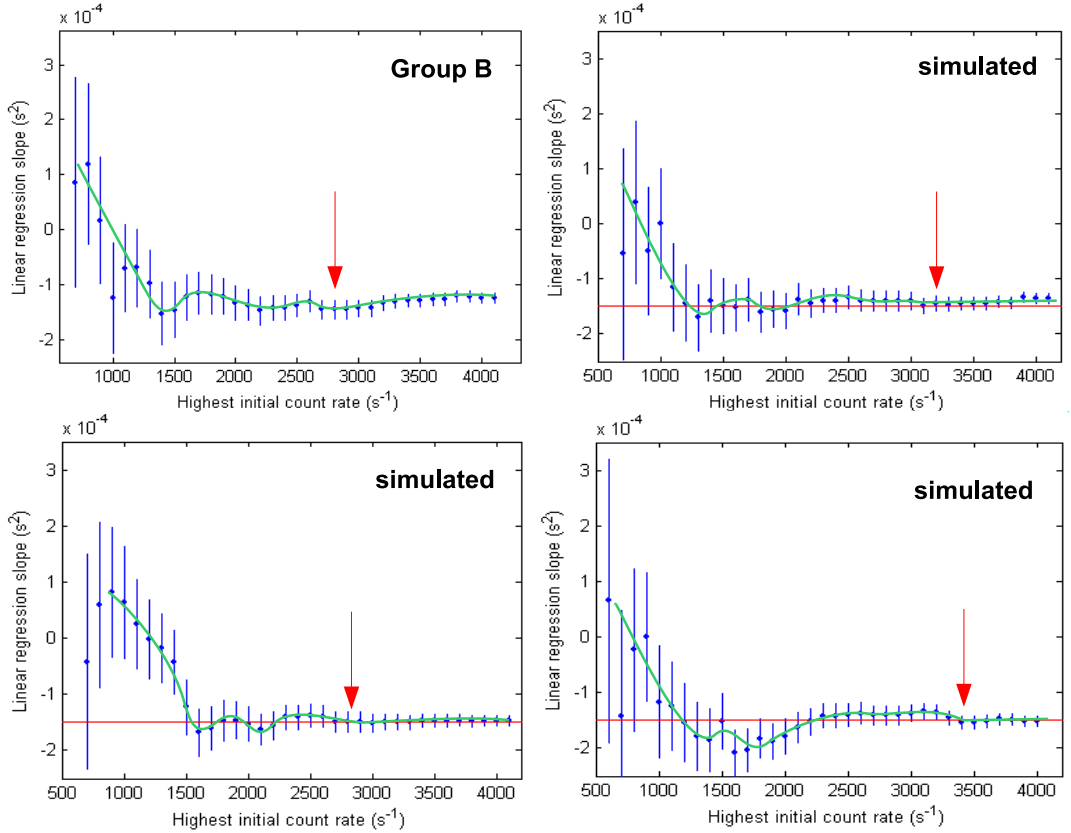

Figure 5.7: Extrapolation slope versus the highest initial count range used for group B.

| group | half-life (s) | slope ($\times 10^{-3}$s$^2$) | count rate range |
|:-----:|:-------------:|:-----------------------------:|:----------------:|
| $B$ | 19.285 ± 0.030 s | -0.146 ± 0.017 | 50-2,800s$^{-1}$ |
| $C$ | 19.308 ± 0.020 s | -0.139 ± 0.012 | 50-2,500s$^{-1}$ |
| $D$ | 19.306 ± 0.022 s | -0.151 ± 0.014 | 50-2,400s$^{-1}$ |

Table 5.3: Half-lives found by linear extrapolation. The extrapolation slope and initial count rates range used for each group are also shown.

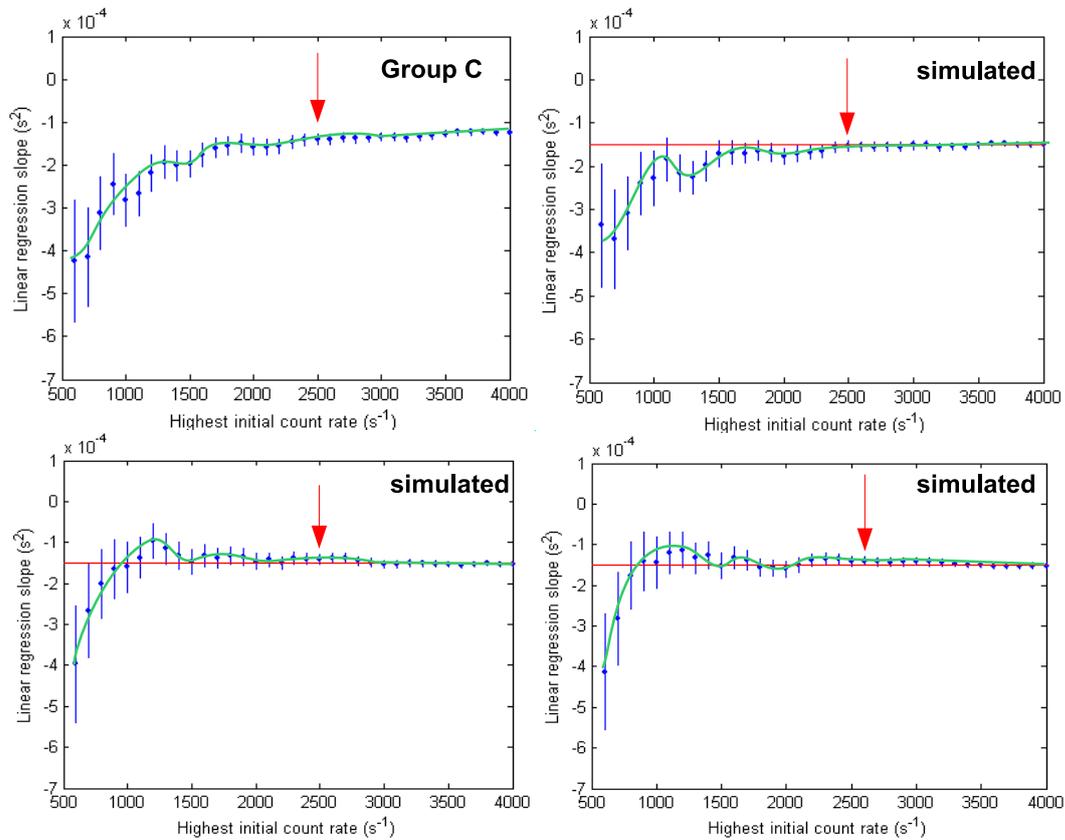

Figure 5.8: Extrapolation slope versus the highest initial count range used for group C.

## Discussion

Although in the end a set of consistent half-life values were extracted, the process used suffers from too many possible pit-falls. Many assumptions were made about the form of the CRD which may turn out to be false. Therefore there is not a great deal of confidence placed in the determined half-life value.

This experiment highlights the significance of CRD systematic errors due to high count rates in precision half-life experiments. It was discovered that the slope of the fitted half-life versus initial count rate is the steepest for low values of the count rate. The value of the slope also appears to be intrinsic to the detection system from the constancy of the extrapolation slope values in table 5.4.2. Therefore, any system that does not adequately deal with reducing

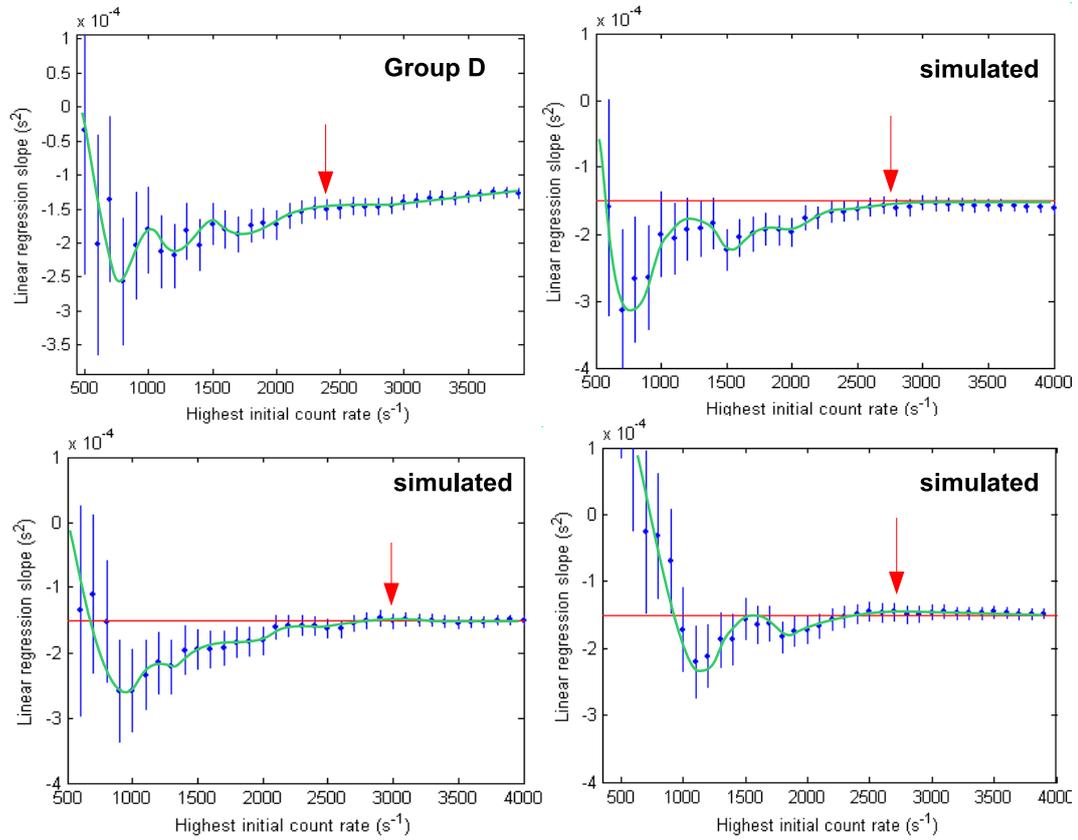

Figure 5.9: Extrapolation slope versus the highest initial count range used for group D.

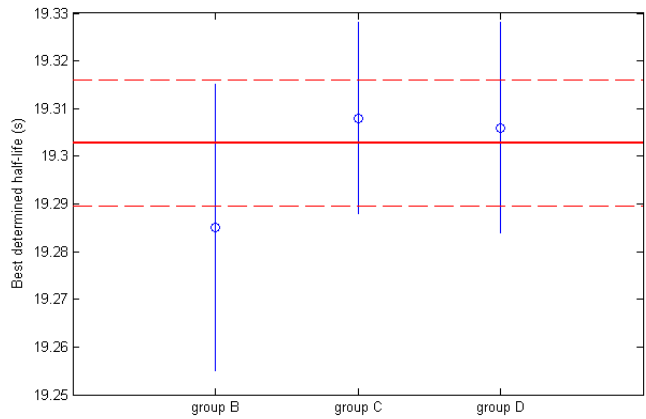

Figure 5.10: Values from Table 5.4.2. The red solid line and dotted lines indicate the value of the weighted mean and associated error.

CRD effects will suffer from systematic effects in the fitted half-life values for all count rates used. Thus, the recommended procedure for future half-life experiments is to collect data at the largest count rates possible, even if this data may be discarded later, it can still be used in determining the magnitude of CRD effects for small count rates.

# Chapter 6

# Conclusion

## 6.1 Comparison with Existing Results

The $^{10}$C half-life of 19.303 $\pm$ 0.013 s obtained in the current experiment is in good agreement with the set of two previously accepted values[1] of 19.280 $\pm$ 0.020 s (Ref.[11]) and 19.295 $\pm$ 0.015 s (Ref.[12]). A visual comparison of the values shown in the form of an "ideograph" is given in figure 6.1. To compare this ideograph to that of the $^{14}$O half-life measurements (p.16), it should be noted that both the previous $^{10}$C measurements were made with germanium detectors.

The previous value of the $^{10}$C half-life obtained from a weighted average was 19.290 $\pm$ 0.012 s. If the current measurement is included, the $^{10}$C half-life becomes 19.296 $\pm$ 0.009 s. This is still in agreement with the previous value. However, I propose to omit the value from Ref.[11] from the weighted mean for two reasons. Firstly, neither the maximum likelihood, nor the recursive least-squares, fitting procedure was used to extract the half-life. Secondly, no acknowledgment of the pile-up problem was ever made. And indeed, the lack of information about the count rates observed makes it very difficult to estimate to what extent pile-up has affected the result. The omission of this value then gives the value of the $^{10}$C half-life to be:

$$t_{1/2} = 19.300 \pm 0.010s$$

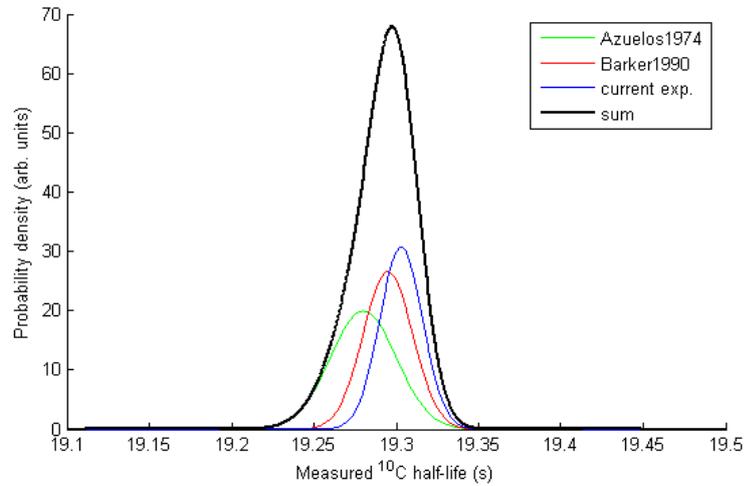

Figure 6.1: Ideograph of the $^{10}$C half-life values currently accepted along with the value obtained in this experiment.

## 6.2   Implications

With this value of the half-life, the $ft$-value, which gives the strength of the $^{10}$C superallowed $0^+ \rightarrow 0^+$ $\beta^+$-decay changes from $3039.5 \pm 4.7$ s [1] to:

$$ft = 3041.0 \pm 4.6 \text{s}$$

and the $\mathcal{F}t$ changes from $3073.0 \pm 4.9$ s [1] to:

$$\mathcal{F}t = 3074.5 \pm 4.8 \text{s}$$

The increase in the experimentally determined $ft$ value ($\sim0.5\%$ but still within uncertainty) may be large enough for a comparison with the systematic uncertainties in the theoretical corrections required in calculating $\mathcal{F}t$. The change in the $^{10}$C $\mathcal{F}t$ value does not change the overall $\overline{\mathcal{F}t}$ value by an amount noticeable at the current precision.

## 6.3  Future Directions

The confidence of the half-life value determined in this experiment is greatly reduced by the large count rate dependent effect observed. The technique in correcting for the (non-linear) dependence of the fitted half-life with initial count rate in Chapter 5 is based on too many assumptions for it to be fully justified in producing a result of high precision. Instead, it may be possible to calculate the count rate dependence from a detailed Monte Carlo simulation of the 4-detector 3-fold coincidence system. Although this would require fewer phenomenological assumptions, the precise size of the count rate dependence effect would rely heavily on the precision of the parameters used in the simulation such as the solid angle of the detectors with respect to the target, the relative geometry between the detectors, and so forth.

A much better approach is to reduce the size of pile-up in the current system. This can be done by optimising the sensitivity of the system to true triple coincidence events by performing tests (such as those described in Ref.[36]) with a $^{22}$Na source. If this turns out not to be enough the use of the 3-fold coincidence system might have to be abandoned and the half-life of $^{10}$C might have to come from other methods such as from energy-cuts in a single detector plastic scintillator $\gamma$-spectrum or detecting the positrons with E-$\Delta$E coincidence systems, such as that in Ref.[55].

## 6.4  Conclusion

Although in the end, a reasonable value of the $^{10}$C half-life was extracted, it was done in the midst of large pile-up effects. There, the confidence of presenting this value as a truly precise determination of the $^{10}$C is low. Nonetheless, this experiment high-lights the importance of searching for count rate dependent effects in high precision half-life measurements.

The 3-fold plastic scintillator coincidence system used in this experiment was used in the hope of achieving two things:

(a) reducing the amount of contaminant activity in the decay curves which reduces statistical uncertainty in the extracted half-life. This was hoped to be achieved by the 3-fold coincidence system detecting only the unique triple-photon decay signature from $^{10}$C and thus filtering out the double-photon signature from the known $^{11}$C contaminant.

(b) reducing pile-up which is known to cause a systematic error in the extracted half-life. This was hoped to be achieved by using plastic scintillators due to the short pulse durations.

In the current setup, (a) was believed to have been achieved satisfactorily. However, it was found that very large rates in the individual detectors were required to achieve reasonable triple coincidence count rates because of the poor efficiency for these events. This canceled out the advantage offered by the fast pulses from plastic scintillators and caused a failure in achieving (b). Any measurements are required to satisfy both (a) and (b) to within an acceptable standard. This has been shown by the current and previous measurement to be difficult for the case of $^{10}$C which is a good illustration of the importance of the "*only count what is going to be used*" maxim for high-precision half-life measurements [39].

The current status of the $^{10}$C half-life is therefore not good. With the low confidence in the current value and the proposed rejection of another (see section 6.1), all that remains is the value from Ref.[12]. However, pile-up was observed in that experiment which had used a germanium detector. The resolution of this problem was simply to discard decay curves with large initial count rates. But there are flaws in doing this as discussed in Chapter 5. Hence, a indisputable measurement of the half-life of $^{10}$C has not yet been achieved. It is imperative that this is done so that an accurate $\mathcal{F}t$ value can be obtained for this decay.

# Appendix A

# E-Lifetime Toolsheet Source Code

The Java source code of the E-Lifetime Toolsheet described in Chapter 2. Requires Kmax v8 to run.

```java
/**
* The 'Runtime' class provides Kmax runtime behavior.
* Check the documentation for 'Kmax Java Interfaces' for details.
*/

import kmax.ext.*;
import java.awt.event.*; //for ActionListener
import javax.swing.*;
import javax.swing.filechooser.*;
import java.io.*;                                                      10
import java.util.Date;
import java.text.DateFormat;

public class Runtime implements KmaxRuntime {

    /** Essential Kmax Objects */
    KmaxToolsheet tlsh; // Store a reference to the toolsheet environment
    KmaxDevice dev; // Declares the CMC100 device
                                                                       20
    /** Widgets in the parameters sub-pane in the Control Panel*/
    KmaxWidget beamOnTimeTextfield;
    KmaxWidget delayTimeTextfield;
    KmaxWidget beamOffTimeTextfield;
    KmaxWidget numOfPassesTextfield;
    KmaxWidget convGainCombobox;
```

```
KmaxWidget  hexReadTypeCombobox;
KmaxWidget  timeIntervalCombobox;
KmaxWidget  minBeamOnProtonCountTextfield;
KmaxWidget  maxBeamOffProtonCountTextfield;                              30
KmaxWidget  saveDataCheckbox;
KmaxWidget  chooseSavePathButton;
KmaxWidget  savePathLabel;

/** Widgets in the status sub-pane*/
KmaxWidget  beamOnRadiobutton;
KmaxWidget  delayRadiobutton;
KmaxWidget  beamOffRadiobutton;
KmaxWidget  currentPassNumLabel;                                         40
KmaxWidget  numOfPassesLabel;
KmaxWidget  lastTransferCountLabel;
KmaxWidget  lastBeamOnProtonCountLabel;
KmaxWidget  lastBeamOffProtonCountLabel;

/** Other widgets on Control Panel*/
KmaxWidget  report;
KmaxHist  histTime;
KmaxHist  histEnergy;
KmaxWidget  startButton;                                                 50
KmaxWidget  stopButton;
KmaxWidget  loadDefaultParamButton;

/** Widgets in the Default Parameters Panel*/
KmaxWidget  defaultBeamOnTimeTextfield;
KmaxWidget  defaultDelayTimeTextfield;
KmaxWidget  defaultBeamOffTimeTextfield;
KmaxWidget  defaultNumOfPassesTextfield;
KmaxWidget  defaultConvGainCombobox;
KmaxWidget  defaultHexReadTypeCombobox;                                  60
KmaxWidget  defaultTimeIntervalCombobox;
KmaxWidget  defaultMinBeamOnProtonCountTextfield;
KmaxWidget  defaultMaxBeamOffProtonCountTextfield;
KmaxWidget  defaultSavePathLabel;
KmaxWidget  defaultSaveDataCheckbox;

/** Global variables for our program*/
final int  SINGLE_READ = 0; // hexReadType = single read
final int  DOUBLE_READ = 1; // hexReadType = double read
                                                                        70
int  beamOnTime;
int  delayTime;
int  beamOffTime;
int  numOfPasses;
int  currentPassNum;
int  convGain;
```

```
int hexReadType; // single or double hexcounter read method for time
int timeInterval; // time interval per pulse into the hexcounter
int minBeamOnProtonCount;
int maxBeamOffProtonCount;                                                    80
boolean saveData;

Timer beamOnTimer;
Timer delayTimer;
Timer beamOffTimer;

File tempEventDataFile;
BufferedWriter tempEventData;

int slotAdc = 2; // the slot of the LeCroy 3511 ADC                          90
int slotHex = 4; // the slot of the KSC 3615 hex counter
int slotSequencer = 7;
int hexTimeChan = 0; // hexcounter channel used for time information
int hexProtonChan = 2; // hexcounter channel used for proton count information
int timeEventID = 10;
int energyEventID = 11;
int numHistTimeChans = 4096;

String defaultFilePath = "/Users/phb/Kmax_Stuff v8.2.2/E-Lifetime parameters file.txt";
String savePath;                                                            100

// The bit patterns to write to the homemade Sequencer
int[] sequencerDisable = {0};
int[] sequencerDataCollect = {(int)Math.pow(2,17)}; //chan 2
int[] sequencerTurnWheelBeamOff = {(int)Math.pow(2,21)}; //chan 6
int[] sequencerTurnWheelBeamOn = {(int)Math.pow(2,22)}; //chan 7

boolean isWheelInBeamOff;
boolean needInitialisation = true;
                                                                            110
/**
* The 'init' method is executed at compile time.
*/
public void init(KmaxToolsheet toolsheet) {

    /** Essential Kmax Objects */
    tlsh = toolsheet; // Save this reference for use in the toolsheet
    dev = tlsh.getKmaxDevice("CMC100");

    /** Widgets in the Parameters sub-pane */                               120
    beamOnTimeTextfield = tlsh.getKmaxWidget("BEAM_ON_TIME");
    delayTimeTextfield = tlsh.getKmaxWidget("DELAY_TIME");
    beamOffTimeTextfield = tlsh.getKmaxWidget("BEAM_OFF_TIME");
    numOfPassesTextfield = tlsh.getKmaxWidget("NUM_OF_PASSES");
    convGainCombobox = tlsh.getKmaxWidget("CONV_GAIN");
    hexReadTypeCombobox = tlsh.getKmaxWidget("HEX_READ_TYPE");
```

```
timeIntervalCombobox = tlsh.getKmaxWidget("TIME_INTERVAL");
minBeamOnProtonCountTextfield = tlsh.getKmaxWidget("MIN_BEAM_ON_PROTON_COUNT");
maxBeamOffProtonCountTextfield = tlsh.getKmaxWidget("MAX_BEAM_OFF_PROTON_COUNT");
saveDataCheckbox = tlsh.getKmaxWidget("SAVE_DATA");                                    130
chooseSavePathButton = tlsh.getKmaxWidget("CHOSE_SAVE_PATH");
savePathLabel = tlsh.getKmaxWidget("SAVE_PATH");

/** Widgets in the Status panel */
beamOnRadiobutton = tlsh.getKmaxWidget("BEAM_ON_STATUS");
delayRadiobutton = tlsh.getKmaxWidget("DELAY_STATUS");
beamOffRadiobutton = tlsh.getKmaxWidget("BEAM_OFF_STATUS");
currentPassNumLabel = tlsh.getKmaxWidget("CURRENT_PASS_STATUS");
numOfPassesLabel = tlsh.getKmaxWidget("NUM_OF_PASSES_STATUS");
lastBeamOnProtonCountLabel = tlsh.getKmaxWidget("LAST_BEAM_ON_PROTON_COUNT");          140
lastTransferCountLabel = tlsh.getKmaxWidget("LAST_TRANSFER_COUNT");
lastBeamOffProtonCountLabel = tlsh.getKmaxWidget("LAST_BEAM_OFF_PROTON_COUNT");

/** Other widgets on Control Panel */
report = tlsh.getKmaxWidget("REPORT");
histTime = (KmaxHist)tlsh.getKmaxWidget("HIST_TIME");
histEnergy = (KmaxHist)tlsh.getKmaxWidget("HIST_ENERGY");
startButton = tlsh.getKmaxWidget("START");
stopButton = tlsh.getKmaxWidget("STOP");
loadDefaultParamButton = tlsh.getKmaxWidget("LOAD_DEFAULT_PARAM");                     150

/** Widgets in the Default Parameters panel*/
defaultBeamOnTimeTextfield = tlsh.getKmaxWidget("DEFAULT_BEAM_ON_TIME");
defaultDelayTimeTextfield = tlsh.getKmaxWidget("DEFAULT_DELAY_TIME");
defaultBeamOffTimeTextfield = tlsh.getKmaxWidget("DEFAULT_BEAM_OFF_TIME");
defaultNumOfPassesTextfield = tlsh.getKmaxWidget("DEFAULT_NUM_OF_PASSES");
defaultConvGainCombobox = tlsh.getKmaxWidget("DEFAULT_CONV_GAIN");
defaultHexReadTypeCombobox = tlsh.getKmaxWidget("DEFAULT_HEX_READ_TYPE");
defaultTimeIntervalCombobox = tlsh.getKmaxWidget("DEFAULT_TIME_INTERVAL");
defaultMinBeamOnProtonCountTextfield = tlsh.getKmaxWidget("DEFAULT_MIN_BEAM_ON_PROTON_ 160
COUNT");
defaultMaxBeamOffProtonCountTextfield = tlsh.getKmaxWidget("DEFAULT_MAX_BEAM_OFF_PROTON_
COUNT");
defaultSaveDataCheckbox = tlsh.getKmaxWidget("DEFAULT_SAVE_DATA");
defaultSavePathLabel = tlsh.getKmaxWidget("DEFAULT_SAVE_PATH");
}

/**
 * The 'GO' method is executed when the GO button is clicked.
 */                                                                                    170
public void GO(KmaxToolsheet toolsheet) {

    /** Clears and setup histograms and report textpane*/
    report.setProperty("TEXT","");

    histTime.setEventID(timeEventID);
```

```
    histEnergy.setEventID(energyEventID);

    histTime.clear();
    histEnergy.clear();                                              180
    histTime.update();
    histEnergy.update();

    /** Default Parameters Panel */
    LOAD_PARAM_FROM_FILE(null);

    /**Parameters sub-pane*/
    LOAD_DEFAULT_PARAM(null);

    /**Status sub-pane*/                                             190
    beamOnRadiobutton.setProperty("VALUE","0");
    delayRadiobutton.setProperty("VALUE","0");
    beamOffRadiobutton.setProperty("VALUE","0");
    currentPassNumLabel.setProperty("LABEL","0");
    numOfPassesLabel.setProperty("LABEL","0");
    lastBeamOnProtonCountLabel.setProperty("LABEL","");
    lastTransferCountLabel.setProperty("LABEL","");
    lastBeamOffProtonCountLabel.setProperty("LABEL","");

    /** Control Panel*/                                              200
    stopButton.setEnabledState(false);

}

/**
 * The 'SRQ' method is executed when a device requests service.
 */
public void SRQ(KmaxDevice device) {
}
                                                                    210
/**
 * The 'HALT' method is executed when the HALT button is clicked.
 */
public void HALT(KmaxToolsheet toolsheet) {
    if(beamOnTimer!=null && delayTimer!=null && beamOffTimer!=null )
        STOP(null);
}

/**
 * The ActionListeners for when the Timer clicks. We have to use slightly different error    220
 * checking since we want the DAQ to stop immediately when an error is encountered and not
 * only when the user clicks OK to the error warning.
 */

/**When Beam On Timer Clicks*/
ActionListener beamOnTimerPerformer = new ActionListener(){
```

```java
public void actionPerformed(ActionEvent evt){

    /**Setup the status sub-pane*/
    beamOnRadiobutton.setProperty("VALUE","0");                              230
    delayRadiobutton.setProperty("VALUE","1");
    beamOffRadiobutton.setProperty("VALUE","0");

    /**Turn wheel to move target to Beam Off position*/
    int error1 = dev.writeInt(slotSequencer,2,16,sequencerTurnWheelBeamOff,0,1);
    int error2 = dev.writeInt(slotSequencer,2,16,sequencerDisable,0,1);
    if((error1 != 0) || (error2 !=0)){
        STOP(null);
        errorCheck(−1,"Error turning wheel away from beam.");
        return;                                                             240
    }
    isWheelInBeamOff=true;

    /**Check the Beam On Proton Count*/
    int[] protonCount = new int[1];
    int error = dev.readInt(slotHex,hexProtonChan,2,protonCount,0,1);
    if(error != 0){
        STOP(null);
        errorCheck(error,"Error reading hex counter in beamOnTimerPerformer.");
        return;                                                             250
    }

    if(protonCount[0] < minBeamOnProtonCount){
        java.awt.Toolkit.getDefaultToolkit().beep(); //Makes the computer beep
        //put a * next to the problematic proton count
        report.setProperty("INSERT", protonCount[0]+"*\t");
    }
    else
        report.setProperty("INSERT", protonCount[0]+"\t");
                                                                            260
    /**Display last Beam On time & proton count - clearing the other fields*/
    lastBeamOnProtonCountLabel.setProperty("LABEL",""+protonCount[0]);
    lastBeamOffProtonCountLabel.setProperty("LABEL","");
    lastTransferCountLabel.setProperty("LABEL","");

    delayTimer.start(); //start the delay Timer
    }
};
                                                                            270
/** When Delay Timer Clicks*/
ActionListener delayTimerPerformer = new ActionListener(){
    public void actionPerformed(ActionEvent evt){
        int error;
        /**Setup the status sub-pane*/
        beamOnRadiobutton.setProperty("VALUE","0");
```

```
        delayRadiobutton.setProperty("VALUE","0");
        beamOffRadiobutton.setProperty("VALUE","1");

        /**enables the DAQ voltage gate from the sequencer*/                    280
        error = dev.writeInt(slotSequencer,2,16,sequencerDataCollect,0,1);
        if(error != 0){
            STOP(null);
            errorCheck(error,"Error enabling Data Collect Voltage on Sequencer.");
            return;
        }

        /**Clearing the hexcounter's proton count and time channels*/
        int error1 = dev.writeInt(slotHex,hexProtonChan,9,null,0,0);
        int error2 = dev.writeInt(slotHex,hexTimeChan,9,null,0,0);             290
        if(error1 != 0 || error2!=0){
            STOP(null);
            errorCheck(error,"Error clearing hex counter in delayTimerPerformer.");
            return;
        }

        /**Start Data acquisition*/
        error = dev.writeInt(slotAdc,0,26,null,0,0); //Enable DAQ - enable LAMs on ADC
        if(error != 0){
            STOP(null);                                                       300
            errorCheck(error,"Error enabling LAMs on ADC in delayTimerPerformer.");
            return;
        }

        /**Start Beam Off Timer*/
        beamOffTimer.start();
    }
};

ActionListener beamOffTimerPerformer = new ActionListener(){             310
    public void actionPerformed(ActionEvent evt){

        int error;

        /**Stop Data Acquisition*/
        error = dev.writeInt(slotAdc,0,24,null,0,0); //disable DAQ - disable LAMs on ADC
        if(error != 0){
            STOP(null);
            errorCheck(error,"Error disabling LAMs on ADC in beamOffTimerPerformer.");
            return;                                                          320
        }

        //disables the DAQ voltage gate
        error = dev.writeInt(slotSequencer,2,16,sequencerDisable,0,1);
        if(error != 0){
            STOP(null);
```

```
        errorCheck(error,"Error disabling all voltages on sequencer.");
        return;
}
                                                                    330
/**Check Proton Count during Beam Off period*/
int[] protonCount = new int[1];
error = dev.readInt(slotHex,hexProtonChan,2,protonCount,0,1);
if(error != 0){
        STOP(null);
        errorCheck(error,"Error reading hex counter in beamOnTimerPerformer.");
        return;
}

if(protonCount[0] > maxBeamOffProtonCount){                          340
        java.awt.Toolkit.getDefaultToolkit().beep(); //Makes the computer beep
        //put a * next the problematic proton count
        report.setProperty("INSERT", protonCount[0]+"*\t");
}
else
        report.setProperty("INSERT", protonCount[0]+"\t");

/**Display the last Beam Off Proton Count*/
lastBeamOffProtonCountLabel.setProperty("LABEL",""+protonCount[0]);

                                                                    350
/**Download and return the contents of the buffer*/
int[] bufferData = downloadBuffer();
if(bufferData == null){
        return;
}

/**Setup the status sub-pane*/
beamOnRadiobutton.setProperty("VALUE","0");
delayRadiobutton.setProperty("VALUE","0");
beamOffRadiobutton.setProperty("VALUE","0");                         360
int lastTransferCount;
if(hexReadType == SINGLE_READ)
        lastTransferCount = bufferData.length/2;
else
        lastTransferCount = bufferData.length/3;

lastTransferCountLabel.setProperty("LABEL",""+lastTransferCount);

/** Write Pass detail to Report window*/
report.setProperty("INSERT",lastTransferCount+"\n");                 370

/**Catch IOExceptions from processData(), write() & creating file commands*/
int numOfAdditionalPasses = -1; //remains -1 if the currentPassNum < numOfPasses
try{
        /**If reached required number of passes*/
```

```
if(currentPassNum >= numOfPasses){

    processData(bufferData,currentPassNum);
```

```
    /**Ask if user wants to have additional passes*/
    java.awt.Toolkit.getDefaultToolkit().beep(); //makes computer beep
    while(true){
        try{
            numOfAdditionalPasses = Integer.parseInt(tlsh.showPrompt("Do you want
            additional passes? If so, enter number (0 = none).","0"));
            if(numOfAdditionalPasses >0){
                numOfPasses = numOfPasses + numOfAdditionalPasses;
                report.setProperty("INSERT","Passes added by the user: "
                +numOfAdditionalPasses+"\n");
```

```
                break;
            }
            if(numOfAdditionalPasses == 0)
                break;
        }
        catch(NumberFormatException e){}
}
/**If none required*/
if(numOfAdditionalPasses == 0){
```

```
    if(saveData)
        tempEventData.write("-99 -99");

    STOP(null);

    /**If save data is clicked, ask the user for a name to save the data to, if
    an empty string is returned signifying do not save, ask to confirm*/
    String saveName;
    if(saveData){
        while(true){
```

```
            saveName = tlsh.showPrompt("Please enter a name to save the data to.
            (empty = don't save)","<Run Name>");
            if(saveName != null){
                if(saveName.equals("")){
                    if(tlsh.showAlert("Are you sure you do not want to save the data?
                    ")==0)
                        return;
                }
                else if(saveName.equals("<Run Name>")){
                }
```

```
                else
                    break;
            }
        }

        /**If the saved name entered already exists, ask user to overwrite or
```

```
enter a different name*/
File reportDataFile;
File eventDataFile;
File timeDataFile;                                                          430
File energyDataFile;
while(true){
    reportDataFile = new File(savePath+File.separator+saveName+"
    -report.txt");
    eventDataFile = new File(savePath+File.separator+saveName+"
    -eventdata.txt");
    timeDataFile = new File(savePath+File.separator+saveName+"
    -timehistogram.txt");
    energyDataFile = new File(savePath+File.separator+saveName+"
    -energyhistogram.txt");                                                 440
    if(reportDataFile.exists() || eventDataFile.exists() ||
    timeDataFile.exists() || energyDataFile.exists())
        if(tlsh.showAlert("Files under the save name '"+saveName+"' already
        exists. Overwrite?")==0){
            reportDataFile.delete();
            eventDataFile.delete();
            break;
        }
        else
            saveName = tlsh.showPrompt("Please enter another name to save   450
            data to.",saveName);
    else
        break;
}

/**Save the report data into a new textfile, to save the event data the
tempEventData file is just renamed, the histogram data is saved using the
export histogram data function within Kmax*/
reportDataFile.createNewFile();
BufferedWriter reportData = new BufferedWriter(new FileWriter             460
(reportDataFile));
reportData.write(report.getProperty("TEXT"));
reportData.close();

if(!tempEventDataFile.renameTo(eventDataFile))
    tlsh.showWarning("'"+saveName+"-eventdata.txt' was not saved properly.
    The event data is still stored in 'temp-eventdata.txt'.
    Please save manually!");

histTime.exportData(savePath+File.separator+saveName+                     470
"-timehistogram.txt",true);
histEnergy.exportData(savePath+File.separator+saveName+
"-energyhistogram.txt",true);
report.setProperty("INSERT","\nData saved successfully under: "+saveName+
"\n");
}
```

```
                return;
            }
        }
        /**Reaches here if still passes required or additional passes requested*/          480
        currentPassNum++;
        /**Update the Status sub-pane*/
        numOfPassesLabel.setProperty("LABEL",""+numOfPasses);
        currentPassNumLabel.setProperty("LABEL",""+currentPassNum);
        beamOnRadiobutton.setProperty("VALUE","1");
        delayRadiobutton.setProperty("VALUE","0");
        beamOffRadiobutton.setProperty("VALUE","0");

        /**Turn Wheel to move target to Beam On Position*/
        int error1 = dev.writeInt(slotSequencer,2,16,sequencerTurnWheelBeamOn,0,1);       490
        int error2 = dev.writeInt(slotSequencer,2,16,sequencerDisable,0,1);
        if((error1 != 0) || (error2 !=0)){
            STOP(null);
            errorCheck(−1,"Error turning wheel into beam.");
            return;
        }
        isWheelInBeamOff = false;

        /**Write Current Pass # to report*/
        report.setProperty("INSERT", currentPassNum+"\t");                                 500

        /**Clear Proton Count*/
        error = dev.writeInt(slotHex,hexProtonChan,9,null,0,0);
        if(error != 0){
            STOP(null);
            errorCheck(error,"Error clearing hex counter in beamOnTimerPerformer.");
            return;
        }

        beamOnTimer.start();                                                               510

        /**process the Data while beamOnTimer is running only if not when additional passes
        were first added.*/
        if(numOfAdditionalPasses == −1) //-1 if currentPassNum>=numOfPasses never entered
            processData(bufferData,currentPassNum−1);
    }
    catch(IOException e){
        tlsh.showWarning("Error output data!");
        STOP(null);
        return;                                                                            520
    }
```

```
    }
};

/**
```

```
* The START Button - starts a run with the current parameters. All parameters are checked
*and CAMAC system reinitialised before every start
*/
public void START(KmaxWidget widget){                                        530

    /** Set all buttons & textfield to suitable state no accidential clicks*/
    stopButton.setEnabledState(true); //enable the Stop button & disable the rest
    startButton.setEnabledState(false);
    beamOnTimeTextfield.setEnabledState(false);
    delayTimeTextfield.setEnabledState(false);
    beamOffTimeTextfield.setEnabledState(false);
    numOfPassesTextfield.setEnabledState(false);
    convGainCombobox.setEnabledState(false);
    hexReadTypeCombobox.setEnabledState(false);                              540
    timeIntervalCombobox.setEnabledState(false);
    minBeamOnProtonCountTextfield.setEnabledState(false);
    maxBeamOffProtonCountTextfield.setEnabledState(false);
    saveDataCheckbox.setEnabledState(false);
    chooseSavePathButton.setEnabledState(false);
    loadDefaultParamButton.setEnabledState(false);

    /**Clear histogram*/
    histTime.clear();
    histEnergy.clear();                                                      550
    histTime.update();
    histEnergy.update();

    /**Check Parameters & Initialise System*/
    if(!assignAndCheckParam()){ //assign and check parameters (also sets up the Timers)
        startButton.setEnabledState(true); //if error in parameters
        stopButton.setEnabledState(false);

        beamOnTimeTextfield.setEnabledState(true);
        delayTimeTextfield.setEnabledState(true);                           560
        beamOffTimeTextfield.setEnabledState(true);
        numOfPassesTextfield.setEnabledState(true);
        convGainCombobox.setEnabledState(true);
        hexReadTypeCombobox.setEnabledState(true);
        timeIntervalCombobox.setEnabledState(true);
        minBeamOnProtonCountTextfield.setEnabledState(true);
        maxBeamOffProtonCountTextfield.setEnabledState(true);
        saveDataCheckbox.setEnabledState(true);
        loadDefaultParamButton.setEnabledState(true);
        chooseSavePathButton.setEnabledState(true);                         570
        return;
    }
    if(!initialise()){ //initialise() returns true if successful
        STOP(null);
        return;
    }
```

```
/** PROMPT TO CONFIRM PARAMETERS*/
String paramString = beamOnTime+"ms\tBEAM ON TIME \n"+delayTime+"ms\tDELAY TIME\n"+beamOffTime
+"ms\tBEAM OFF TIME\n"+numOfPasses+"\tPASSES PER RUN\n"+convGain+"chans\tCONV GAIN\n"; 580
if(hexReadType == SINGLE_READ)
    paramString = paramString + "Single\tHEX READS\n";
else
    paramString = paramString + "Double\tHEX READS\n";
paramString = paramString+timeInterval+"ms\tTIME INTERVAL ON PULSER BOX\n"+minBeamOnProtonCount+
"\tMIN BEAM ON PROTON COUNT\n"+maxBeamOffProtonCount+"\tMAX BEAM OFF PROTON COUNT\n";
if(saveData)
    paramString = paramString + "yes\tSAVE DATA\n";
else
    paramString = paramString + "no\tSAVE DATA\n";                                      590

if(tlsh.showAlert(paramString+"\nACCEPT THESE PARAMETERS AND START?")==2){
//returns 2 if cancel is clicked, 0 if OK is clicked
    STOP(null);
    return;
}

/**Create the temp file for storing Event Data*/
try{
    if(saveData){                                                                        600
        tempEventDataFile = new File(savePath+File.separator+"Temp-eventdata.txt");
        if(tempEventDataFile.exists())
            tempEventDataFile.delete();
        tempEventDataFile.createNewFile();

        tempEventData = new BufferedWriter(new FileWriter(tempEventDataFile));
    }
}
catch(IOException e){
    tlsh.showWarning("Error creating temp file for event data at: "+               610
    tempEventDataFile.getAbsolutePath());
    STOP(null);
    return;
}

/**Write Run Details to Report Window*/
report.setProperty("TEXT",""); //clears the report window before every run
//write the current time & date
Date nowDate = new Date();
String currentTime = DateFormat.getTimeInstance().format(nowDate);                  620
String currentDate = DateFormat.getDateInstance().format(nowDate);
report.setProperty("INSERT",currentTime+", "+currentDate+"\n\n");
//Parameters used for the run
report.setProperty("INSERT",paramString+"\n");
report.setProperty("INSERT","PASS\tBM_ON_P\tBM_OFF_P\tNUM_EVENTS\n\n");
```

```
currentPassNum = 1; //A pass is defined to begin at the start of the Beam On Period

/**Set up status sub-pane*/
beamOnRadiobutton.setProperty("VALUE","1");                                           630
delayRadiobutton.setProperty("VALUE","0");
beamOffRadiobutton.setProperty("VALUE","0");
numOfPassesLabel.setProperty("LABEL",""+numOfPasses);
currentPassNumLabel.setProperty("LABEL",""+currentPassNum);

/**Write Current Pass # to report*/
report.setProperty("INSERT", currentPassNum+"\t");

/**Clear Proton Count*/
int error = dev.writeInt(slotHex,hexProtonChan,9,null,0,0);                           640
if(error != 0){
    STOP(null);
    errorCheck(error,"Error clearing hex counter in beamOnTimerPerformer.");
    return;
}

/**Turn Wheel move target to Beam On*/
int error1 = dev.writeInt(slotSequencer,2,16,sequencerTurnWheelBeamOn,0,1);
int error2 = dev.writeInt(slotSequencer,2,16,sequencerDisable,0,1);
if((error1 != 0) || (error2 !=0)){                                                    650
    STOP(null);
    errorCheck(−1,"Error turning wheel into beam.");
    return;
}
isWheelInBeamOff = false;

/**Start Beam on Timer*/
beamOnTimer.start();
}
                                                                                     660
/**
* The STOP button - Can only be pressed when going through a run (otherwise disabled)
*/
public void STOP(KmaxWidget widget){
    /**Stop all Timers*/
    beamOnTimer.stop();
    delayTimer.stop();
    beamOffTimer.stop();

    /**Inhibit CAMAC & clear response buffer*/                                        670
    int error = dev.writeInt(30,9,26,null,0,0); //CAMAC Inhibit
    error = dev.readInt(27,2,0,new int[1],0,1); //reading a value from buffer will clear it

    /**Move Wheel to Beam Off position if required*/
    if(isWheelInBeamOff == false){
        int error1 = dev.writeInt(slotSequencer,2,16,sequencerTurnWheelBeamOff,0,1);
```

```
        int error2 = dev.writeInt(slotSequencer,2,16,sequencerDisable,0,1);
        if((error1 != 0) || (error2 !=0)){
            STOP(null);
            errorCheck(−1,"Error turning wheel away from beam.");          680
            return;
        }
        isWheelInBeamOff=true;
    }

    /**Close the temp Event Data Writer if required*/
    try{
        if(saveData && tempEventData!=null){
            tempEventData.close();
        }                                                                  690
    }
    catch(IOException e){
    }

    /** reset the buttons interface to Stopped Mode*/
    startButton.setEnabledState(true);
    stopButton.setEnabledState(false);
    beamOnTimeTextfield.setEnabledState(true);
    delayTimeTextfield.setEnabledState(true);
    beamOffTimeTextfield.setEnabledState(true);                            700
    numOfPassesTextfield.setEnabledState(true);
    convGainCombobox.setEnabledState(true);
    hexReadTypeCombobox.setEnabledState(true);
    timeIntervalCombobox.setEnabledState(true);
    minBeamOnProtonCountTextfield.setEnabledState(true);
    maxBeamOffProtonCountTextfield.setEnabledState(true);
    saveDataCheckbox.setEnabledState(true);
    loadDefaultParamButton.setEnabledState(true);
    chooseSavePathButton.setEnabledState(true);
}                                                                          710

/**
 * assignAndCheckParam() method assigns the global variables to the values in the parameters
 * specified in the 'Parameters' sub-pane it also checks whether the entered parameters are
 * valid for use. If an error is detected a prompt with details is displayed to the user and
 * FALSE is returned if no error detected then 'paramErrorString' is empty, TRUE is returned.
 */
public boolean assignAndCheckParam(){

    String paramErrorString = ""; //The String contain error messages to be returned  720

    /**Check parameters from the parameters sub-pane are valid by inserting error messages*/
    try{
        if((beamOnTime = Integer.parseInt(beamOnTimeTextfield.getProperty("TEXT"))) < 0)
            paramErrorString = paramErrorString + "-ve Beam On Time, ";
        else if(beamOnTime < 500) // Check to pick up mistakes of using seconds and not ms
```

```
        paramErrorString = paramErrorString + "Beam on Time too short";
    if((delayTime = Integer.parseInt(delayTimeTextfield.getProperty("TEXT"))) < 0)
        paramErrorString = paramErrorString + "-ve delayTime, ";
    if((beamOffTime = Integer.parseInt(beamOffTimeTextfield.getProperty("TEXT"))) < 0)    730
        paramErrorString = paramErrorString + "-ve Beam Off Time, ";
    else if(beamOffTime < 500) // Check to pick up mistakes of using seconds and not ms
        paramErrorString = paramErrorString + "Beam Off Time too short, ";
    // Do not want -ve or zero number of passes
    if((numOfPasses = Integer.parseInt(numOfPassesTextfield.getProperty("TEXT"))) <= 0)
        paramErrorString = paramErrorString + "-ve or 0 Number of Passes, ";

    convGain = Integer.parseInt(convGainCombobox.getProperty("ITEM"));
    timeInterval = Integer.parseInt(timeIntervalCombobox.getProperty("ITEM"));
    minBeamOnProtonCount = Integer.parseInt(minBeamOnProtonCountTextfield.getProperty("TEXT"));   740
    maxBeamOffProtonCount = Integer.parseInt(maxBeamOffProtonCountTextfield.getProperty("TEXT"));

    if(saveDataCheckbox.getProperty("VALUE").equals("1"))
        saveData = true;
    else
        saveData = false;
    savePath = savePathLabel.getProperty("LABEL");
    File saveDirectory = new File(savePath);
    if(!saveDirectory.exists())                                                                    750
        paramErrorString = paramErrorString + "The Output Files Save Path does not exist!
        Please create a directory, ";

}
catch(NumberFormatException e){
// This exception is thrown when using Integer.parseInt on a string containing characters
    paramErrorString = paramErrorString + "Only numbers valid in time, number of passes &
    proton count fields, ";
}
                                                                                                   760
if(hexReadTypeCombobox.getProperty("ITEM").equals("Single"))
    hexReadType = SINGLE_READ;
else if(hexReadTypeCombobox.getProperty("ITEM").equals("Double"))
    hexReadType = DOUBLE_READ;
else
    paramErrorString = paramErrorString + "Hex Read Type invalid, ";

if(beamOffTime/timeInterval > numHistTimeChans){
    paramErrorString = paramErrorString + "Currently require "+beamOffTime/timeInterval+
    " time bins. Histogram only supports "+numHistTimeChans+". ";                                  770
    paramErrorString = paramErrorString + "Please lower beam off time or raise time interval.";

}

/**Handling of Parameter Error String*/
if(!paramErrorString.equals("")){
```

```
        tlsh.showWarning("Invalid Parameters: "+paramErrorString);
        return false;
    }
```

```
    /** Setup the Timers with new Times*/
    beamOnTimer = new Timer(beamOnTime,beamOnTimerPerformer);
    beamOnTimer.setRepeats(false);
    delayTimer = new Timer(delayTime,delayTimerPerformer);
    delayTimer.setRepeats(false);
    beamOffTimer = new Timer(beamOffTime,beamOffTimerPerformer);
    beamOffTimer.setRepeats(false);
```

```
    /**Setup Histogram display ranges*/
```

```
    //display from 0 to beamOffTime + 10%
    int[] histTimeDispRange = {0,(int)(Math.ceil(1.1*beamOffTime/timeInterval)),0,0};
    histTime.setRanges(histTimeDispRange);
    int[] histEnergyDispRange = {0,convGain,0,0}; //set the ranges for the energy histogram
    histEnergy.setRanges(histEnergyDispRange);
```

```
    return true;
}
```

```
/**
 * Initialise all the CAMAC Electronics
 */
public boolean initialise(){
```

```
    /** Initialise CAMAC*/
    int error;
```

```
    error = dev.writeInt(28, 8, 26, null, 0, 0); // Initialises the CAMAC (Z)
    if(errorCheck(error,"Error after initialising CAMAC.")) return false;
```

```
    error = dev.writeInt(28, 9, 26, null, 0, 0); // CAMAC clear
    if(errorCheck(error, "Error after CAMAC clear.")) return false;
```

```
    error = dev.writeInt(30,9,26,null,0,0); //CAMAC Inhibit
    if(errorCheck(error, "Error after CAMAC Inhibit.")) return false;
```

```
    int[] cmc100ControlRegister = {4}; //Enable LAMs as List trigger with bit pattern '0100'
    error = dev.writeInt(30,0,17,cmc100ControlRegister,0,1); //write to control register
    if(errorCheck(error,"Error writing to control register of CMC100.")) return false;
```

```
    /** Write list of commands to List Sequencer*/
    if(!writeList()) return false;
```

```
    /** Initialise the ADC */
    int[] adcControlRegister = new int[1];
    //word to be written to ADC to begin acquisition & sets conversion gain
```

```
switch(convGain){
    case 256:
        adcControlRegister[0] = Integer.parseInt("0E00",16);
        break;                                                          830
    case 512:
        adcControlRegister[0] = Integer.parseInt("0500",16);
        break;
    case 1024:
        adcControlRegister[0] = Integer.parseInt("0400",16);
        break;
    case 2048:
        adcControlRegister[0] = Integer.parseInt("0300",16);
        break;
    case 4096:                                                          840
        adcControlRegister[0] = Integer.parseInt("0200",16);
        break;
    default:
        tlsh.showWarning("ADC Conv Gain setting not valid.");
        return false;
}

error = dev.writeInt(slotAdc,0,16,adcControlRegister,0,1); //write to ADC control register
if(errorCheck(error,"Error writing control word to begin acquisition.")) return false;
                                                                        850
int[] lamMask = { 1 << (slotAdc − 1) }; //only enable LAM on the CMC100 for the ADC
error = dev.writeInt(30,0,16,lamMask, 0, 1); // writes and enables the above Lam Mask
if(errorCheck(error,"Error writing Lam Mask to enable LAMs on CMC100 for the ADC"))
return false;

/** Have everything ready for Data Acqusition to begin when ADC LAM become enabled */

//disable LAM on ADC. This is the mechanism for DAQ switch.
error = dev.writeInt(slotAdc,0,24,null,0,0);
if(errorCheck(error,"Error disabling LAM on the ADC")) return false;   860

// Clear all voltages on Sequencer
error = dev.writeInt(slotSequencer,2,16,sequencerDisable,0,1);
if(errorCheck(error,"Error disabling all voltages on sequencer.")) return false;

//remove CAMAC Inhibit. Nothing will happen Since DAQ switch mechanism not enabled.
error = dev.writeInt(30,9,24,null,0,0);
if(errorCheck(error,"Error removing CAMAC Inhibit")) return false;

return true;                                                           870

}

/**
 * Separate method to write the commands into the list sequencer. It is only called within
 * initialise(), and it's mainly there to make the code look tidier.
```

```
*/
public boolean writeList(){

    int error;                                                                          880
    int lengthOfList; // the number of commands or length of the list to be written.
    int[] listArray; // the array containing the commands

    int[] data = {0};
    error = dev.writeInt(99,0,4,data,0,1); // write current list address to 0
    if(errorCheck(error,"Error writing list address.")) return false;

    /** The list if single read of hexcounter requested*/
    if(hexReadType == SINGLE_READ){
        lengthOfList = 5;                                                               890
        listArray = new int[lengthOfList];

        listArray[0] = 0x08000002; // On NIM Trigger, jump to location 2
        listArray[1] = 0x08000002; // On LAM Trigger, jump to location 2
        listArray[2] = (slotHex << 9) + (0 << 4) + hexTimeChan; //read hexcounter (response)
        listArray[3] = (slotAdc << 9) + (2 << 4) + 0; //read value & clear LAM on ADC (response)
        listArray[4] = 0x1F000000; // Quit Program Mode
    }
    /** The list if double read of hexcounter requested*/
    else if(hexReadType == DOUBLE_READ){                                               900
        lengthOfList = 6;
        listArray = new int[lengthOfList];

        listArray[0] = 0x08000002; // On NIM Trigger, jump to location 2
        listArray[1] = 0x08000002; // On LAM Trigger, jump to location 2
        listArray[2] = (slotHex << 9) + (0 << 4) + hexTimeChan; //read hexcounter (response)
        listArray[3] = (slotHex << 9) + (0 << 4) + hexTimeChan; //read hexcounter (response)
        listArray[4] = (slotAdc << 9) + (2 << 4) + 0; //read & clear LAM on ADC (response)
        listArray[5] = 0x1F000000; // Quit Program Mode
    }                                                                                   910
    else{
        tlsh.showWarning("Hexcounter read type not valid.");
        return false;
    }

    error = dev.writeInt(27,0,16,listArray, 0,lengthOfList); //write the list
    if(errorCheck(error, "Error writing List!")) return false;

    return true;
}                                                                                       920

/**
 * Returns an array containing only the FIFO/Response buffer contents. Non-useful values are
 * not returned.
 */
```

```
public int[] downloadBuffer(){
    int error;
    int[] FIFOcount = {0}; //Temporary Array for storing the FIFO/buffer Count
    error = dev.readInt(27,2,1,FIFOcount,0,1); //read the FIFO count                    930
    if(errorCheck(error,"Error reading the FIFO count")){
        STOP(null);
        return null;
    }
    int bufferCount = FIFOcount[0];
    int[] bufferData= new int[bufferCount];

    /** Buffer full check*/
    if(bufferCount>950000){
        tlsh.showWarning("Response Buffer full! Must halt");                             940
        STOP(null);
        return null;
    }

    error = dev.readInt(27,2,0,bufferData,0,bufferCount); //read the buffer
    //error produced when amount of data transferred above is less than length of bufferData
    if(error == 2013){
        int[] FIFOxferCount = {0};
        error = dev.readInt(99,0,5,FIFOxferCount,0,1);
        if(errorCheck(error,"Error reading the FIFO xfer count")){                        950
            STOP(null);
            return null;
        }
        int[] bufferDataTemp = bufferData;
        bufferCount = FIFOxferCount[0]−1; //bufferCount now contains the amount of useful data
        bufferData = new int[bufferCount];
        System.arraycopy(bufferDataTemp,0,bufferData,0,bufferCount); //truncates to useful data
    }
    else if(errorCheck(error,"Error reading buffer data")){
        STOP(null);                                                                      960
        return null;
    }
    return bufferData;
}

/**
* This method processes the data
*/
public void processData (int[] bufferData, int currentPassNumToOutput) throws IOException{
                                                                                         970
    int result;
    int[] timeData;
    int[] energyData;
    if(hexReadType == SINGLE_READ){
        if(bufferData.length%2 != 0){
            tlsh.showWarning("Buffer Contents not even in size from single hex read!");
```

```
            STOP(null);
            return;
        }
        timeData = new int[bufferData.length/2]; // 2 values per event          980
        energyData = new int[bufferData.length/2];
        for(int i=0; i < timeData.length; i++){
            timeData[i] = bufferData[2*i];
            energyData[i] = bufferData[2*i+1];
        }
    }
    else if(hexReadType == DOUBLE_READ){
        if(bufferData.length%3 != 0){
            tlsh.showWarning("Buffer Contents size not divisble by 3 from double hex read!");
            STOP(null);                                                          990
            return;
        }
        timeData = new int[bufferData.length/3]; // 3 values per event
        energyData = new int[bufferData.length/3];
        for(int i=0; i < timeData.length; i++){
            if(bufferData[3*i+1] > (bufferData[3*i] + 1))
                timeData[i] = bufferData[3*i+1]−1;
            else
                timeData[i] = bufferData[3*i];
            energyData[i] = bufferData[3*i+2];                                   1000
        }
    }
    else{
        tlsh.showWarning("Hexcounter read type invalid!");
        STOP(null);
        return;
    }

    result = tlsh.addEventBlock(tlsh.PRIMARY_BUFFER,timeEventID,1,timeData.length,timeData);  1010
    result = tlsh.sortEventBuffer(tlsh.PRIMARY_BUFFER); // update the histograms
    result = tlsh.clearEventBuffer(tlsh.PRIMARY_BUFFER); //clear the Event buffer
    histTime.update();

    result = tlsh.addEventBlock(tlsh.PRIMARY_BUFFER,energyEventID,1,energyData.length,energyData);
    result = tlsh.sortEventBuffer(tlsh.PRIMARY_BUFFER); // update the histograms
    result = tlsh.clearEventBuffer(tlsh.PRIMARY_BUFFER); //clear the Event buffer
    histEnergy.update();

    /**Write Event Mode Data to the temp Event Data File*/                       1020
    if(saveData){
        tempEventData.write("-1 "+currentPassNumToOutput);
        tempEventData.newLine();
        for(int i=0;i<energyData.length;i++){
            tempEventData.write(energyData[i]+" "+timeData[i]);
            tempEventData.newLine();
```

```
            }
        }

    }                                                                    1030

    /**
    * Returns TRUE if error found and FALSE if no error. A prompt with appropiate error/success
    * message is then displayed to the user.
    */
    public boolean errorCheck(int error, String errorOutput){
        if (error != 0){
            tlsh.showWarning(errorOutput+" [msg = " + dev.getErrorMessage(error) + "]");
            return true;
        }                                                                1040
        return false;
    }

    /**
    * Loads the 'Current Default Parameters' on the 'Default Parameters Panel' from the file
    * specified by 'defaultFilePath' String
    */
    public void LOAD_PARAM_FROM_FILE(KmaxWidget widget){
        try{
            BufferedReader defaultFileBReader = new BufferedReader(new FileReader(defaultFilePath)); 1050

            int lineCount = 0;
            String line;

            /**Reads the defaultFile until it is empty. Certain line numbers correspond to
            parameters which are written to the current default parameters.*/
            while((line = defaultFileBReader.readLine()) != null){
                lineCount++;
                if(lineCount == 2)
                    defaultConvGainCombobox.setProperty("ITEM",line);            1060
                if(lineCount == 4)
                    defaultBeamOnTimeTextfield.setProperty("TEXT",line);
                if(lineCount == 6)
                    defaultDelayTimeTextfield.setProperty("TEXT",line);
                if(lineCount == 8)
                    defaultBeamOffTimeTextfield.setProperty("TEXT",line);
                if(lineCount == 10)
                    defaultNumOfPassesTextfield.setProperty("TEXT",line);
                if(lineCount == 12)
                    defaultHexReadTypeCombobox.setProperty("ITEM",line);        1070
                if(lineCount == 14)
                    defaultTimeIntervalCombobox.setProperty("ITEM",line);
                if(lineCount == 16)
                    defaultMinBeamOnProtonCountTextfield.setProperty("TEXT",line);
                if(lineCount == 18)
                    defaultMaxBeamOffProtonCountTextfield.setProperty("TEXT",line);
```

```
            if(lineCount == 20)
                defaultSaveDataCheckbox.setProperty("VALUE",line);
            if(lineCount == 22)
                defaultSavePathLabel.setProperty("LABEL",line);                    1080
        }
        if(lineCount == 0){
            tlsh.showWarning("Empty Default Parameters File at: "+defaultFilePath+
            ". Creating new file...\n");
            SAVE_PARAM_TO_FILE(null);
        }
        if(lineCount != 22)
            tlsh.showWarning("Invalid Number of lines for a Default Parameters File at: "
            +defaultFilePath+". \n");

                                                                                   1090
        defaultFileBReader.close();
    }
    catch(IOException e){
        tlsh.showWarning("Error reading Current Default Parameters from "+defaultFilePath+
        ". \n\n Please Choose another...");
        defaultFilePath = tlsh.showFileChooser("Choose Default File Path","txt","OK");
        LOAD_PARAM_FROM_FILE(null);
        return;
    }
                                                                                   1100
}

/**
 * Saves the 'Current Default Parameters' on the 'Default Parameters Panel' to the file
 * specified by 'defaultFilePath' String
 */
public void SAVE_PARAM_TO_FILE(KmaxWidget widget){

    File defaultFile = new File(defaultFilePath);
                                                                                   1110
    try{
        defaultFile.delete(); //clears the existing defaultFile by creating a new one
        defaultFile.createNewFile();

        BufferedWriter defaultFileBWriter = new BufferedWriter(new FileWriter(defaultFile));

        /** Writes the parameters in the current parameters panel to the file */
        defaultFileBWriter.write("DEFAULT_CONV_GAIN");
        defaultFileBWriter.newLine();
        defaultFileBWriter.write(defaultConvGainCombobox.getProperty("ITEM"));         1120
        defaultFileBWriter.newLine();

        defaultFileBWriter.write("DEFAULT_BEAM_ON_TIME");
        defaultFileBWriter.newLine();
        defaultFileBWriter.write(defaultBeamOnTimeTextfield.getProperty("TEXT"));
        defaultFileBWriter.newLine();
```

```
defaultFileBWriter.write("DEFAULT_DELAY_TIME");
defaultFileBWriter.newLine();
defaultFileBWriter.write(defaultDelayTimeTextfield.getProperty("TEXT"));          1130
defaultFileBWriter.newLine();

defaultFileBWriter.write("DEFAULT_BEAM_OFF_TIME");
defaultFileBWriter.newLine();
defaultFileBWriter.write(defaultBeamOffTimeTextfield.getProperty("TEXT"));
defaultFileBWriter.newLine();

defaultFileBWriter.write("DEFAULT_NUM_OF_PASSES");
defaultFileBWriter.newLine();
defaultFileBWriter.write(defaultNumOfPassesTextfield.getProperty("TEXT"));          1140
defaultFileBWriter.newLine();

defaultFileBWriter.write("DEFAULT_HEX_READ_TYPE");
defaultFileBWriter.newLine();
defaultFileBWriter.write(defaultHexReadTypeCombobox.getProperty("ITEM"));
defaultFileBWriter.newLine();

defaultFileBWriter.write("DEFAULT_TIME_INTERVAL");
defaultFileBWriter.newLine();
defaultFileBWriter.write(defaultTimeIntervalCombobox.getProperty("ITEM"));          1150
defaultFileBWriter.newLine();

defaultFileBWriter.write("DEFAULT_MIN_BEAM_ON_PROTON_COUNT");
defaultFileBWriter.newLine();
defaultFileBWriter.write(defaultMinBeamOnProtonCountTextfield.getProperty("TEXT"));
defaultFileBWriter.newLine();

defaultFileBWriter.write("DEFAULT_MAX_BEAM_OFF_PROTON_COUNT");
defaultFileBWriter.newLine();
defaultFileBWriter.write(defaultMaxBeamOffProtonCountTextfield.getProperty("TEXT"));          1160
defaultFileBWriter.newLine();

defaultFileBWriter.write("DEFAULT_SAVE_DATA");
defaultFileBWriter.newLine();
defaultFileBWriter.write(defaultSaveDataCheckbox.getProperty("VALUE"));
defaultFileBWriter.newLine();

defaultFileBWriter.write("DEFAULT_SAVE_PATH");
defaultFileBWriter.newLine();
defaultFileBWriter.write(defaultSavePathLabel.getProperty("LABEL"));          1170

defaultFileBWriter.close();
}
catch(IOException e){
    tlsh.showWarning("Error writing Current Default Parameters to "+defaultFilePath+". \n");
    return;
```

```
        }
}

/**                                                                            1180
 * Loads the parameters in the 'Current Default Parameters' sub-pane into the 'Parameters'
 * sub-pane in the Control Panel
 */
public void LOAD_DEFAULT_PARAM(KmaxWidget widget){

    convGainCombobox.setProperty("ITEM",defaultConvGainCombobox.getProperty("ITEM"));
    beamOnTimeTextfield.setProperty("TEXT",defaultBeamOnTimeTextfield.getProperty("TEXT"));
    delayTimeTextfield.setProperty("TEXT",defaultDelayTimeTextfield.getProperty("TEXT"));
    beamOffTimeTextfield.setProperty("TEXT",defaultBeamOffTimeTextfield.getProperty("TEXT"));
    numOfPassesTextfield.setProperty("TEXT",defaultNumOfPassesTextfield.getProperty("TEXT"));   1190
    hexReadTypeCombobox.setProperty("ITEM",defaultHexReadTypeCombobox.getProperty("ITEM"));
    saveDataCheckbox.setProperty("VALUE",defaultSaveDataCheckbox.getProperty("VALUE"));
    timeIntervalCombobox.setProperty("ITEM",defaultTimeIntervalCombobox.getProperty("ITEM"));
    minBeamOnProtonCountTextfield.setProperty("TEXT",defaultMinBeamOnProtonCountTextfield
    .getProperty("TEXT"));
    maxBeamOffProtonCountTextfield.setProperty("TEXT",defaultMaxBeamOffProtonCountTextfield
    .getProperty("TEXT"));
    savePathLabel.setProperty("LABEL",defaultSavePathLabel.getProperty("LABEL"));
}
/**                                                                            1200
 * This Choose Save Path button in the Default Parameters Pane. Displays a choose File Prompt
 * for directories/folders. The user selected Directory's path is then set as the default
 * save path label
 */
public void CHOOSE_DEFAULT_SAVE_PATH(KmaxWidget widget){

    JFileChooser fc = new JFileChooser();
    fc.setFileSelectionMode(JFileChooser.DIRECTORIES_ONLY);
    if(fc.showOpenDialog(null) == JFileChooser.APPROVE_OPTION)
        defaultSavePathLabel.setProperty("LABEL",fc.getSelectedFile().getAbsolutePath());    1210
    else
        return;

}

/**
 * The Choose Save Path button in the Parameters sub-pane. Sets the save path label to the
 * user selected directory
 */
public void CHOOSE_SAVE_PATH(KmaxWidget widget){                               1220

    JFileChooser fc = new JFileChooser();
    fc.setFileSelectionMode(JFileChooser.DIRECTORIES_ONLY);
    if(fc.showOpenDialog(null) == JFileChooser.APPROVE_OPTION)
        savePathLabel.setProperty("LABEL",fc.getSelectedFile().getAbsolutePath());
    else
```

```
        return;
}

/**                                                                    1230
 * Stops the user from changing the state of the status sub-pane Radiobuttons. When a user
 * clicks on a radiobutton the state changes, but this method is then called which
 * immediately changes it back again. We do this trick because if we use .setEnabled(false),
 * the widget becomes gray and we do not see it as clearly.
 */
public void BEAM_ON_STATUS(KmaxWidget widget){
    if(beamOnRadiobutton.getProperty("VALUE").equals("1"))
        beamOnRadiobutton.setProperty("VALUE","0");
    else
        beamOnRadiobutton.setProperty("VALUE","1");                    1240
}
public void DELAY_STATUS(KmaxWidget widget){
    if(delayRadiobutton.getProperty("VALUE").equals("1"))
        delayRadiobutton.setProperty("VALUE","0");
    else
        delayRadiobutton.setProperty("VALUE","1");
}
public void BEAM_OFF_STATUS(KmaxWidget widget){
    if(beamOffRadiobutton.getProperty("VALUE").equals("1"))
        beamOffRadiobutton.setProperty("VALUE","0");                   1250
    else
        beamOffRadiobutton.setProperty("VALUE","1");
}

/**
 * Functions of the ROI features
 */
public void INTEGRAL_ALL(KmaxWidget widget){
    report.setProperty("INSERT","\nIntegral of whole spectrum = "+histEnergy.getTotal()+"\n");
                                                                        1260
}

public void INTEGRAL_REGION(KmaxWidget widget){
    report.setProperty("INSERT","\nIntegral of selected region = "
    +histEnergy.getRegionSum(histEnergy.getSelectedRegion())+"\n");
}
} // End of the Runtime object
```